\documentclass[a4paper,11pt,oneside]{book}
\usepackage{hyperref}
\usepackage[T1]{fontenc}
\usepackage[utf8]{inputenc}
\usepackage[english]{babel}
\usepackage{color}
\usepackage{amsmath,amscd,amsbsy,amssymb,url,epstopdf}
\usepackage{graphicx}
\usepackage{enumitem}
\usepackage{tabulary}
\usepackage{tikz}
\usepackage{makeidx}  
\usepackage{array}
\usepackage{breakcites}
\newcolumntype{N}{@{}m{0pt}@{}}

\usetikzlibrary{arrows}
\usetikzlibrary{fit}

\newcommand{\bs}{\boldsymbol}

\newcommand{\bx}{\bs{x}}
\newcommand{\bidx}{b_{\bx}}
\newcommand{\ctr}{r}
\DeclareMathOperator*{\var}{Var}

\DeclareMathOperator*{\argmin}{arg \, min}

\newcommand{\bt}{\bs{\theta}}

\newcommand{\citep}{\cite}

\newcommand{\tweb}{\text{c}} 
\newcommand{\tads}{\text{r}} 

\makeatletter
\renewcommand{\@chapapp}{}

\makeatother

\title{\Huge \textbf{Display Advertising with Real-Time Bidding (RTB) and Behavioural Targeting}}
 \author{
	\textsc{Jun Wang}\\
	University College London\\
	\textsf{j.wang@cs.ucl.ac.uk}\\ \\
	\textsc{Weinan Zhang}\\
	Shanghai Jiao Tong University\\
	\textsf{wnzhang@sjtu.edu.cn}\\ \\
	\textsc{Shuai Yuan}\\
	MediaGamma Ltd\\
	\textsf{shuai.yuan@mediagamma.com}
}

\begin{document}

\frontmatter
\maketitle


\tableofcontents

\mainmatter

\chapter*{Abstract}

  The most significant progress
  in recent years in online display advertising is  what is known as the Real-Time Bidding (RTB) mechanism to buy and sell ads. RTB essentially facilitates buying an individual ad impression in real time while it is still being generated from a user's visit. RTB not only scales up the buying process by aggregating a large amount of available
  inventories across publishers but, most importantly, enables direct targeting of individual users. As such, RTB has fundamentally changed the landscape of digital marketing.
  Scientifically, the demand for automation, integration and optimisation in RTB also brings new
  research opportunities in information retrieval, data mining,
  machine learning and other related fields. In this monograph, an overview is given of the fundamental infrastructure, algorithms, and
  technical solutions of this new frontier of computational
  advertising. The covered topics include user response prediction, bid landscape forecasting, bidding algorithms, revenue optimisation, statistical
  arbitrage, dynamic pricing, and ad fraud detection.

\chapter{Introduction}
\label{c-intro} 
An advertisement is a marketing message intended to encourage potential customers to purchase a product or to subscribe to a service. Advertising is also a way to establish a brand image through the repeated presence of an advertisement (ad) associated with the brand in the media. Television, radio, newspaper, magazines, and billboards are among the major channels that traditionally place ads, however, the advancement of the Internet enables users to seek information online. Using the Internet, users are able to express their information requests, navigate specific websites and perform e-commerce transactions. Major search engines have continued to improve their retrieval services and users' browsing experience by providing relevant results. Since many more businesses and services are transitioning into the online space, the Internet is a natural choice for advertisers to widen their strategy, reaching potential customers among Web users \citep{yuan2012internet}.

As a result, online advertising is now one of the fastest advancing areas in the IT industry. In display and mobile advertising, the most significant technical development in recent years is the growth of Real-Time Bidding (RTB), which facilitates a real-time auction for a display opportunity. Real-time means the auction is per impression and the process  usually occurs less than 100 milliseconds before the ad is placed. RTB has fundamentally changed the landscape of the digital media market by scaling the buying process across a large number of available inventories among publishers in an automatic fashion. It also encourages \emph{user behaviour targeting}, a significant shift towards buying focused on user data rather than contextual data \citep{Yuan:2013:RBO:2501040.2501980}.
 
 
Scientifically, the further demand for automation, integration and optimisation in RTB opens new research opportunities in the fields such as Information Retrieval (IR), Data Mining (DM), Machine Learning (ML), and Economics. IR researchers, for example, are facing the challenge of defining the relevancy of underlying audiences given a campaign goal, and consequently, developing techniques to find and filter them out in the real-time bid request data stream \citep{zhang2016implicit,perlich2012bid}. For data miners, a fundamental task is identifying repeated patterns over the large-scale streaming data of bid requests, winning bids and ad impressions \citep{cui2011bid}.  For machine learners, an emerging problem is telling a machine to react to a data stream, i.e., learning to bid cleverly on behalf of advertisers and brands to maximise conversions while keeping costs to a minimum \citep{xu2016lift,ran2016cikm,cai2017real}.

It is also of great interest to study learning over multi-agent systems and consider the incentives and interactions of each individual learner (bidding agent). For economics researchers, RTB provides a new playground for micro impression-level auctions with various bidding strategies and macro multiple marketplace competitions with different pricing schemes, auction types and floor price settings, etc.

More interestingly, per impression optimisation allows advertisers and agencies to maximise effectiveness based on their own, or the 3rd party, user data across multiple sources. Advertisers buy impressions from multiple publishers to maximise certain Key Performance Indicators (KPIs) such as clicks or conversions, while publishers sell their impressions through multiple advertisers to optimise their revenue \citep{yuan2012sequential}. This brings the online advertising market a step closer to the financial markets, where marketplace unity is strongly promoted. A common objective, such as optimising clicks or conversions across webpages, advertisers, and users, calls for significant multi-disciplinary research that combines statistical Machine Learning, Data Mining, Information Retrieval, and behavioural targeting with game theory, economics and optimisation.
 
Despite its rapid growth and huge potential, many aspects of RTB remain unknown to the research community for a variety of reasons. In this monograph, we aim to offer insightful knowledge of real-world systems, to bridge the gaps between industry and academia, and to provide an overview of the fundamental infrastructure, algorithms, and technical and research challenges of the new frontier of computational advertising.
 
\section{A short history of online advertising}
\label{s-files} 
The first online ad appeared in 1994 when there were only around 30 million people on the Web. The Web version of the Oct. 27, 1994 issue of \emph{HotWired} was the first to run a true banner ad for AT\&T.
 
\subsection{The birth of sponsored search and contextual advertising}
Online advertising has been around for over a decade. The \textit{sponsored search} paradigm was created in 1998 by Bill Gross of Idealab with the founding of \url{Goto.com}, which became Overture in October 2001, was acquired by Yahoo! in 2003 and is now Yahoo! Search Marketing \citep{Jansen2007Sponsored}. Meanwhile, Google started its own service \textit{AdWords} using \textit{Generalized Second Price Auction} (GSP) in February 2002, adding quality-based bidding in May 2002 \citep{Karp2008Google}. In 2007, Yahoo! Search Marketing followed, added quality-based bidding as well \citep{Dreller2010Brief}. It is worth mentioning that Google paid 2.7 million shares to Yahoo! to solve the patent dispute, as reported by \cite{Google2004Dispute}, for the technology that matches ads with search results in sponsored search.
Web search has now become an integral part of daily life, vastly reducing the difficulty and time once associated with satisfying an information necessity. Sponsored search allows advertisers to buy certain keywords to promote their business when users use such a search engine and greatly contributes to its continuing a free service.
 
On the other hand, in 1998, \emph{display} advertising began as a concept \textit{contextual advertising} \citep{Anagnostopoulos:2007:JCA:1321440.1321488,Broder:2007:SAC:1277741.1277837}. Oingo, started by Gilad Elbaz and Adam Weissman, developed a proprietary search algorithm based on word meanings and built upon an underlying lexicon called WordNet. Google acquired Oingo in April 2003 and renamed the system \textit{AdSense} \citep{Karp2008Google}. Later, Yahoo! Publish Network, Microsoft adCenter and Advertising.com Sponsored Listings amongst others were created to offer similar services \citep{Kenny2011Contextual}. The contextual advertising platforms evolved to adapt to a richer media environment, including video, audio and mobile networks with geographical information. These platforms allowed publishers to sell blocks of space on their webpages, video clips and applications to make money. Usually such services are called an \textit{advertising network} or a \textit{display network} and are not necessarily run by search engines, as they can consist of huge numbers of individual publishers and advertisers. Sponsored search ads can also be considered a form of contextual ad that matches with simple context: query keywords; but it has been emphasised due to its early development, large market volume and  research attention.
 
\subsection{The arrival of ad exchange and real-time bidding}
 
Around 2005, new platforms focusing on real-time bidding (RTB) based buying and selling impressions were created. Examples include ADSDAQ, AdECN, DoubleClick Advertising Exchange, adBrite, and Right Media Exchange, which are now known as \textit{ad exchanges}. Unlike traditional ad networks, these ad exchanges aggregate multiple ad networks together to balance the demand and supply in marketplaces and use auctions to sell an ad impression in real time when it is generated by a user visit \citep{Yuan:2013:RBO:2501040.2501980}. Individual publishers and advertising networks can both benefit from participating in such businesses. Publishers sell impressions to advertisers who are interested in associated user profiles and context while advertisers, on the other hand, can contact more publishers for better matching and buy impressions in real-time together with their user data. At the same time, other similar platforms with different functions emerged \citep{Graham2010Brief} including (i) \textit{demand side platform} (DSP), which serves advertisers managing their campaigns and submits real-time bidding responses for each bid request to the ad exchange via algorithms, and (ii) \textit{supply side platform} (SSP), created to serve publishers managing website ad inventory. However, real-time bidding (RTB) and multiple ad networks aggregation do not change the nature of such marketplaces (where buying and selling impressions happen), but only make the transactions in real-time via an auction mechanism.  For simplicity, we may use the term ``ad exchange'' in this monograph to better represent the wider platforms where trading happens.
 
 

\section{The major technical challenges and issues}
Real-time advertising generates large amounts of data over time. Globally, DSP Fikisu claims to process 32 billion ad impressions daily \citep{zhang2017managing} and DSP Turn reports to handle 2.5 million per second at peak time \citep{shen20150}. The New York Stock Exchange, to better envision the scale, trades around 12 billion shares daily.\footnote{According to Daily NYSE group volume, \url{http://goo.gl/2EflkC}, accessed: 2016-02.} It is fair to say the volume of transactions from display advertising has already surpassed that of the financial market. Perhaps even more importantly, the display advertising industry provides computer scientists and economists a unique opportunity to study and understand the Internet traffic, user behaviour and incentives, and online transactions. Only this industry aggregates nearly all the Web traffic, in the form of ads transactions, across websites and users globally.

With real-time per-impression buying established together with the cookie-based user tracking and syncing (the technical details will be explained in Chapter \ref{c-how}), the RTB ecosystem provides the opportunity and infrastructure to fully unleash the power of user behavioural targeting and personalisation \citep{zhang2016implicit,wang2006unifying,zhao2013interactive} for that objective. It allows machine driven algorithms to automate and optimise the relevance match between ads and users \citep{Raeder:2012:DPM:2339530.2339740,Zhang:2014:ORB:2623330.2623633,Zhang:2015,ran2016cikm}.
 
RTB advertising has become a significant battlefield for Data Science research, acting as a test bed and application for many research topics, including
user response (e.g. click-through rate, CTR) estimation \citep{chapelle2014simple,chapelle2015offline,he2014practical,ran2016cikm}, behavioural targeting \citep{ahmed2011scalable,perlich2012bid,zhang2016implicit}, knowledge extraction \citep{ahmed2011scalable,yan2009much}, relevance feedback \citep{chapelle2014modeling}, fraud detection \citep{stone2011understanding,Alrwais:2012:DGC:2420950.2420954,crussell2014madfraud,stitelman2013using}, incentives and economics \citep{balseiro2015repeated,balseiro2015optimal}, and recommender systems and personalisation \citep{juanfield,zhang2016implicit}.

 \subsection{Towards information general retrieval (IGR)}
A fundamental technical goal in online advertising is to automatically deliver the right ads to the right users at the right time with the right price agreed by the advertisers and publishers. As such, RTB based online advertising is strongly correlated with the field of Information Retrieval (IR), which traditionally focuses on building relevance correspondence between information needs and documents \citep{baeza1999modern}. The IR research typically deals with textual data but has been extended to multimedia data including images, video and audio signals \citep{smeulders2000content}. It also covers categorical and rating data, including Collaborative Filtering and Recommender Systems \citep{wang2008unified}. In all these cases, the key research question of IR is to study and model the relevance between the queries and documents in the following two distinctive tasks: retrieval and filtering. The retrieval tasks are those in which information needs (queries) are \emph{ad hoc}, while the document collection stays relatively static. By contrast, information filtering tasks are defined when information needs stay static, whereas documents keep entering the system. A rich literature can be found from the probability ranking principle \citep{robertson1977probability}, the RSJ and BM25 model \citep{jones2000probabilistic}, language models of IR \citep{ponte1998language}, to the latest development of learning to rank \citep{joachims2002optimizing,liu2009learning}, results diversity \citep{wang2009portfolio,agrawal2009diversifying,zhu2009risky} and novelty \citep{clarke2008novelty}, and deep learning of information retrieval \citep{lideep,deng2013deep}.

We, however, argue that IR can broaden its research scope by going beyond the applications of Web search and enterprise search, turning towards \emph{general retrieval} problems derived from many other applications. Essentially, as long as there is concern with building a correspondence between two information objects, under various objectives and criteria \citep{gorla2013probabilistic}, we would consider it a general retrieval problem. Online advertising is one of the application domains, and we hope this monograph will shed some light on new information general retrieval (IGR) problems. For instance, the techniques presented on real time advertising are built upon the rich literature of IR, data mining, machine learning and other relevant fields, to answer various questions related to the relevance matching between ads and users. But the difference and difficulty, compared to a typical IR problem, lies in its consideration of various economic constraints. Some of the constraints are related to incentives inherited from the auction mechanism, while others relate to disparate objectives from the participants (advertisers and publishers). In addition, RTB also provides a useful case for relevance matching that is bi-directional and unified between two matched information objects \citep{robertson1982unified,gorla2016bi,gorla2013probabilistic}. In RTB, there is an inner connection between ads, users and publishers \citep{yuan2012internet}:  advertisers would want the matching between the underlying users and their ads to eventually lead to conversions, whereas publishers hope the matching between the ads and their webpage would result in a high ad payoff. Both objectives, among others, require fulfilment when the relevancy is calculated.

\section{The organisation of this monograph}
The targeted audience of this monograph is academic researchers and industry practitioners in the field. The intention is to help the audience acquire domain knowledge and to promote research activities in RTB and computational advertising in general.

The content of the monograph is organised in four folds. Firstly, chapters 2 and 3 provide a general overview of the RTB advertising and its mechanism. Specifically, in Chapter \ref{c-how}, we explain how RTB works, as well as the mainstream user tracking and synchronising techniques that have been popular in the industry; In Chapter \ref{c-auct} we introduce the RTB auction mechanism and the resulting forecasting techniques in the auction market. 
Next, we cover the problems from the view of advertisers: in Chapter \ref{c-ctr}, we explain various user response models proposed in the past to target users and make ads more fit to the underlying user's patterns, while in Chapter \ref{c-bid}, we present bid optimisation from advertisers' viewpoints with various market settings. After that, 
in Chapter \ref{c-dync}, we focus on the publisher's side and explain dynamic pricing of reserve price, programmatic direct, and new type of advertising contracts. The monograph then concludes with attribution models in Chapter \ref{c-attri} and ad fraud detection in Chapter \ref{c-fraud}, two additional important subjects in RTB.

There are several read paths depending on reader's technical backgrounds and interests. For academic researchers, 
chapters 2 and 3  shall help them understand the real-time online advertising systems currently deployed in the industry. The later chapters shall help industry practitioners grasp the research challenges, the state of the art algorithms and potential future systems in this field. As these later chapters are on specialised topics, they can be read in independently at a deeper level.

\chapter{How RTB Works}
\label{c-how} 
 
 
In this chapter, we explore how real-time bidding functions. We start with an introduction of the key players in the ecosystem and then illustrate the mechanism for targeting individual users. We then introduce common user tracking and cookie synchronising methods that play an important role in aggregating user behaviour information across the entire Web. This chapter focuses more on engineering rather than scientific aspects, but nonetheless it serves to provide domain knowledge and context of the RTB system necessary for the mathematical models and algorithms introduced in later chapters. For readability, the complete terminology is enumerated in Appendix~\ref{c-rtb-glossary}.

\section{RTB ecosystem}
\begin{figure}[t]
	\centering
	\includegraphics[width=\textwidth]{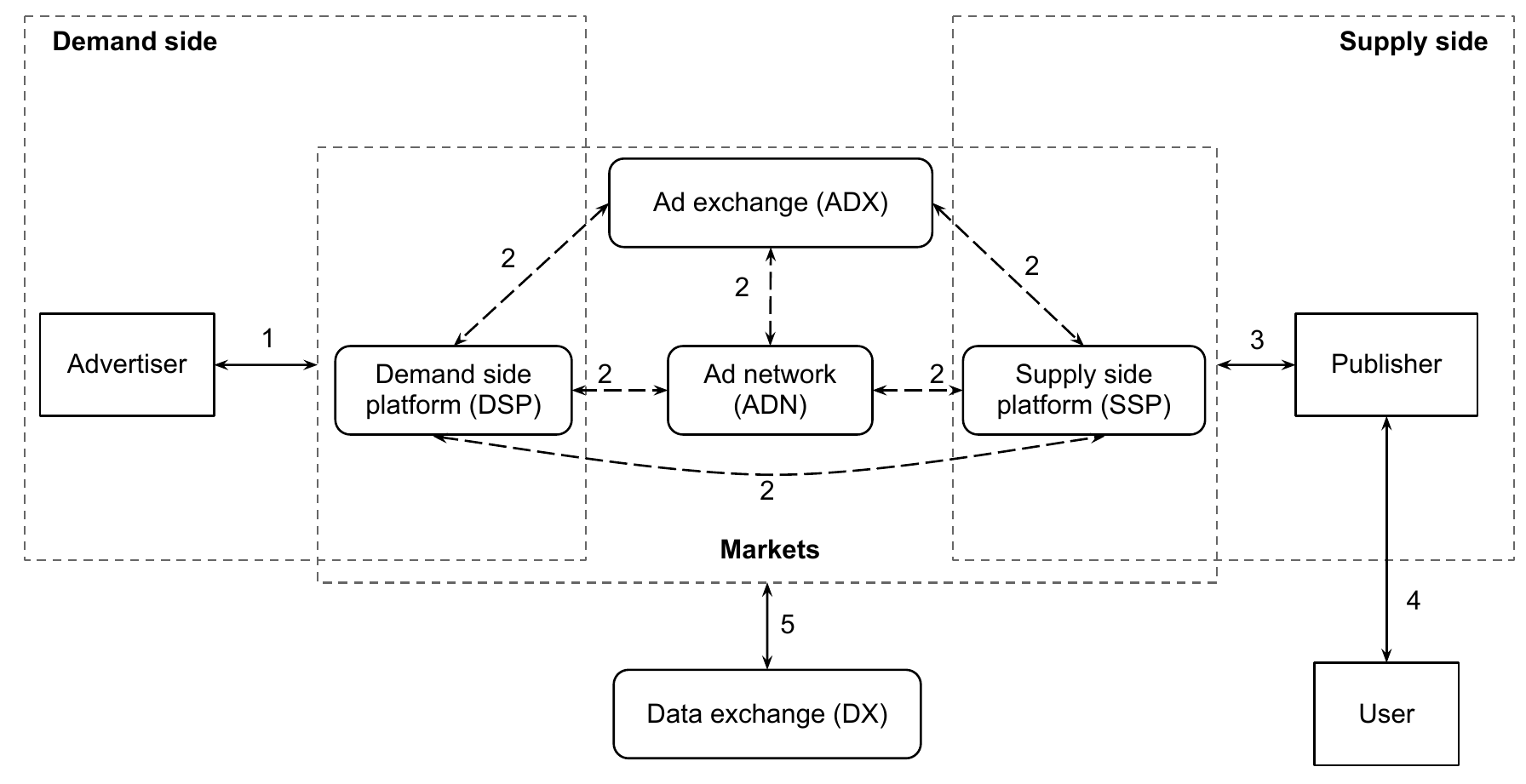}
	\caption{The various players of online display advertising and the ecosystem: 1. The advertiser creates campaigns in the market. 2. The market trades campaigns and impressions to balance the demand and supply for better efficiency. 3. The publisher registers impressions with the market. 4. The user issues queries or visits webpages. 5. The markets can query data exchanges, a.k.a. data management platform (DMP), user profiles in real-time. Source: \citep{yuan2012internet}.}
	\label{fig-players}
\end{figure}

Apart from four major types of players: \textit{advertisers}, \textit{publishers}, \textit{ad networks/ad exchanges} and \textit{users}, RTB has created new tools and platforms which are unique and valuable. Figure \ref{fig-players} illustrates the ecosystem:
 
\begin{itemize}
	\item \textbf{Supply side platforms} (SSP) serve publishers by registering their inventories (impressions) from multiple ad networks and accepting bids and placing ads automatically;
 
	\item \textbf{Ad exchanges} (ADX) combine multiple ad networks together \citep{muthukrishnan2009ad}. When publishers request ads with a given context to serve users, the ADX contacts candidate Ad Networks (ADN) in real-time for a wider selection of relevant ads;
 
 \item \textbf{Demand side platforms} (DSP) serve advertisers or ad agencies by bidding for their campaigns in multiple ad networks automatically;
 	
	\item \textbf{Data exchanges} (DX), also called Data Management Platforms (DMP), serve DSP, SSP and ADX by providing user historical data (usually in real-time) for better matching.
\end{itemize}

The emergence of DSP, SSP, ADX and DX resulted from the presence of thousands of ad networks available on the Internet, a barrier for advertisers as well as publishers entering the online advertising business. Advertisers had to create and edit campaigns frequently for better coverage as well as analyse data across many platforms for better impact. Publishers had to register with and compare several ad networks carefully to achieve optimal revenue. The ADX arose as an aggregated marketplace of multiple ad networks for the purpose of alleviating such problems. Advertisers could then create their campaigns and set desired targeting a single time and analyse the performance data stream in a single place, while publishers could register with ADX and collect optimal profit without any manual interference.
 
The ADX is normally connected with two platforms --- DSP and SSP --- for their different emphases on customers. The DSP works as the agency of advertisers by bidding and tracking in selected ad networks, while the SSP works as the agency of publishers by selling impressions and selecting optimal bids. However, the goal of these platforms is the same: they attempt to create a uniform marketplace for customers. On one hand, this reduces human labour, and on the other hand, balances demand and supply in various small markets for better economic efficiency.
 
Due to the opportunities for profit in such business, the line between these platforms is becoming less tangible. In this monograph, we choose to use the term ``ad exchange'' to boardly describe the marketplace where impression trading happens.
 
The DX collects user data and sells it anonymously to DSP, SSP, ADX and sometimes to advertisers directly in real-time bidding (RTB) for better matching between ads and users. This technology is usually referred to as \textit{behavioural targeting}. Intuitively, if a user's past data shows interest in an advertiser' products or services, then the advertiser would have a higher chance of securing a transaction by displaying its ads, which results in higher bidding for the impression. Initially the DX was a component of other platforms, but more individual DXs are now operating alongside analysing and tracking services.

\begin{figure}[t!]
	\centering
	\includegraphics[width=\textwidth]{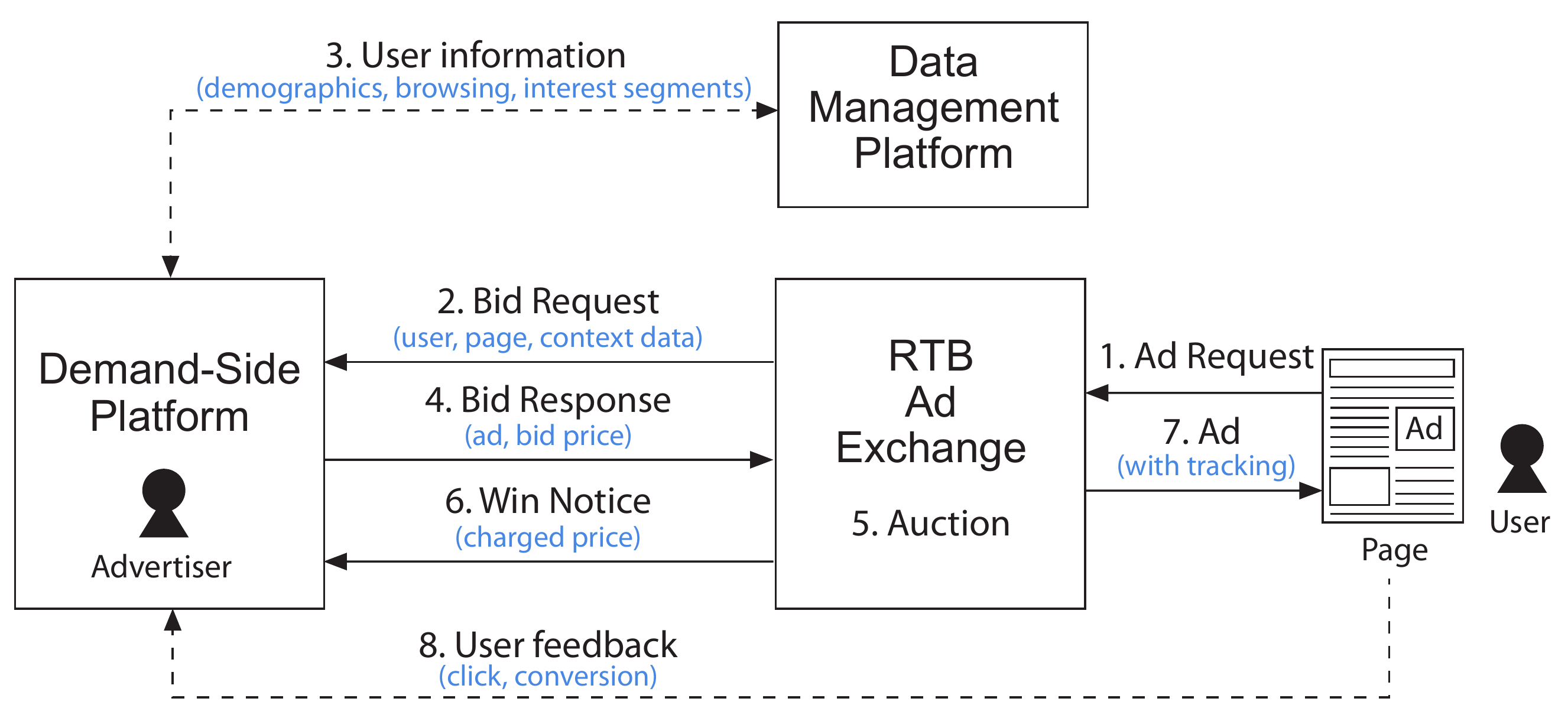}
	\caption{How RTB works for behavioural targeting. Source: \citep{zhang2014real}.}
	\label{fig-rtb}
\end{figure}
 
\section{User behavioural targeting: the steps}
With the new RTB ecosystem, advertisers can target the underlying users based on their previously observed behavior. Here we explain how RTB works with behavioural targeting --- from a user visiting a website, a bid request sent, to the display of the winning ad usually within 100ms, as illustrated in Figure \ref{fig-rtb}:

\begin{enumerate}[start=0]
\item When a user visits a webpage, an impression is created on publisher's website. While the page loads,
\item An ad request is sent to an ad exchange through an ad network or a SSP;
\item The ad exchange queries DSPs for advertisers' bids;
\item The DSP can contact data exchanges for the 3rd party
user data;
\item If the advertiser decides to bid, the bid is generated and submitted (for example, the user is interested in travel, a travel related advertiser, e.g. \url{booking.com}, could expect the user to convert to their campaign and may be willing to bid higher);
\item The winner is selected at ad exchanges (largely based on the second price auction), then selected at SSP if the SSP sends the bid request to multiple ad exchanges;
\item The winning notice is sent to the advertiser;
\item Following the reversed path, the winner's ad (creative, i.e., the text, picture, or video that the advertiser wants to show to users) is displayed on the webpage for the specific user;
\item The tracker collects the user's feedback, determining whether the user clicked the ad and whether the ad led to any conversion.
\end{enumerate}
 
\begin{figure}[t!]
	\centering
	\includegraphics[width=\textwidth]{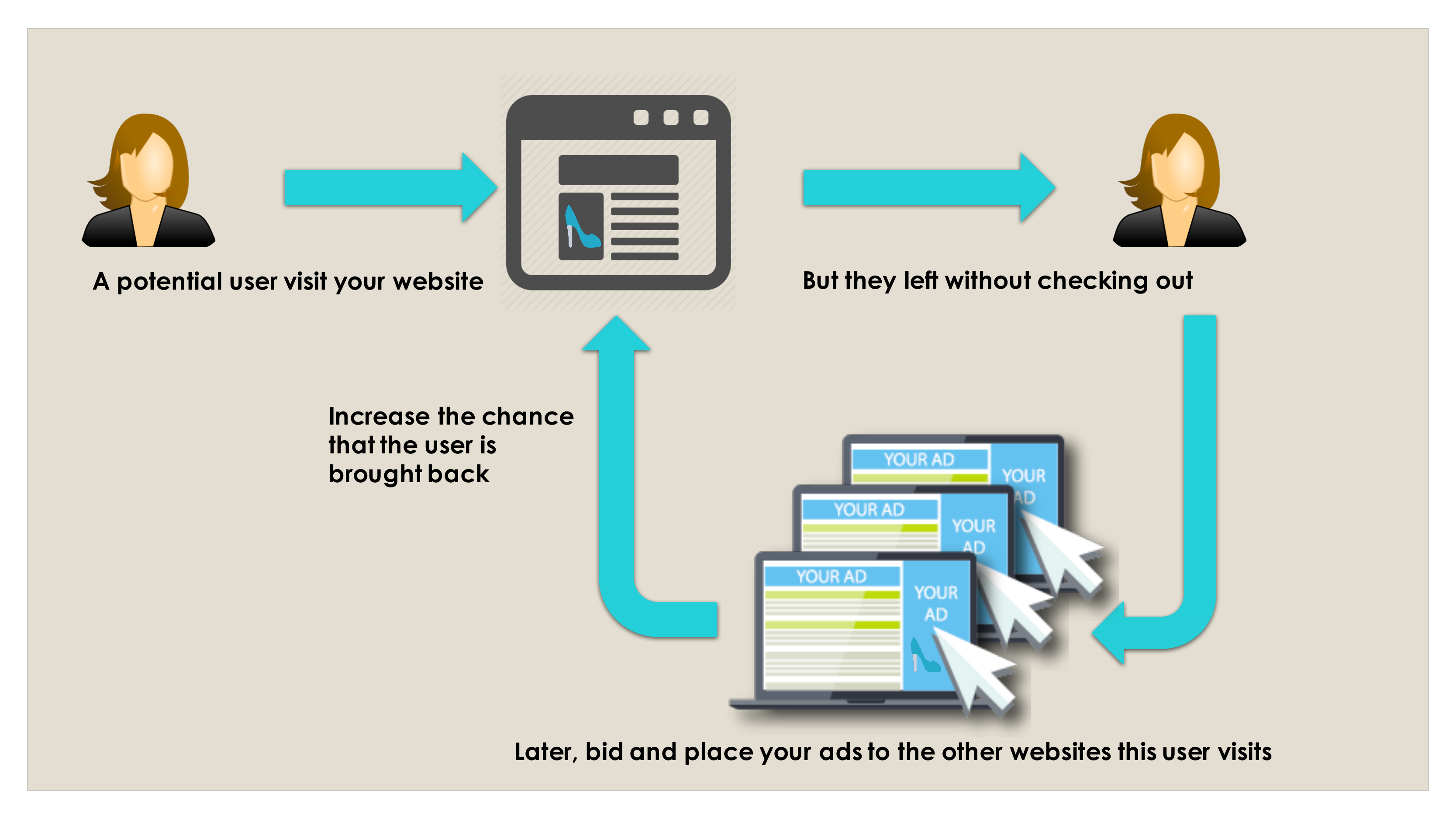}
	\caption{Personalised, retargeting ads: keep ads in front of the users even after they leave the advertiser's website.}
	\label{retargeting}
\end{figure}

The above process marks a fundamental departure from contextual advertising
\citep{Anagnostopoulos:2007:JCA:1321440.1321488,Broder:2007:SAC:1277741.1277837} as it puts more focus on the underlying audience data rather than the contextual data from the webpage itself. In the simplest form of behavioural targeting, advertisers (typically e-commerce merchandisers) would \emph{re-target} users who have previously visited their websites but have not initially converted. As illustrated in Figure \ref{retargeting}, suppose that you run a Web store \url{www.ABC.com}, and user \texttt{abc123} comes and saves a pair of \$150 shoes to their shopping cart but never checks out. You can trace and serve them an ad when they visit other websites later, directing them back to your store to close the sale.
 
One of the reasons to employ the impression level per audience buying is to make \emph{retargeting} such users possible. However, the problem of identifying users across the Internet remains, e.g, recognising in other domains (websites) from the RTB exchange the exact user \texttt{abc123} the advertiser recorded in their own domain.

\section{User tracking}
A user is typically identified by an HTTP cookie, designed to allow websites to remember the status of an individual user, including remembering shopping items added in the cart in an online store or recording the user's previous browsing activities for generating personalised and dynamical content. A cookie, in the form of a small piece of data, is sent from a website and stored in the user's Web browser the first time the user browses a website. Every time the user loads that website again, the browser sends the cookie back to the server to identify the user. Note that a cookie is tied to a specific domain name. If a browser makes an HTTP request to \url{www.ABC.com}, \url{www.ABC.com} can place a cookie in the user's browser, specifying their own user ID, say, \texttt{abc123}. In a later session, when the browser makes another HTTP request to \url{www.ABC.com}, \url{www.ABC.com} can read the cookie and determine the ID of the user is \texttt{abc123}. Another domain, \url{DEF.com} for example, cannot read a cookie set by \url{ABC.com}. This behaviour is the result of cross-origin policy set in \citep{barth2011web} to protect user privacy.
 
\begin{figure}[t!]
	\centering
	\includegraphics[width=\textwidth]{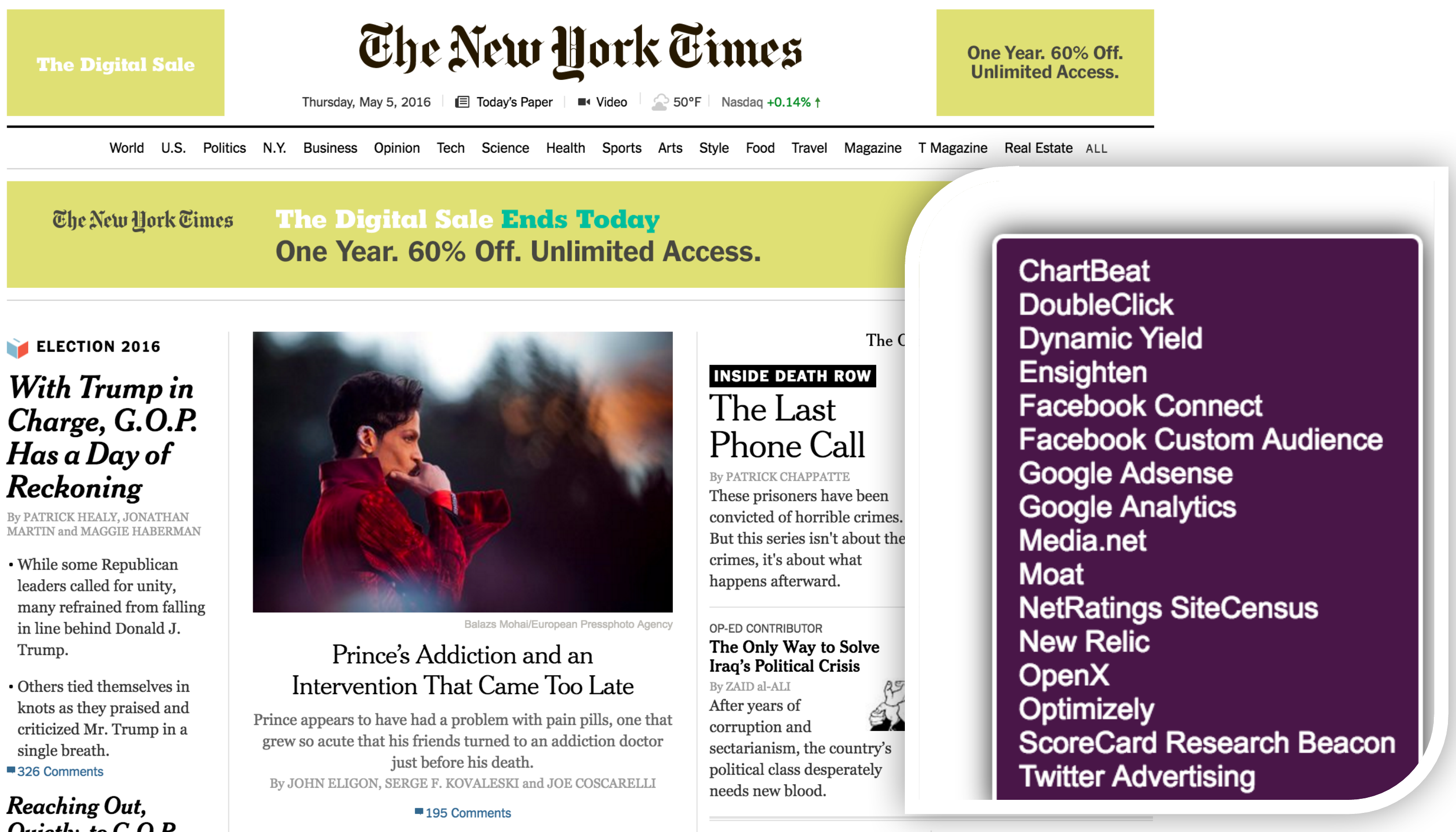}
	\caption{The \emph{New York Times} frontpage referred to as many as 16 third parties that delivered ads and installed cookies, as reported by Ghostery.}
	\label{user-tracking}
\end{figure}

In the context of display advertising, each service provider (including ad exchanges, DSP bidders or DMP user trackers) would act as a single domain to build up their own user ID systems across a number of their client websites (either for advertisers or publishers). The service provider would insert their code snippet\footnote{Usually called tracking code for DSP bidders, and ad serving code or ad tags for SSPs or ad exchanges.} under their own domain name to the HTML code of a managed webpage. In fact, there are quite a few third parties who drop cookies to track users. For instance, in a single webpage from \emph{New York Times}, there are as many as sixteen user trackers, as reported in Figure \ref{user-tracking}.
 
As all these tracking systems only have a local view of their users, and there are enormous numbers of Web users and webpages, the observed user behaviours within individual domains would not adequately create effective targeting. For example, suppose when a browser requests an ad from an ad exchange, it only passes along the cookie data that's stored inside that domain name, say \url{ad.exchange.com}.\footnote{\url{ad.exchange.com} is just a toy example of a domain name here.} This means that the exchange has no knowledge about whatever data the bidder, as well as other third party cookie systems, might have collected under \url{ad.bidder.com}, and vice versa. Therefore, when the exchange sends a bid request to a bidder with the ad exchange's own user ID, the bidder has no knowledge about that ID and thus cannot accurately decide what ad to serve. Their ID systems need to be linked together in order to identify users across the entire Internet. This is done by a technique called \emph{Cookie Syncing}, which shall be covered next.

\section{Cookie syncing}
Cookie syncing, a.k.a. cookie matching or mapping, is a technical process enabling user tracking platforms to link separate IDs given to the same user. \cite{acar2014web} shows that nearly 40\% of all tracking IDs are synced between at least two entities in the Internet. A study of Web search queries in \citep{gomer2013network} showed a 99.5\% chance that a user will be tracked by all top 10 trackers within 30 clicks on search results. A network analysis from the study further indicated that a network constructed by third party tracking exhibits the property of small world, implying it is efficient in spreading the user information and delivering targeted ads.
 
Cookie syncing is commonly achieved by employing \emph{HTTP 302 Redirect} protocol\footnote{An HTTP response with this 302 status code is a common way of performing URL redirection. The Web server also provides a URL in the location header field. This code invites the user agent (e.g. a web browser) to make a second, otherwise identical, request to the new URL specified in the location field.} to make a webpage available under more than one URL address. The process begins when a user visits a marketer's website, such as \url{ABC.com}, which includes a \emph{tag} from a third-party tracker such as \url{ad.bidder.com}. The tag is commonly implemented through an embedded 1x1 image (known as pixel tags, 1x1 pixels, or web bugs).  Pixel tags are typically single pixel, transparent GIF images that are added to a webpage by, for instance,
 
\vspace{10pt}
\texttt{<img src="\url{http://ad.bidder.com/pixel?parameters=xxx"/>}}.
\vspace{10pt}

Even though the pixel tag is virtually invisible, it is served just like any other image online. The difference is the webpage is served from the site's domain while the image is served from the tracker's domain. This allows the user tracker to read and record the cookie's unique ID and the extended information it needs. The trick of cookie sync using a pixel is that, instead of returning the required 1x1 pixel immediately, one service redirects the browser to another service to retrieve the pixel. During the redirect process, the two services exchange the information and sync the user's ID.
 
\begin{figure}[t!]
	\centering
	\includegraphics[width=\textwidth]{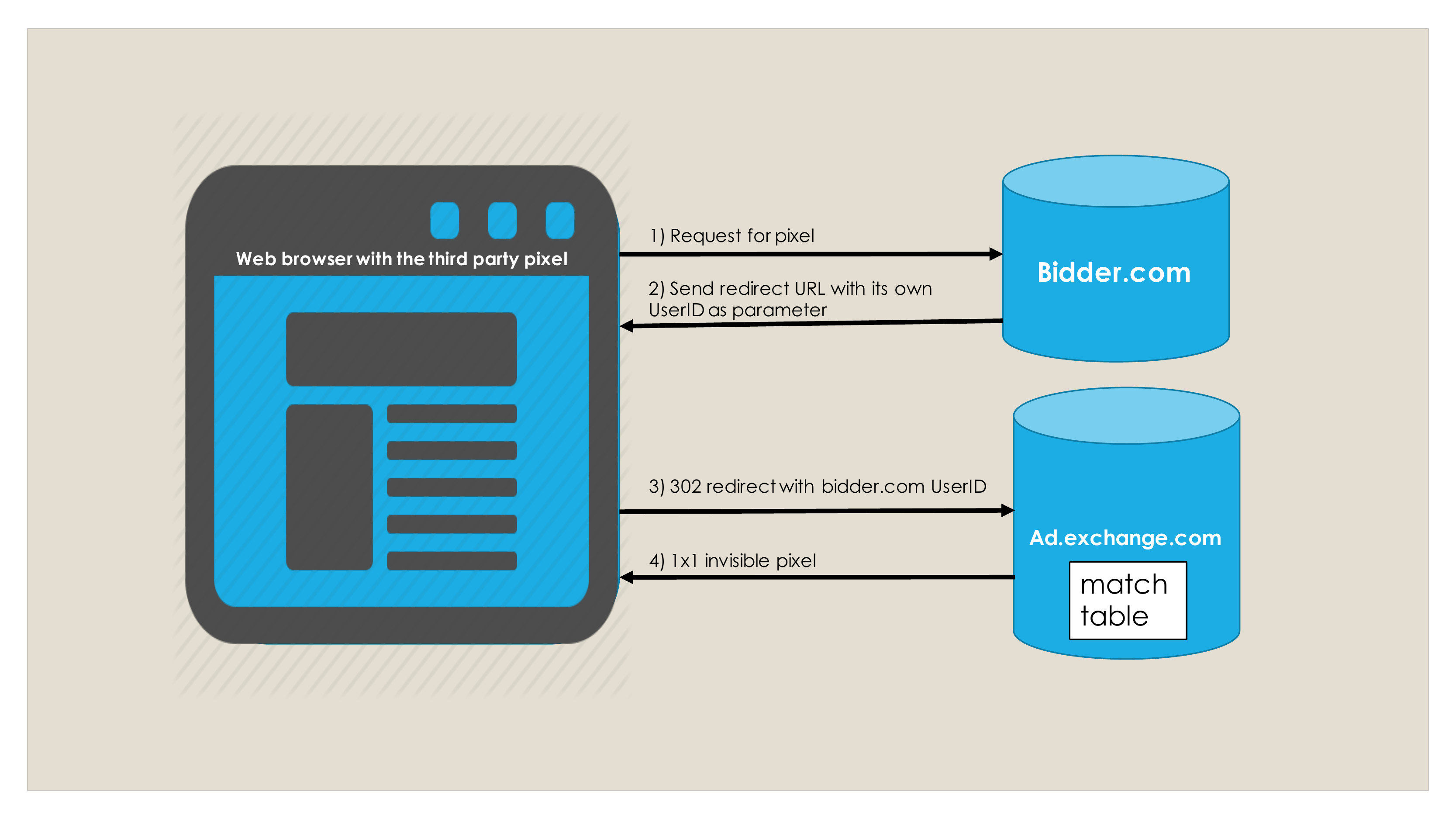}
	\caption{Cookie syncing in a marketer's website managed by \url{ad.bidder.com}.}
	\label{cookie-sync}
\end{figure}
 
As illustrated in Figure \ref{cookie-sync}, in step (1), the browser makes a request from the pixel tag to \url{ad.bidder.com} and includes in this request any tracking cookies set by \url{ad.bidder.com}. If the user is new to \url{ad.bidder.com}, it sets its \url{ad.bidder.com} cookie. In step (2), the tracker from \url{ad.bidder.com} retrieves its tracking ID from the cookie and, instead of returning the required 1x1 pixel, redirects the browser to \url{ad.exchange.com} using http 302 redirect and encoding the tracking ID into the URL as a parameter. (3) The browser then makes a request to \url{ad.exchange.com}, which includes the full URL \url{ad.bidder.com} redirected to as well as \url{ad.exchange.com}'s own tracking cookie (if one exists). (4) \url{ad.exchange.com} returns the required 1x1 pixel and can now link its own ID associated with the user to \url{ad.bidder.com}'s ID and create a record in its match table.
 
The above cookie sync takes place when \url{ad.bidder.com} is managing its Web properties, but the cookie sync process can also be completed alongside served ads when the bidder wins an impression in RTB.
If the sync is bidirectional, the \url{ad.exchange.com} makes the redirect back to the \url{ad.bidder.com}, passing its own ID in the URL parameter. The \url{ad.bidder.com} receives this request, reads its own cookie, and stores the ad exchange user ID along with its own ID in the cookie-matching table.

Using cookies to track users does have drawbacks. Only applicable to browsers, cookies can also be easily deleted by clearing the browser's cache. Users can even choose to disable cookies completely (i.e., private or incognito mode). This makes alternative tracking techniques preferable. For example, device or browser fingerprinting uses information collected about the remote computing device for the purpose of identifying the user. Fingerprints can be used to fully or partially identify individual users or devices even when cookies are turned off.  One example, canvas fingerprinting, uses the browser's Canvas API to draw invisible images and extract
a persistent, long-term fingerprint without the user's knowledge. \cite{acar2014web} found that over 5\% of the top 100,000 websites have already employed canvas fingerprinting. Trackers can also abuse Flash cookies for regenerating previously removed HTTP cookies, a technique referred to as respawning or Flash cookies \citep{boda2011user}. A study by \cite{eckersley2010unique} showed 94.2\% of browsers with Flash or Java were unique, while \cite{soltani2010flash} found 54 of the 100 most popular sites stored Flash cookies.

\chapter[RTB Auction \& Bid Landscape]{RTB Auction Mechanism and Bid Landscape Forecasting}
\label{c-auct}
In online advertising, sellers (typically publishers) may gain access to partial information about the market demand of their ad impressions from historic transactions. However, they do not usually have knowledge about how much an individual ad impression is worth on the market. Different advertisers may have different (private) valuations of a given ad impression. The valuation is typically based on the prediction of the underlying user's likelihood to convert should their ad be placed in the impression.
 
In such a situation, \emph{auction} is generally regarded as a fair and transparent way for advertisers and publishers to agree with a price quickly, whilst enabling the best possible sales outcome. Specifically, an auction is the process of selling an item by offering it up for bids, taking bids, and then selling it to the highest bidder.  As demonstrated in sponsored search by \cite{edelman2005internet}, auctions have become an effective tool to sell search impressions by posting a bid for underlying query keywords --- for a general introduction and discussion about keyword auction and ad exchange, we refer to \citep{mcafee2011design}. Subsequently, display advertising has followed suit and employed auctions in real-time to sell an ad impression each time when it is being generated from a user's visit \citep{Yuan:2013:RBO:2501040.2501980}. In this chapter, we briefly introduce the auction mechanisms in RTB display advertising, explaining how the second price auction has been utilised. It is worth mentioning that the second price auction is a sealed-bid auction and advertisers are notified of the winning price only if they have actually won the auction. Thus, each advertiser only presents a local view of the market data. At the end of this chapter, we address the unique data censorship problem and explain how to estimate the winning probability in an unbiased way in the RTB auction market.

\section{The second price auction in RTB}
 
As introduced in Chapter \ref{c-how} (also see Figure \ref{fig-rtb}), in RTB, advertisers would be able to bid an individual impression immediately when it is still being generated from a user visit. RTB exchanges typically employ the second price auction. In this type of auction, instead of paying for the bid offered, the bidder pays the price calculated from the second highest bid. 

The reason behind paying the second highest bid is that impressions with same or similar user profiles will continuously appear in ad exchanges. If advertisers pay what they bid (i.e., the first price auction),  they  would not state their true valuations, but rather keep adjusting their bids in response to other bidders' behaviours. Unstable bidding behaviours have been well observed in continuously repeated first price auctions, such as in the search markets  \citep{edelman2007strategic}.  To understand this in RTB, suppose there are two advertisers and the impressions associated with the same targeted user group are worth \$6 CPMs (cost per mille impressions, with \emph{mille} being Latin for thousand) to the first advertiser and \$8 CPMs to the second. The floor price (the lowest acceptable bid) is assumed to be \$2 CPMs. Thus, when they bid between \$2 CPMs and \$6 CPMs, each of them tries to outbid each other with a small amount. As a result, the winning price increases continuously until reach \$6 CPMs where the second advertiser stops the bidding. Then, without the competition from the second advertiser, the first advertiser would drop back to the minimum bid \$2 CPMs. At that point, the second advertiser comes back and the competition restarts again and the \emph{cycling} behaviour will continue indefinitely.

\begin{figure}[t!]
	\centering
	\includegraphics[width=\textwidth]{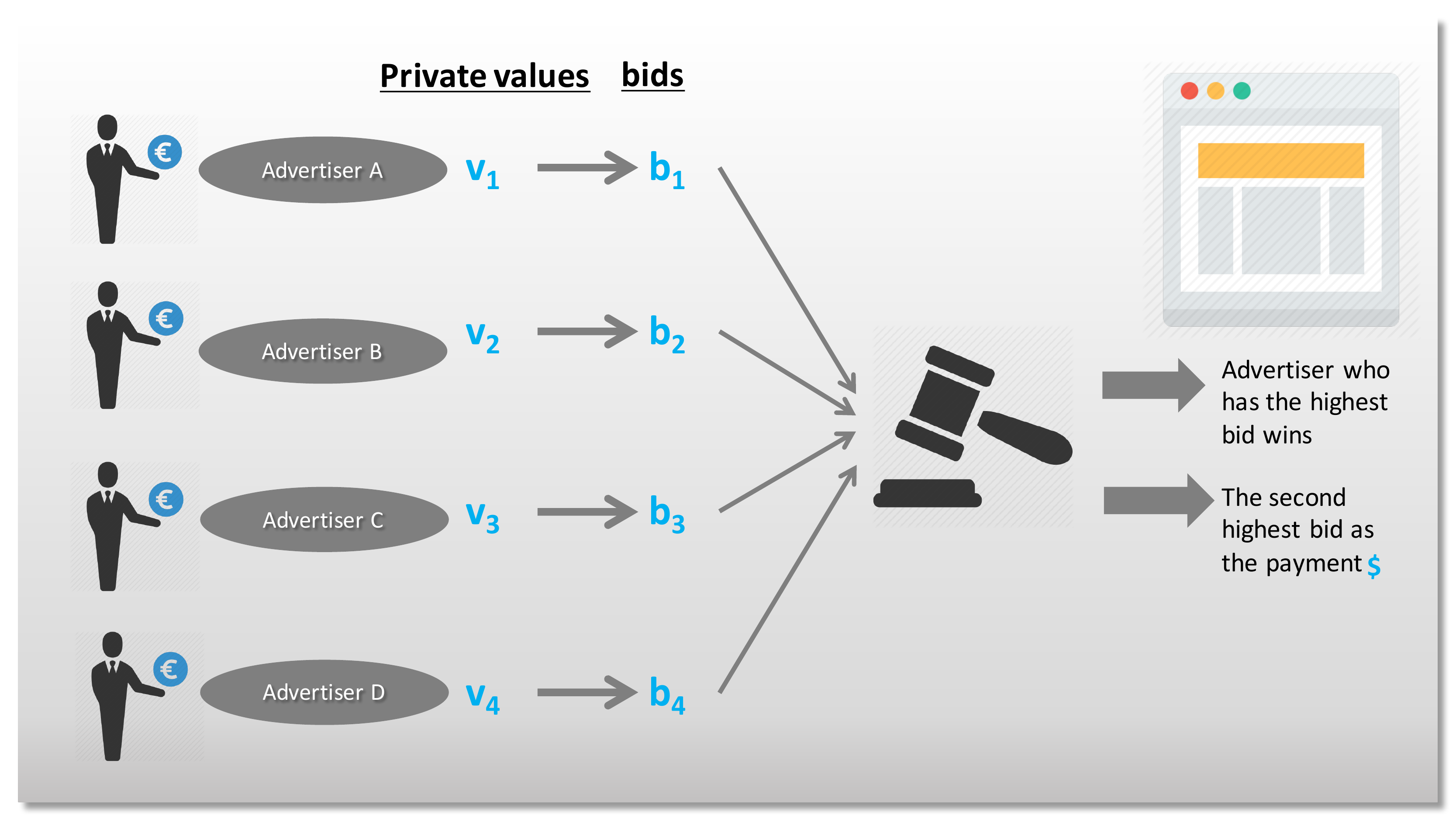}
	\caption{A simple illustration of the second price auction in RTB exchange. }
	\label{spa}
\end{figure}

Thus, our focus in this chapter will be on the second price auction. Figure \ref{spa} illustrates a simple second price auction for an ad impression offered to four advertisers. When receiving the bid requests (consisting of features to describe the impression), advertisers will have their own assessment about the value of the impression. For direct targeting campaigns, it is estimated based on the likelihood the user is going to convert and also the value of that conversion. For branding campaigns, it is mostly bounded by the average budget per targeted impression. Note that the valuation is private, as it relies on the direct sales or advertising budget, which is only known by advertisers themselves. Suppose Advertisers A, B, C, and D, based on their valuations, place bids as \$10, \$8, \$12, \$6 CPMs respectively. Advertiser C would win the auction with the actual payment price \$10 CPM.\footnote{The actual cost for this impression is $\$10\text{CPM}/1000=\$0.01$.} It is worth noticing that unlike sponsored search where CPC (cost per click) is typically used for the payment, RTB mainly employs CPM. Thus, it is the advertisers' responsibility to optimise the effectiveness of the campaigns such as clicks or conversions on top of the CPM costs. This makes sense in RTB as advertisers typically have more information about their campaigns and the user intents (e.g., via their own data collection or a third party DMP etc) and as such know better about the potential conversions and clicks.

In practice, RTB auctions are done in a hierarchical manner because the auction process may start from SSPs (supply side platforms) who manage the inventory by auctioning off and gathering the best bids from their connected ad exchanges and ad networks. Those connected ad exchanges and ad networks then subsequently run their own auctions in order to pick up the highest bids to send over. By the same token, the DSPs (demand side platforms) that further follow up the auction may also have an internal auction to pick up the highest bids from their advertisers and send it back to the connected ad exchanges or ad networks.

\subsection{Truthful bidding is the dominant strategy}
 
An auction, as a market clearing mechanism, imposes competition among advertisers. We intend to incentivise them to reveal their private valuations when placing their bids. The second price auction would ensure that advertisers are better off when they tell the truth by specifying their bid exactly as their private value \citep{milgrom2004putting}.

To see this, we follow a simplified case in \citep{menezes2005introduction}: suppose an ad impression, represented by its vectorised feature $\bx$,  is to be sold to one of $n$ advertisers (bidders) in an ad exchange. Advertisers submit their bids simultaneously without observing the bids made by others. Each advertiser $i$, $i = \{1,...,n\}$, privately, estimates its click-through rate as $c_i(x)$.

Without loss of generality, we assume the value of a click for all the advertisers is the same and set it as 1. The private valuation $v_i$ is thus equal to $c_i(x)$ for each advertiser $i$. We also consider that all the advertisers are \emph{risk-neutral}. In other words, they are indifferent between an expected value from a random event and receiving the same value for certain.
 
Each advertiser knows his own private valuation $c_i$, but does not know their opponent advertisers' valuations. They however can have a belief, and therefore an estimation about these valuations.  The belief is commonly represented by a distribution. To make our discussion simple, the opponent' valuations are assumed to be drawn independently from the cumulative distribution function $F(\cdot)$ with its density $f(\cdot)$ in
the interval $[0, +\infty]$. That is, $F_V(v)=P(V\le v)$ denotes the probability that the random variable $V$ is less than
or equal to a certain value $v$. For simplicity, we assume that the ad exchange (the seller) sets
the reserve price at zero (we will discuss the reserve price in Chapter~\ref{c-dync}).

Without loss of generality, we look at bidder 1, who
has a value equal to $v_1$, and chooses a bid $b_1$ to maximise his expected profits
given that bidders $2,...,n$ follow some strategy $b(\cdot)$. Assume bidder $i$ places the highest bid among bidders $2,...,n$. Bidder 1's expected profit $\pi_1$
can be written as
\begin{align}
\pi_1 (v_1,b_i,b(\cdot)) = \hfill
  \begin{cases}
    v_1-b_i       & \text{if } b_1>b_i>\max\{b(v_2),...,\\&~~~~~b(v_{i-1}),b(v_{i+1}),...,b(v_{n})\}\\
    0  & \text{if } b_1<\max\{b(v_2),...,b(v_{n})\},\\
  \end{cases}
   \label{eq:spa}
\end{align}
where $b_1>b_i>\max\{b(v_2),...,b(v_{i-1}),b(v_{i+1}),...,b(v_{n})\}$ holds with probability $\int_{0}^{b_1}dF(x)^{n-1} =\int_{0}^{b_1}(n-1)f(x)F(x)^{n-2}dx$. As such, we have
\begin{align}
\pi_1 (v_1,b_i,b(\cdot)) = \int_{0}^{b_1}(v_1-x)(n-1)f(x)F(x)^{n-2}dx.\hfill
     \label{eq:expected reward}
\end{align}

The advertiser 1 is to choose $b_1$ in such a way that the above expected reward is maximised. When $b_1>v_1$, we have
 
\begin{align}
\pi_1 (v_1,b_i,b(\cdot)) = \int_{0}^{v_1}(v_1-x)(n-1)f(x)F(x)^{n-2}dx + \nonumber\\ \int_{v_1}^{b_1}(v_1-x)(n-1)f(x)F(x)^{n-2}dx,\hfill
     \label{eq:expected-reward-1}
\end{align}
where the second integration is negative. Thus, the expected revenue will increase when $b_1$ reaches $v_1$. By the same token, when $b_1<v_1$, the second integration is positive. If $b_1$ reaches $v_1$, the expected reward will increase by the amount:
\begin{align}
 \int_{b_1}^{v_1}(v_1-x)(n-1)f(x)F(x)^{n-2}dx.\hfill
     \label{eq:expected-reward-2}
\end{align}
 
Thus, the expected reward is maximised when $b_1=v_1$. So truth-telling is a \emph{dominant} strategy, where dominance occurs when one strategy is better than another strategy for one bidder (advertiser), no matter how other opponents may bid.

In practice, however, advertisers might join the auction with a fixed budget and be involved in multiple second-price auctions over the lifetime of a campaign. As such, the truth-telling might not be a dominant strategy.
\cite{balseiro2015repeated} studied a fluid mean-field equilibrium (FMFE) that approximates the rational behaviours of the advertisers in such a setting.
 
While a game theoretical view provides insights into the advertisers' and publishers' strategic behaviours, a practically more useful approach is to take a statistical view of the market price and the volume, which will be introduced next.
 

\section{Winning probability}
 
A bid request can be represented as a high dimensional feature vector \citep{lee2012estimating}. As before, we denote the vector as $\bx$, which encodes much information about the impression. An example feature vector includes,
 
\vspace{10pt}
\begin{center}
\center \texttt{Gender=Male \&\& Hour=18 \&\& City=London \&\& \\Browser=firefox \&\&  URL=www.abc.com/xyz.html}.
\end{center}
\vspace{10pt}
 
Without loss of generality, we regard the bid requests as generated from an i.i.d. (independent and identical distribution) $\bx \sim p_x(\bx)$, where the time-dependency is modelled by considering week/time as one of the features.  Based on the bid request $\bx$, the ad agent (or demand-side platform, a.k.a. DSP) will then provide a bid $\bidx$ following a bidding strategy. If such bid wins the auction, the corresponding labels, i.e., user response $y$ (either click or conversion) and market price $z$, are observed. Thus, the probability of a data instance $(\bx, y, z)$ being observed relies on whether the bid $b_x$ would win or not and we denote it as $P(\text{win}|\bx, \bidx)$. Formally, this generative process of creating observed training data $D = \{(\bx, y, z)\}$ is summarised as:
 
\begin{align}
\underbrace{q_x(\bx)}_{\text{winning impression}} \equiv \underbrace{P(\text{win}|\bx, \bidx)}_{\text{prob. of winning the auction}} \cdot \underbrace{p_x(\bx)}_{\text{bid request}},
\label{eq:pq}
\end{align}
 
where probability $q_x(\bx)$ describes how feature vector $\bx$ is distributed within the training data. The above equation indicates the relationship (bias) between the p.d.f. of the pre-bid full-volume bid request data (prediction) and the post-bid winning impression data (training); in other words, the predictive models would be trained on $D$, where $\bx \sim q_x(\bx)$, and be finally operated on prediction data $\bx \sim p_x(\bx)$.
 
As explained, the RTB display advertising uses the second price auction \citep{milgrom2004putting,Yuan:2013:RBO:2501040.2501980}.
In the auction, the market price $z$ is defined as the second highest bid from the competitors for an auction. In other words, it is the lowest bid value one should have in order to win the auction. The form of its distribution is unknown unless one has a strong assumption as given in the previous section for theoretical analysis. In practice, one can assume the market price $z$ is a random variable generated from a fixed yet unknown p.d.f. $p_z^{\bs{x}}(z)$; then the auction winning probability is the probability when the market price $z$ is lower than the bid $\bidx$:
 
\begin{align}
w(\bidx) \equiv P(\text{win}|\bx, \bidx) = \int_0^{\bidx} p_z^{\bs{x}}(z)dz,
\label{eq:win-mp}
\end{align}
 
where to simplify the solution and reduce the sparsity of the estimation, the market price distribution (a.k.a., the bid landscape) is estimated on a campaign level rather than per impression $\bx$ \citep{cui2011bid}. Thus, for each campaign, there is a $p_z(z)$ to estimate, resulting in the simplified winning function $w(\bidx)$, as formulated in \citep{amin2012budget}. \cite{zhang2016bidaware} proposed a solution of estimating the winning probability $P(\text{win}|\bx, \bidx)$ and then using it for creating bid-aware gradients to solve CTR estimation and bid optimisation problems.
 
 
 
\section{Bid landscape forecasting}
\label{sec:landscape}
 
 
Estimating the winning probability and the volume (bid landscape forecasting) is a crucial component in online advertising framework. However, it is paid less attention than the components of user response prediction (Chapter~\ref{c-ctr}) and bid optimisation (Chapter~\ref{c-bid}). 
Researchers proposed several heuristic forms of functions to model the winning price distribution. \cite{Zhang:2014:ORB:2623330.2623633} provided two forms of winning probability w.r.t. the bid price, which is based on the observation of an offline dataset. However, this derivation has many drawbacks since the winning price distribution in real world data may deviate largely from a simple functional form. \cite{cui2011bid} proposed a log-normal distribution to fit the winning price distribution. \cite{chapelle2015offline} used a Dirac conditioned distribution to model the winning price under the condition of given historical winning price. The main drawback of these distributional methods is that they may lose the effectiveness of handling various dynamic data and they also ignore the real data divergence \citep{cui2011bid}.
 
Basically, from an advertiser's perspective, given the winning price distribution $p_z(z)$ and the bid price $b$, the probability of winning the auction is
\begin{align}
w(b) = \int_0^{b} p_z(z) dz,
\label{eq:winning}
\end{align}
 
If the bid wins the auction, the ad is displayed and the utility and cost of this ad impression are then observed.
The expected cost of winning with the bid $b$ is denoted as $c(b)$. With the second price auction, the expected cost is given as
 
\begin{align}
c(b) = \frac{\int_0^{b}z p_z(z) dz}{\int_0^{b} p_z(z) dz},
\end{align}
 
In summary, estimating the winning price distribution (p.d.f. $p(z)$) or the winning probability given a bid (c.d.f. $w(b)$) is the key for bid landscape forecasting, which will be explained next.
 
\subsection{Tree-based log-normal model}
 
In view of forecasting, a template-based method can be used to to fetch the corresponding winning price distribution w.r.t. the given auction request, as demonstrated by \cite{cui2011bid}.
 
The targeting rules of different advertising campaigns may be quite different. As such, the bid requests received by each campaign may follow various distributions. \cite{cui2011bid} proposed to partition the campaign targeting rules into mutually exclusive \emph{samples}. Each sample refers to a unique combination of targeted attributes, such as \texttt{Gender=Male \&\& Hour=18 \&\& City=London}. Then the whole data distribution following the campaign's targeting rules, e.g. \texttt{Gender=Male \&\& Hour=18-23 \&\& Country=UK}, is aggregated from the samples. Thus the campaign's targeting rules can be represented as a set of samples $S_c=\{s\}$.
 
Furthermore, for the bid landscape modelling of each sample $s$, \cite{cui2011bid} first assumed the winning price $z$ follow a log-normal distribution
 
\begin{equation}
p_s(z; \mu, \sigma) = \frac{1}{z \sigma \sqrt{2 \pi}} e^{\frac{-(\ln z - \mu)^2}{2 \sigma^2}},
\end{equation}
 
where $\mu$ and $\sigma$ are two parameters of the log-normal distribution.
 
\cite{cui2011bid} proposed to adopt gradient boosting decision trees (GBDT) \citep{friedman2002stochastic} to predict the mean $\mathbb{E}[s]$ and standard deviation $\text{Std}[s]$ of winning prices of each sample $s$ based on the features extracted from the targeting rules of the sample, and then with a standard transformation the (predicted) log-normal distribution parameters are obtained
 
\begin{align}
\mu_s &= \ln \mathbb{E}[s] - \frac{1}{2} \ln \Big(1 + \frac{\text{Std}[s]^2}{\mathbb{E}[s]^2} \Big),\\
\sigma_s^2 &= \ln \Big( 1 + \frac{\text{Std}[s]^2}{\mathbb{E}[s]^2} \Big).
\end{align}
 
With the winning price distribution $p_s(z; \mu_s, \sigma_s)$ of each sample $s$ modelled, the campaign-level $c$ winning price distribution is calculated from the weighted summation of its targeting samples' $p_s(z; \mu, \sigma)$
 
\begin{equation}
p_c(z) = \sum_{s \in S_c} \pi_s \frac{1}{z \sigma_s \sqrt{2 \pi}} e^{\frac{-(\ln z - \mu_s)^2}{2 \sigma_s^2}},
\end{equation}
 
where $\sum_{s\in S_c}\pi_s = 1$ and $\pi_s$ is the prior probability (weight) of sample $s$ in the campaign's targeting rules $S_c$, which can be defined as the proportion of the sample instances in the campaign's whole instances.
 
\subsection{Censored linear regression}
 
The drawback of the log-normal model above is that the feature vector of a bid request is not fully utilised. A simple way for estimating the winning price from the feature vector is to model it as a regression problem, e.g., linear regression
 
\begin{equation}
\hat{z} = \bs{\beta}^T \bs{x},
\end{equation}
 
where $\bs{x}$ is the feature vector of the bid request and $\bs{\beta}$ is the model coefficient vector. \cite{wu2015predicting} learned the regression model using a likelihood loss function with a white Gaussian noise
 
\begin{equation}
z = \bs{\beta}^T \bs{x} + \epsilon,
\end{equation}
 
where $\epsilon \sim \mathcal{N}(0, \sigma^2)$.
 
Furthermore, since the winning price of a bid request can be observed by the DSP only if it wins the auction, the observed $(\bs{x}, z)$ instances are right-censored and, therefore, biased.
 
\cite{wu2015predicting} adopt a censored linear regression \citep{greene2005censored} to model the winning price w.r.t. auction features. For the observed winning price $(\bs{x}, z)$, the data log-likelihood is
 
\begin{equation}
p(z) = \log \phi \Big( \frac{z - \bs{\beta}^T \bs{x}}{\sigma} \Big),
\end{equation}
 
where $\phi$ is the p.d.f. of a standard normal distribution, i.e., with 0 mean and 1 standard deviation.
 
For the losing auction data $(\bs{x}, b)$ with the bid price $b$, we only know the underlying winning price is higher than the bid, i.e., $z>b$, the partial data log-likelihood is defined as the probability of the predicted winning price being higher the bid, i.e.,
 
\begin{equation}
P(\hat{z}>b) = \log \Phi \Big( \frac{\bs{\beta}^T \bs{x} - b}{\sigma} \Big),
\end{equation}
 
where $\Phi$ is the c.d.f. of a standard normal distribution.
 
Therefore, with the observed data $W=\{(\bs{x}, z)\}$ and censored data $L=\{\bs{x}, b\}$, the training of censored linear regression is
 
\begin{equation}
\min_{\bs{\beta}} - \sum_{(\bs{x}, z) \in W} \log \phi \Big( \frac{z - \bs{\beta}^T \bs{x}}{\sigma} \Big) - \sum_{(\bs{x}, b) \in L} \log \Phi \Big( \frac{\bs{\beta}^T \bs{x} - b}{\sigma} \Big).
\end{equation}
 
\subsection{Survival Model}
 
\cite{amin2012budget} and \cite{zhang2016bidaware} went further from a counting perspective to address the problem of censored data. To see this, suppose there is no data censorship, i.e., the DSP wins all the bid requests and observes all the winning prices. The winning probability $w_o(\bidx)$ can be obtained directly from the observation counting:
 
\begin{align}
w_o(\bidx) = \frac{\sum_{(\bx', y, z) \in D} \delta(z < \bidx)}{|D|},
\label{eq:wo}
\end{align}
 
where $z$ is the historic winning price of the bid request $\bx'$ in the training data, the indicator function $\delta(z < \bidx)=1$ if $z < \bidx$ and 0 otherwise. This is a baseline of $w(\bidx)$ modelling.
 
However, the above treatment is rather problematic. In practice there are always a large portion of the auctions the advertiser loses ($z \geq \bidx $),\footnote{For example, in the iPinYou dataset \citep{liao2014ipinyou}, the overall auction winning rate of 9 campaigns is 23.8\%, which is already a very high rate in practice.} in which the winning price is not observed in the training data. Thus, the observations of the winning price are \emph{right-censored}: when the DSP loses, it only knows that the winning price is higher than the bid, but does not know its exact value. In fact, $w_o(\bidx)$ is a biased model and overestimates the winning probability. One way to look at this is that it ignores the counts for losing auctions where the historic bid price is higher than $\bidx$ (in this situation, the winning price should have been higher than the historic bid price and thus higher than $\bidx$) in the denominator of Eq.~(\ref{eq:wo}).
 
\cite{amin2012budget} and \cite{zhang2016bidaware} used survival models \citep{johnson1999survival} to handle the bias from the censored auction data. Survival models were originally proposed to predict patients' survival rate for a given time after certain treatment. As some patients might leave the investigation, researchers do not know their exact final survival period, but only know the period is longer than the investigation period. Thus the data is right-censored. The auction scenario is quite similar, where the integer winning price\footnote{The mainstream ad exchanges require integer bid prices. Without a fractional component, it is reasonable to analogise bid price to survival days.} is regarded as the patient's underlying survival period from low to high and the bid price as the investigation period from low to high. If the bid $b$ wins the auction, the winning price $z$ is observed, which is analogous to the observation of the patient's death on day $z$. If the bid $b$ loses the auction, one only knows the winning price $z$ is higher than $b$, which is analogous to the patient's left from the investigation on day $b$.
 
Specifically, \cite{amin2012budget} and \cite{zhang2016bidaware} leveraged the non-parametric Kaplan-Meier Product-Limit method \citep{kaplan1958nonparametric} to estimate the winning price distribution $p_z(z)$ based on the observed impressions and the losing bid requests.
 
Suppose there is a campaign that has participated $N$ RTB auctions. Its bidding log is a list of $N$ tuples $\langle b_i, w_i, z_i \rangle_{i=1\ldots N}$, where $b_i$ is the bid price of this campaign in the auction $i$, $w_i$ is the boolean value of whether this campaign won the auction $i$, and $z_i$ is the corresponding winning price if $w_i=1$. The problem is to model the probability of winning an ad auction $w(\bidx)$ with bid price $\bidx$.
 
If the data is transformed into the form of $\langle b_j, d_j, n_j \rangle_{j=1\ldots M}$, where the bid price $b_j < b_{j+1}$. $d_j$ denotes the number of ad auction winning cases with the winning price exactly valued $b_j - 1$ (in analogy to patients die on day $b_j$). $n_j$ is the number of ad auction cases which cannot be won with bid price $b_j - 1$ (in analogy to patients survive to day $b_j$), i.e., the number of winning cases with the observed winning price no lower than $b_j - 1$\footnote{Assume that the campaign will not win if it is a tie in the auction.} plus the number of losing cases when the bid is no lower than $b_j - 1$. Then with bid price $\bidx$, the probability of losing an ad auction is
 
\begin{align}
l(\bidx) = \prod_{b_j < \bidx} \frac{n_j - d_j}{n_j},
\end{align}
 
which just corresponds to the probability a patient survives from day 1 to day $\bidx$.
Thus the winning probability will be
 
\begin{align}
w(\bidx) = 1- \prod_{b_j < \bidx} \frac{n_j - d_j}{n_j}.
\label{eq:km}
\end{align}
 
\begin{table}[t!]
\caption{An example of data transformation of 8 instances with bid price between 1 and 4. Left: tuples of bid, win and cost $\langle b_i, w_i, z_i \rangle_{i=1\ldots 8}$. Right: transformed survival model tuples $\langle b_j, d_j, n_j \rangle_{j=1\ldots 4}$ and the calculated winning probabilities. Here we also provide a calculation example of $n_3=4$ shown as blue in the right table. The counted cases of $n_3$ in the left table are 2 winning cases with $z\geq3-1$ and the 2 losing cases with $b\geq3$, shown highlighted in blue colour. Source \citep{zhang2016bidaware}.}\label{tab:survival-data-trans}
\small
  \begin{minipage}[b]{0.25\columnwidth}
  \resizebox{0.98\columnwidth}{!}{
  \begin{tabular}{|ccc|} \hline
    $b_i$ & $w_i$ & $z_i$ \\ \hline
    2 & win & 1 \\
    {\color{blue}3} & {\color{blue}win} & {\color{blue}2} \\
    2 & lose & $\times$ \\
    3 & win & 1 \\
    {\color{blue}3} & {\color{blue}lose} & {\color{blue}$\times$}\\
    {\color{blue}4} & {\color{blue}lose} & {\color{blue}$\times$}\\
    {\color{blue}4} & {\color{blue}win} & {\color{blue}3}\\
    1 & lose & $\times$ \\ \hline
    \end{tabular}}
    \end{minipage}
  \hfill
  \begin{minipage}[b]{0.75\columnwidth}
    \resizebox{0.98\columnwidth}{!}{
    \begin{tabular}{|cccclc|N} \hline
    $b_j$ & $n_j$ & $d_j$ & $\frac{n_j-d_j}{n_j}$ & $w(b_j)$ & $w_o(b_j)$ & \\[7pt] \hline
    1 & 8 & 0 & 1 & $1-1=0$ & 0 &\\[4pt]
    2 & 7 & 2 & $\frac{5}{7}$ & $1-\frac{5}{7}=\frac{2}{7}$ & $\frac{2}{4}$ &\\[4pt]
    3 & {\color{blue}4} & 1 & $\frac{3}{4}$ & $1-\frac{5}{7}\frac{3}{4}=\frac{13}{28}$ & $\frac{3}{4}$ &\\[4pt]
    4 & 2 & 1 & $\frac{1}{2}$ & $1-\frac{5}{7}\frac{3}{4}\frac{1}{2}=\frac{41}{56}$ & $\frac{4}{4}$ &\\[4pt] \hline
    \end{tabular}}
  \end{minipage}
\end{table}
 
Table~\ref{tab:survival-data-trans} gives an example of transforming the historic $\langle b_i, w_i, z_i \rangle$ data into the survival model data $\langle b_j, d_j, n_j \rangle$ and the corresponding winning probabilities calculated by Eqs.~(\ref{eq:km}) and (\ref{eq:wo}). It can be observed that the Kaplan-Meier Product-Limit model, which is a non-parametric maximum likelihood estimator of the data \citep{dabrowska1987non}, makes use of all winning and losing data to estimate the winning probability of each bid, whereas the observation-only counting model $w_o(\bidx)$ does not. As we can see in the table $w_o(\bidx)$ is consistently higher than $w(\bidx)$.
 
While developing accurate bid landscape forecasting models is a worthwhile research goal, \cite{lang2012handling}, however, pointed out that in practice it can be more effective to handle the forecasting error by frequently rerunning the offline optimisation, which updates the landscape model and the bidding strategy with the real-time information. This would link to the feedback control and pacing problems, which will be discussed in Section~\ref{sec:pacing}.

\chapter{User Response Prediction}
\label{c-ctr}
Learning and predicting user response is critical for personalising tasks, including content recommendation, Web search and online advertising. The goal of the learning is to estimate the probability that the user will respond with, e.g. clicks, reading, conversions in a given context \citep{menon2011response}. The predicted probability indicates the user's interest on the specific information item, such as a news article, webpage, or an ad, which shall influence the subsequent decision making, including document ranking and ad bidding. Taking online advertising as an example, click-through rate (CTR) estimation has been utilised later for calculating a bid price in ad auctions \citep{perlich2012bid}.  It is desirable to obtain an accurate prediction not only to improve the user experience, but also to boost the volume and profit for the advertisers. For the performance-driven RTB display advertising, user response prediction, i.e. click-through rate (CTR) and conversion rate (CVR) prediction \citep{lee2012estimating,ahmed2014scalable}, is a crucial building box directly determining the follow-up bidding strategy to drive the performance of the RTB campaigns.

This chapter will start with a mathematical formulation of user response prediction. We will then discuss the most widely used linear models including logistic regression and Bayesian probit regression, and then move it to non-linear models including factorisation machines, gradient tree models, and deep learning. Our focus will be on the techniques that have been successful in various CTR prediction competitions for RTB advertising.\footnote{Some CTR prediction models, such as FTRL logistic regression \citep{mcmahan2013ad} and Bayesian profit regression \citep{graepel2010web}, though are originally proposed for sponsored search, are sufficiently generic for user response prediction in various scenarios, which are also presented in this chapter.}

\section{Data sources and problem statement}

\begin{figure}[t]
	\centering
	\includegraphics[width=0.8\columnwidth]{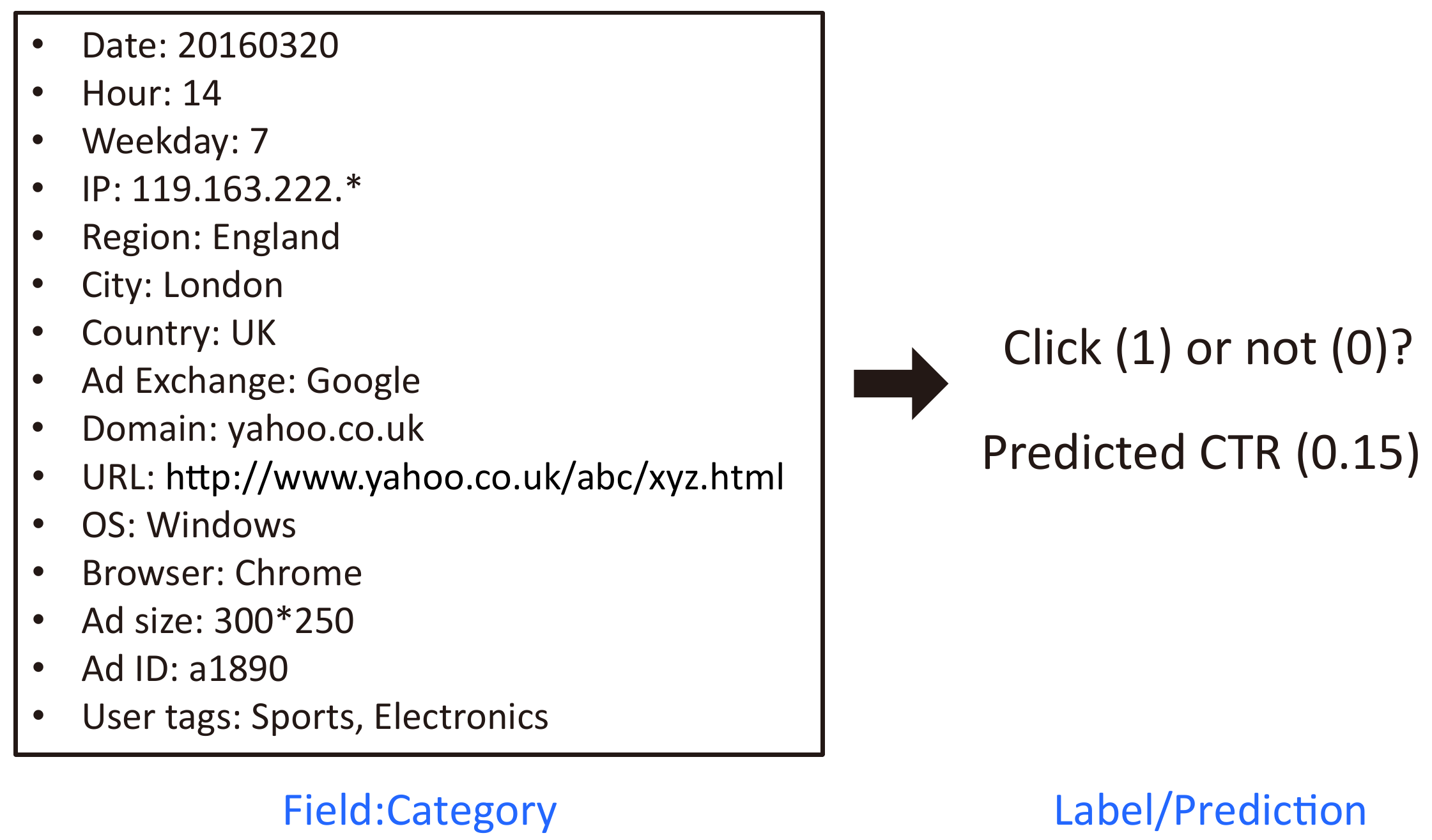}
	\caption{An example of multi-field categorical data instance of ad display context and its click label and CTR prediction.}\label{fig:ctr-prob-def}
\end{figure}

Figure~\ref{fig:ctr-prob-def} provides an illustration of a data instance of user response prediction problem. A data instance can be denoted as a $(\bx, y)$ pair, where $y$ is the user response label, usually binary, such as whether there is a user click on the ad (1) or not (0), $\bx$ is the input feature vector describing the user's context and the candidate ad.
As shown in the figure, the raw data of feature $\bx$ is normally in a multi-field categorical form.
For example, the field \texttt{Weekday} consists of 7 categories, i.e. \texttt{Monday}, \texttt{Tuesday},..., \texttt{Sunday}; the field \texttt{Browser} consists of several categories, e.g. \texttt{Chrome}, \texttt{IE}, \texttt{Firefox} etc.; the field \texttt{City} consists of tens of thousands of cities such as \texttt{London}, \texttt{Paris}, \texttt{Amsterdam}.

Advertisers or DSPs collect billions of such data instances daily to learn user response patterns. They may also collect information forming extra fields to extend the representation of the training data, such as joining and syncing with user's demographic data from a third-party data provider by the user cookie or device ID etc.

Typical feature engineering over such multi-field categorical data is \emph{One-Hot encoding} \citep{he2014practical}. In One-Hot encoding, each field is modelled as a high dimensional space and each category of this field is regarded as one dimension. Only the dimension with the field category is set as 1, while all others are 0. The encoded binary vectors for an example with three fields will be like
\[ \underbrace{[0,1,0,0,0,0,0]}_{\texttt{Weekday=Tuesday}}, \underbrace{[1,0,0,0,0]}_{\texttt{Browser=Chrome}}, \underbrace{[0,0,1,0,\ldots,0,0]}_{\texttt{City=London}}.\]

The binary encoded vector of the data instance is then created by concatenating the binary vectors of the field as
\[ \bx = [\underbrace{0,1,0,0,0,0,0}_{\texttt{Weekday=Tuesday}}, \underbrace{1,0,0,0,0}_{\texttt{Browser=Chrome}}, \underbrace{0,0,1,0,\ldots,0,0}_{\texttt{City=London}}].\]

Since the dimension of each field depends on the number of categories of this field, the resulted binary vector $\bx$ is extremely sparse. In some practical applications, the hash tricks are applied to reduce the vector dimensions \citep{chapelle2014simple,weinberger2009feature}


\section{Logistic regression with stochastic gradient descent}

We denote $\bs{x}\in \mathbb{R}^N$ to describe the binary bid request features as discussed previously.  A straightforward solution, the logistic regression model, to predict the CTR is given as
\begin{align}
\hat{y} = P(y=1|\bx) = \sigma(\bs{w}^T \bx) = \frac{1}{1 + e^{-\bs{w}^T \bx}}, \label{eq:sigmoid}
\end{align}
and the non-click probability is
\begin{align}
1 - \hat{y} = P(y=0|\bx) = \frac{e^{-\bs{w}^T \bx}}{1 + e^{-\bs{w}^T \bx}},
\end{align}
where $\bs{w}\in \mathbb{R}^N$ is the model coefficient vector, which represents a set of parameters to be learned over training data.

The cross entropy loss function is commonly used for training the logistic regression model:
\begin{align}
L(y, \hat{y}) = - y \log \hat{y} - (1-y) \log (1-\hat{y}).
\end{align}

In addition, the loss function is normally with a regularisation term to help the model avoid overfitting. With L2-norm regularisation, the loss function becomes:
\begin{align}
L(y, \hat{y}) = - y \log \hat{y} - (1-y) \log (1-\hat{y}) + \frac{\lambda}{2} \|\bs{w}\|_2^2.
\end{align}

Taking the derivation leads to the gradient on the efficient vector $\bs{w}$
\begin{align}
\frac{\partial L(y, \hat{y})}{\partial \bs{w}} = (\hat{y} - y) \bx + \lambda \bs{w}, \label{eq:lr-grad}
\end{align}
and with the learning rate $\eta$, the stochastic gradient descent (SGD) update of $\bs{w}$ is
\begin{align}
\bs{w} \leftarrow (1-\eta \lambda)\bs{w} + \eta (y - \hat{y}) \bx, \label{eq:lr-update}
\end{align}
for an instance $\bx$ randomly sampled from the training data. Note that the input bid request $\bx$ is a sparse vector and only contains a small number of non-zero entries that equals the number of the fields ($M$). Both the calculation of $\hat{y}$ and the update of coefficients $\bs{w}$ are very fast as they are only involved with the non-zero entries.

In SGD learning, the learning rate $\eta$ can be a fixed value, or a decayed value depending on the application. The decayed value of $\eta_t$ at $t$-th iteration can be updated as
\begin{align}
\eta_t = \frac{\eta_0}{\sqrt{t}}, \label{eq:lr-eta-update}
\end{align}
in order to fine-tune the parameters in the later stage. \cite{he2014practical} provided several practical updating schemes of $\eta$, and
the optimal implementation of $\eta$ updating depends on the specific training data.

\section{Logistic regression with follow-the-regularised-leader}

To bypass the problem of tuning the learning rate $\eta$ with various strategies, \cite{mcmahan2013ad} proposed an online learning algorithm of logistic regression, called \emph{follow-the-regularised-leader proximal} (FTRL-proximal).

In the $t$-th iteration, FTRL calculates the new coefficients $\bs{w}$ as
\begin{align}
\bs{w}_{t+1} = \argmin_{\bs{w}} \Big( \bs{w}^T \bs{g}_{1:t} + \frac{1}{2} \sum_{s=1}^t \sigma_s \|\bs{w} - \bs{w}_s\|^2_2 + \lambda_1 \|\bs{w}\|_1 \Big), \label{eq:ftrl-obj}
\end{align}
where $\bs{g}_{1:t} = \sum_{s=1}^{t} \bs{g}_s$ and $\bs{g}_s$ is the $s$-th iteration logistic regression gradient as in Eq.~(\ref{eq:lr-grad}) and $\sigma_s$ is defined as $\sigma_s = \frac{1}{\eta_t} - \frac{1}{\eta_{k-1}}$.

The solution of Eq.~(\ref{eq:ftrl-obj}) is in fact very efficient in time and space: only one parameter per coefficient needs to be stored. Eq.~(\ref{eq:ftrl-obj}) can be rewritten as
\begin{align}
\bs{w}_{t+1} = \argmin_{\bs{w}}  \bs{w}^T \Big(\bs{g}_{1:t} - \sum_{s=1}^t \sigma_s \bs{w}_s\Big) + \frac{1}{\eta_t} \|\bs{w}\|^2_2 + \lambda_1 \|\bs{w}\|_1. \label{eq:ftrl-obj-trans}
\end{align}

Let $\bs{z}_t$ be the number stored in memory at the beginning of each iteration $t$:
\begin{align}
\bs{z}_t = \bs{g}_{1:t} - \sum_{s=1}^t \sigma_s \bs{w}_s = \bs{z}_{k-1} + \bs{g}_t + \Big( \frac{1}{\eta_t} - \frac{1}{\eta_{t-1}} \Big) \bs{w}_t,
\end{align}
where the closed form solution for each dimension $i$ of $\bs{w}$ is
\begin{align}
w_{t+1,i}=\left\{
                \begin{array}{ll}
                  0 & \text{if~} |z_{t,i}| \leq \lambda_1\\
                  - \eta_t (z_{t,i} - \text{sign}(z_{t,i})\lambda_1) & \text{otherwise}
                \end{array}
              \right.
\end{align}
where $\text{sign}(z_{t,i}) = 1$ if $z_{t,i}$ positive and $-1$ otherwise, which is normally used in parameter updating with L1 regularisation.
Practically, it has been shown that logistic regression with FTRL works better than that with SGD.
The experiments on RTB and non-RTB advertising user response prediction have demonstrated the efficiency and effectiveness of the FTRL training \citep{mcmahan2013ad,ta2015factorization}.


\section{Bayesian probit regression}

\cite{graepel2010web} proposed a Bayesian learning model called Bayesian probit regression.  In order to deal with the uncertainty of the parameter estimation, the coefficients $\bs{w}$ is regarded as a random variable with p.d.f. defined as Gaussian distribution
\begin{align}
p(\bs{w}) = \mathcal{N}(\bs{w}; \bs{\mu}, \bs{\Sigma}).
\end{align}

The prediction function is thus modelled as a conditional distribution\footnote{In this section, for mathematical convenience, the binary label is denoted as $y=-1$ for non-click cases and $y=1$ for click cases.}
\begin{align}
P(y_i | \bs{x}_i, \bs{w}) = \Phi (y_i \bs{x}_i^T \bs{w}),
\end{align}
where the non-linear function $\Phi(\theta) = \int_{-\infty}^\theta \mathcal{N}(s; 0, 1)ds$ is the c.d.f. of the standard Gaussian distribution.
The model parameter $w$ is assumed to be drawn from a Gaussian distribution, which originates from a prior and updated with the posterior distribution by data observation.
\begin{align}
p(\bs{w} | \bs{x}_i, y_i) \propto P(y_i | \bs{x}_i, \bs{w}) \mathcal{N}(\bs{w}; \bs{\mu}_{i-1}, \bs{\Sigma}_{i-1})
\end{align}
The posterior is non-Gaussian and it is usually solved via variational methods in practice. Let $\mathcal{N}(\bs{w}; \bs{\mu}_{i}, \bs{\Sigma}_{i})$ be the posterior distribution of $\bs{w}$ after observing the case $(y_i, \bs{x}_i)$. The variational inference aims to minimise the Kullback-Leibler divergence by finding the optimal distribution parameters $\bs{\mu}_i$ and $\bs{\Sigma}_i$.
\begin{align}
(\bs{\mu}_i, \bs{\Sigma}_i) = \argmin_{(\bs{\mu}, \bs{\Sigma})} \mathbf{KL} \Big( \Phi (y_i \bs{x}_i^T \bs{w}) \mathcal{N}(\bs{w}; \bs{\mu}_{i-1}, \bs{\Sigma}_{i-1}) \| \mathcal{N}(\bs{w}; \bs{\mu}_{i}, \bs{\Sigma}_{i}) \Big)
\end{align}

Considering up to the second-order factors gives the closed form solution of this optimisation problem as
\begin{align}
\mu_{i} &= \mu_{i-1} + \alpha \bs{\Sigma}_{i-1} \bs{x}_i\\
\Sigma_{i} &= \Sigma_{i-1} + \beta (\bs{\Sigma}_{i-1} \bs{x}_i)(\bs{\Sigma}_{i-1} \bs{x}_i)^T,
\end{align}
where
\begin{align}
\alpha &= \frac{y_i}{\sqrt{\bs{x}_i^T \bs{\Sigma}_i \bs{x}_i + 1}} \frac{\mathcal{N}(\theta)}{\Phi(\theta)}\\
\beta &= \frac{1}{\sqrt{\bs{x}_i^T \bs{\Sigma}_i \bs{x}_i + 1}} \frac{\mathcal{N}(\theta)}{\Phi(\theta)} \Big( \frac{\mathcal{N}(\theta)}{\Phi(\theta)} + \theta \Big),
\end{align}
where
\begin{align}
\theta = \frac{y_i \bs{x}_i^T \bs{\mu}_{i-1}}{\sqrt{\bs{x}_i^T \bs{\Sigma}_{i-1} \bs{x}_i + 1}}.
\end{align}

Note that \cite{graepel2010web} assumed the independence of features and only focused on the diagonal elements in $\bs{\Sigma}_i$ in practice.

\vspace{20px}
To sum up, the above sections have introduced several linear models and their variations for the response prediction problem. Linear models are easy to implement with high efficiency. However, the regression methods can only learn shallow feature patterns and lack the ability to catch high-order patterns unless an expensive pre-process  of combining features is conducted \citep{cui2011bid}. To tackle this problem, non-linear models such as factorisation machines (FM) \citep{menon2011response,ta2015factorization} and tree-based models \citep{he2014practical} are studied to explore feature interactions.

\section{Factorisation machines}

\cite{rendle2010factorization} proposed the factorisation machine (FM) model to directly explore  features' interactions by mapping them into a low dimensional space:
\begin{equation}
\hat{y}_{\text{FM}}(\bs{x})=\sigma \Big( w_{0}+\sum_{i=1}^{N} w_{i} x_{i} + \sum_{i=1}^{N}\sum_{j=i+1}^{N} x_{i} x_{j} \bs{v}_{i}^T\bs{v}_{j}\Big), \label{eq:fm}
\end{equation}
where each feature $i$ is assigned with a bias weight $w_i$ and a $K$-dimensional vector $\bs{v}_i$; the feature interaction is modelled as their vectors' inner products between $\bs{v}_i$. The activation function $\sigma()$ can be set according to the problem. For CTR estimation, the sigmoid activation function, as in Eq.~(\ref{eq:sigmoid}), is normally used. Note that Eq.~(\ref{eq:fm}) only involves the second-order feature interactions. By naturally introducing tensor product, FM supports higher-order feature interactions.

FM is a natural extension of matrix factorisation that has been very successful in collaborative filtering based recommender systems \citep{koren2009matrix}. Specifically, when there are only two fields in the data: user ID $u$ and item ID $i$, the factorisation machine prediction model will be reduced to
\begin{align}
\hat{y}_{\text{MF}}(\bs{x})=\sigma \Big( w_{0}+ w_{u} + w_{i} + \bs{v}_{u}^T\bs{v}_{i}\Big), \label{eq:mf}
\end{align}
which is the standard matrix factorisation model.

The inner product of feature vectors practically works well to explore the feature interactions, which is important in collaborative filtering.
With the success of FM in various recommender systems \citep{rendle2010pairwise,rendle2011fast,chen2011feature,loni2014cross}, it is natural to explore FM's capability in CTR estimation.

\cite{menon2011response} proposed to use collaborative
filtering via matrix factorisation for the ad CTR estimation task.
Working on the mobile ad CTR estimation problem, \cite{oentaryo2014predicting} extended FM into a hierarchical importance-aware factorisation machine (HIFM) by incorporating importance weights and hierarchical learning into FM.
\cite{ta2015factorization} borrowed the idea of FTRL in \citep{mcmahan2013ad} and applied this online algorithm onto FM and reported significant performance improvement on ad CTR estimation.


\section{Decision trees}
\begin{figure}[t]
	\centering
	\includegraphics[width=0.6\columnwidth]{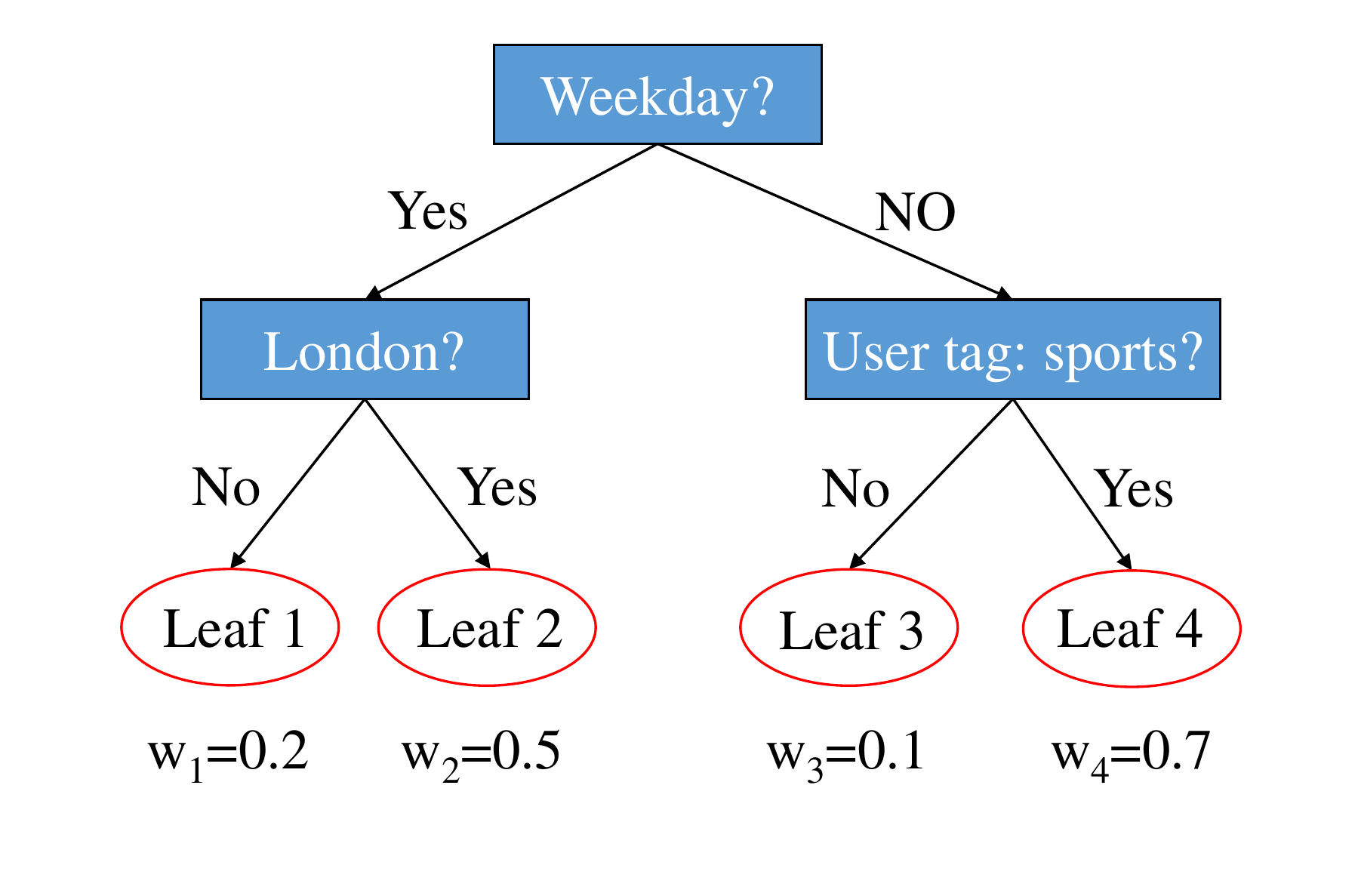}
	\caption{An example decision tree, where a given bid request $\bx$ is mapped to a leaf, and the weight $w$ at the leaf produces the prediction value.}\label{fig:tree}
\end{figure}

Decision tree is a simple non-linear supervised learning method \citep{breiman1984classification} and can be used for CTR estimation. The tree model that predicts the label of a given bid request is done by learning a simple sequential, tree structured (binary), decision rule from the training data.  As illustrated in Figure \ref{fig:tree}, a bid request instance $\bx$ is parsed through the tree on the basis of its attributes, and at the end arrived at one of the leaves. The weight assigned to the leaf is used as the prediction. Mathematically, a decision tree is a function, denoted as $f(\bx)$ below:
\begin{align}
\hat{y}_{\text{DT}}(\bs{x})= f(\bx) = w_{I(\bx)}, \quad I: R^d \to \{1,2,...,T\}, \label{eq:tree}
\end{align}
where  a leaf index function $I(\bx)$ is introduced to map an instance bid request $\bx$ to a leaf $t$. Each leaf is assigned a weight $w_t$, $t=1,...,T$, where $T$ denotes the total number of leaves in the tree, and the prediction is assigned as the weight of the mapped leaf $I(\bx)$.

\section{Ensemble learning}
A major problem of decision trees is their high variance --- a small change in the data could often lead to very different splits in the tree \citep{friedman2001elements}. To reduce the instability  and also help avoiding over-fitting the data, in practice, multiple decision trees can be combined together. Our discussion next focuses on \emph{bagging} and \emph{boosting}, the two most widely-used ensemble learning techniques.

\subsection{Bagging (bootstrap aggregating)}
\emph{Bagging} (Bootstrap aggregating) is a technique averaging over a set of  predictors that have been learned over randomly-generated training sets (bootstrap samples) \citep{breiman1996bagging}.

More specifically, given a standard training set $D$ of $n$ training examples, first generate $K$ new training sets $D_{k}$, each of size $n$, by sampling from $D$ uniformly and with replacement (some examples may be repeated in each dataset). The $K$ (decision tree) models are then fitted using the $K$ bootstrapped training sets. The bagging estimate is obtained by averaging their outputs:
\begin{align}
\hat{y}_{\text{bag}}(\bs{x})=\frac{1}{K}\sum_{k=1}^K \hat f^{\text{bag}}_k(\bx). \label{eq:bag}
\end{align}

A good bagging requires the basis decision trees to be as little correlated as possible. However, if one or a few features (e.g, the field \texttt{Weekday}, \texttt{Browser}, or  \texttt{City}) are strong predictors for the target output, these features are likely to be selected in many of the trees, causing them to become correlated. \emph{Random Forest} \citep{breiman2001random} solves this by only selecting a random subset of the features as a node to split (the tree), leading to more stable and better performance in practice.

\subsection{Gradient boosted regression trees}
By contrast, \emph{boosting} methods aim at combining the outputs of many \emph{weak} predictors (in an extreme case, perform just slightly better than random guessing) to produce a single \emph{strong} predictor \citep{friedman2002stochastic,friedman2001greedy}. Unlike bagging where each basis predictor (e.g., decision tree) is independently constructed using a bootstrap
sample of the data set, boosting works in an iterative manner; it relies on building the successive trees that continuously adjust the prediction that is incorrectly produced by earlier predictors. In the
end, a weighted vote is taken for prediction. More specifically, for a boosted decision tree:
\begin{align}
\hat{y}_{\text{GBDT}}(\bs{x})=\sum_{k=1}^K f_k(\bx), \label{eq:mf2}
\end{align}
where the prediction is a linear additive combination of the outputs from $K$ numbers of basis decision trees $f_k(\bx)$. For simplicity, consider the weight of each predictor is equal.

Thus, the tree learning is to find right decision $f_k$ so that the following objective is minimised:
\begin{align}
\sum_{i=1}^{N} L(y_i,\hat y_i) + \sum_{k=1}^{K} \Omega (f_k),\label{eq:obj-learn}
\end{align}
where the objective consists of a loss function $L$ of the predicted values $\hat y_i$, $i=1,...,N$, against the true labels $y_i$, $i=1,...,N$, and a regularisation term $\Omega$ controlling the complexity of each tree model. One can adopt \emph{Forward Stagewise Additive} training by sequentially adding and training a new basis tree without adjusting the parameters of those that have already been added and trained \citep{friedman2001elements}. Formally, for each $k=1,...,M$ stage
\begin{align}
\hat f_k = \argmin_{f_k} \sum_{i=1}^{N} L\big(y_i,F_{k-1}(\bx_i) +  f_k( \bx_i)\big) + \Omega (f_k)  \label{eq:FSAT}
\end{align}
where
\begin{align}
 F_0(\bx) &= 0, \\
 F_k(\bx) &=  F_{k-1}(\bx) +  \hat f_k( \bx),  \label{eq:FSAT2}
 \end{align}
and the training set is denoted as $\{y_i,\bx_i\}, i=1,...,N$. The complexity of a tree is defined as
\begin{align}
\Omega (f_k) = \gamma T + \frac{1}{2}\lambda\sum_{t=1}^T w^2_t, \label{eq:complexity} \end{align}
where $\lambda$ and $\gamma$ are the hyper-parameters. $t$ is the index of the features and there are $T$ number of features.

To solve Eq.~\ref{eq:FSAT}, one can consider the objective as the function of prediction $\hat y_k$ and make use of Taylor expansion of the objective:
\begin{align}
&\sum_{i=1}^{N} L\big(y_i,F_{k-1}(\bx_i) +  f_k( \bx_i)\big) + \Omega (f_k) \nonumber\\
\simeq &  \sum_{i=1}^{N} \Big(L^{k-1} + \frac{\partial L^{k-1}}{\partial \hat y_i^{k-1}}  f_k( \bx_i) + \frac{1}{2} \frac{\partial^2 L^{k-1}}{\partial^2 \hat y_i^{k-1}}   f_k( \bx_i)^2 \Big) + \Omega (f_k) \nonumber \\
\simeq  & \sum_{i=1}^{N} \Big(\frac{\partial L^{k-1}}{\partial \hat y_i^{k-1}}  f_k( \bx_i) + \frac{1}{2} \frac{\partial^2 L^{k-1}}{\partial^2 \hat y_i^{k-1}}   f_k( \bx_i)^2 \Big) + \Omega (f_k),\label{eq:FSAT3}
\end{align}
where define $L^{k-1}=L\big(y_i,\hat y_i^{k-1}\big)=L\big(y_i,F_{k-1}(\bx_i)\big)$, which can be safely removed as it is independent of the added tree $f_k(\bx_i)$. Replacing Eqs.~(\ref{eq:tree}) and (\ref{eq:complexity}) into the above gives:
\begin{align}
& \sum_{i=1}^{N} \Big(\frac{\partial L^{k-1}}{\partial \hat y_i^{k-1}}  f_k( \bx_i) +  \frac{1}{2}\frac{\partial^2 L^{k-1}}{\partial^2 \hat y_i^{k-1}}   f_k( \bx_i)^2 \Big) + \Omega (f_k) \nonumber \\
= & \sum_{i=1}^{N} \Big(\frac{\partial L^{k-1}}{\partial \hat y_i^{k-1}}  w_{I( \bx_i)} +  \frac{1}{2}\frac{\partial^2 L^{k-1}}{\partial^2 \hat y_i^{k-1}}   w_{I( \bx_i)}^2 \Big) + \gamma T + \frac{1}{2}\lambda\sum_{t}^T w^2_t \\
= & \sum_{t=1}^{T} \Big((\sum_{i \in I_t}\frac{\partial L^{k-1}}{\partial \hat y_i^{k-1}})  w_t +  \frac{1}{2}(\sum_{i \in I_t}  \frac{\partial^2 L^{k-1}}{\partial^2 \hat y_i^{k-1}} + \lambda)  w_t^2 \Big) + \gamma T, \label{eq:FSAT4}
\end{align}
where $I_t$ denotes the set of instance indexes that have been mapped to leaf $t$. The above is the sum of $T$ independent quadratic functions of the weights $w_t$. Taking the derivative and making it to zero obtains the optimal weights for each leaf $t$ as:
\begin{align}
w_t = -\frac{\sum_{i \in I_t}(\partial L^{k-1})/(\partial \hat y_i^{k-1})}{ \Big(\sum_{i \in I_t}( \partial^2 L^{k-1})/(\partial^2 \hat y_i^{k-1})+ \lambda \Big)}. \label{eq:leaf-weight}
\end{align}

Placing back the weights give the optimal objective function as
\begin{align}
& - \frac{1}{2}\sum_{t=1}^{T}  \frac{\big(\sum_{i \in I_t}(\partial L^{k-1}/\partial \hat y_i^{k-1})\big)^2}{ \Big(\sum_{i \in I_t}( \partial^2 L^{k-1})/(\partial^2 \hat y_i^{k-1})+ \lambda \Big)} + \gamma T. \label{eq:FSAT5}
\end{align}

The above solution is a general one where one can plugin many loss functions.
Let us consider a common squared-error loss:
\begin{align}
L\Big(y_i,F_{k-1}(\bx_i) +  f_k( \bx_i)\Big)= &\Big(y_i - F_{k-1}(\bx_i) -  f_k( \bx_i)\Big)^2 \\
= &  \Big(r_{ik} -  f_k( \bx_i)\Big)^2; \\
L^{k-1}=L\Big(y_i,F_{k-1}(\bx_i)\Big)= & \Big(y_i - F_{k-1}(\bx_i))\Big)^2 = (r_{ik})^2,
\label{eq:squared-error}
\end{align}
where $r_{ik}$ is simply the residual of the previous model on the $i$-th training instance. Replacing the first and second derivatives of $L^{k-1}$ into Eq.~(\ref{eq:leaf-weight}) gives:
\begin{align}
w_t = \frac{\sum_{i \in I_t} y_{i} - F_{k-1}(\bx_i)}{ \Big(|I_t| + \lambda/2 \Big)} = \frac{\sum_{i \in I_t} r_{ik}}{ \Big(|I_t| + \lambda/2 \Big)}, \label{eq:leaf-weight-squared-error}
\end{align}
where one can see that the optimal weight at leaf $t$ is the one that best fits the previous residuals in that leaf when the regularisation $\lambda$ is not considered.

With the optimal objective function in Eq.~(\ref{eq:FSAT5}), one can take a greedy approach to grow the tree and learn its tree structure: for each tree node, all features are enumerated. For each feature, sort the instances by feature value. If it is real-valued, use a linear scan to decide the best split along that feature. For categorical data in our case, one-hot-encoding is used. The best split solution is judged based on the optimal objective function in Eq.~(\ref{eq:FSAT5}) and take the best split solution along all the features. A distributed version can be found at \citep{Tyree:2011:PBR:1963405.1963461}.

\subsection{Hybrid Models}
All the previously mentioned user response prediction models are not mutually exclusive and in practice they can be fused in order to boost the performance. Facebook reported a solution that combines decision
trees with logistic regression \citep{he2014practical}. The idea was to non-linearly transform the input features by gradient boosted regression trees (GBRTs) and the newly transformed features were then treated as new categorical input features to a sparse linear classifier. To see this, suppose there is an input bid request with features $\bx$. Multiple decision trees are learned from GBRTs. If the first tree thinks $\bx$ belong to node 4, the second node 7, and the third node 6, then the features \texttt{1:4, 2:7, 3:6} are generated for this bid request and then hash it in order to feed to a regressor.  A follow-up solution from the Criteo Kaggle CTR contest also reported that using a field-aware factorisation machine \citep{juanfield} rather than logistic regression as the final prediction from GBRTs features had also led to improved accuracy of the prediction.

\section{User lookalike modelling}
Compared with sponsored search or contextual advertising, RTB advertising has the strength of being able to directly target specific users --- \emph{explicitly} build up user profiles and detect user interest segments via tracking their online behaviours, such as browsing history, clicks, query words, and conversions.\footnote{Note that in sponsored search, user-level targeting is still possible, namely remarketing lists for search ads (RLSA), but it is more emphasised and more widely used in RTB display advertising. More details about RLSA: \url{https://support.google.com/adwords/answer/2701222?hl=en}.}

Thus, when estimating the user's response (rate), the advertiser would be able to,  on the basis of the learned user profiles, identify and target unknown users who are found in ad exchanges and have the similar interests and commercial intents with the known (converted) customers. Such technology is referred as user \emph{look-alike modelling} \citep{zhang2016implicit,mangalampalli2011feature}, which in practice has proven to be effective in providing high targeting accuracy and thus bringing more conversions to the campaigners \citep{yan2009much}.

The current user profiling methods include building keyword and topic distributions \citep{ahmed2011scalable} or clustering users onto a (hierarchical) taxonomy \citep{yan2009much}. Normally, these inferred user interest segments are then used as target restriction rules or as features leveraged in predicting users' ad response \citep{Zhang:2014:ORB:2623330.2623633}, where those regression techniques introduced previously, e.g., logistic regression, matrix factorisation, boosted decision trees, can be incorporated.

However, a major drawback of the above solutions is that the user interest segments building, is performed independently and has little attention of its use of ad response prediction. In the next section, we shall introduce a particular technique, transfer learning that implicitly transfers the knowledge of user browsing pattens to that of the user response prediction \citep{zhang2016implicit}.

\section{Transfer learning from Web browsing to ad clicks}
Transfer learning deals with a learning problem where the training data of the target task is expensive to get, or easily outdated; the training is helped by transferring the knowledge learned from other related tasks \citep{pan2010survey}. Transfer learning has been proven to work on a variety of problems such as classification \citep{dai2007transferring}, regression \citep{liao2005logistic} and collaborative filtering \citep{li2009transfer}. It is worth mentioning that there is a related technique called multi-task learning, where the data from different tasks are assumed to drawn from the same distribution \citep{taylor2009transfer}. By contrast, transfer learning methods may allow for arbitrary source and target tasks. In real time bidding based advertising, \cite{dalessandro2014scalable} proposed a transfer learning scheme based on logistic regression prediction models, where the parameters of ad click prediction model were restricted with a regularisation term from the ones of user Web browsing prediction model. \cite{zhang2016implicit} extended it by considering matrix factorisation to fully explore the benefit of collaborative filtering.

Specifically, in RTB advertising, there are commonly two types of observations about underlying user behaviours: one from their browsing behaviours (the interaction with webpages) and one from their ad responses, e.g., conversions or clicks, towards display ads (the interactions with the ads) \citep{dalessandro2014scalable}. There are two predictions tasks for understanding user behaviours:

\textit{Web Browsing Prediction (CF Task).} Each user's online browsing behaviour is logged as a list containing previously visited publishers (domains or URLs). A common task of using data is to leverage collaborative filtering (CF) \citep{wang2006unifying,rendle2010factorization} to infer the users' profile, which is then used to predict whether the user is interested in visiting any given new publisher. Formally, denote the dataset for CF as $D_{\tweb}$, which contains $N_{\tweb}$ users and $M_{\tweb}$ publishers and an observation is denoted as $(\bs{x}_{\tweb}, y_{\tweb})$, where $\bs{x}_{\tweb}$ is a feature vector containing the attributes from users and publishers and $y_{\tweb}$ is the label indicating whether the user visits the publisher or not.

\textit{Ad Response Prediction (CTR Task).} Each user's online ad feedback behaviour is logged as a list of pairs of rich-data ad impression event and the corresponding feedback (e.g., click or not). The task is to build a click-through rate (CTR) prediction model \citep{chapelle2014simple} to estimate how likely it is that the the user will click a specific ad impression in the future. Each ad impression event consists of various information, such as user data (cookie ID, location, time, device, browser, OS etc), publisher data (domain, URL, ad slot position etc), and advertiser data (ad creative, creative size, campaign etc.). Mathematically, denote the ad CTR dataset as $D_{\tads}$ and its data instance as $(\bs{x}_{\tads}, y_{\tads})$, where $\bs{x}_{\tads}$ is a feature vector and $y_{\tads}$ is the label indicating whether the user clicks a given ad or not.

Although they are different prediction tasks, two tasks share a large proportion of users and their features. Thus a user interest model can be built jointly from the two tasks. Typically there are large amount of observations about user browsing behaviours and it may be promising to use the knowledge learned from publisher CF recommendation to help infer display advertising CTR estimation.

In the solution proposed by \cite{zhang2016implicit}, the prediction models on the CF task and CTR task are learned jointly. Specifically, the authors build a joint data generation framework by the two prediction tasks.For both predictions, the labels are considered to be binary. More detailed implementation of the CF and CTR prediction modules are provided in \citep{zhang2016implicit}.

\section{Deep learning over categorical data}
Deep neural networks have demonstrated their superior performance on various tasks ranging from image recognition \citep{he2015delving}, speech recognition \citep{hinton2012deep}, and machine translation \citep{bahdanau2014neural} to natural language processing \citep{collobert2008unified}. In some of tasks, the prediction accuracy has arguably reached a comparable human-level \citep{mnih2015human,he2015delving}.

A notable similarity among those tasks is that visual, aural, and textual signals are known to be spatially and/or temporally correlated. The newly introduced unsupervised training on deep structures \citep{bengio2007greedy} or embedding the knowledge as prior would be able to explore such \emph{local} dependency and establish a \emph{sparse} representation of the feature space, making neural network models effective in learning high-order features directly from the raw feature input. For instance, convolutional neural networks employ local filters to explore local dependencies among pixels \citep{lecun1998gradient,lecun1995convolutional}.

With such learning ability, deep learning would be a good candidate to estimate online user response rate such as ad CTR. However, there are two difficulties: firstly, most input features in CTR estimation are discrete categorical features, e.g., the user location city, device type, publisher website, ad category, and their local dependencies (thus the sparsity in the feature space) are unknown.  Thus how deep learning improves the CTR estimation via learning feature representation on such large-scale multi-field discrete categorical features. Secondly, training deep neural networks (DNNs) on a large input feature space requires tuning a huge number of parameters, which is computationally expensive. For instance, unlike the image and video cases, there are about 1 million binary input features and 100 hidden units in the first layer; then it requires 100 million links to build the first layer neural network.

\cite{zhang2016deep} studied deep learning over a large multi-field categorical feature space by using embedding methods in both supervised and unsupervised fashions. Two types of deep learning models were introduced: Factorisation Machine supported Neural Network (FNN) and Sampling-based Neural Network (SNN). Specifically, FNN with a supervised-learning embedding layer using factorisation machines \citep{rendle2012factorization} is proposed to efficiently reduce the dimension from sparse features to dense continuous features. The second model SNN is a deep neural network powered by a sampling-based restricted Boltzmann machine or a sampling-based denoising auto-encoder with a proposed negative sampling method. Based on the embedding layer, multiple layers neural nets with full connections were built to explore non-trivial data patterns. Later, inspired by FNN and FM, \cite{qu2016product} proposed a product-based feature-interaction layer between the feature embedding layer and the full connection layers to induce the feature interaction learning.

In the context of sponsored search, \cite{zhang2014sequential} argued that user's behaviours on ads are highly correlated with historical events of the user's activities, such as what queries they submitted, what ads they clicked or ignored in the past, and how long they spent on the landing pages of clicked ads. They used recurrent neural networks (RNN) to directly model the dependency on user's sequential behaviours into the click prediction process through the recurrent structure in the network.

\section{Dealing with missing data}
As discussed in Chapter \ref{c-auct}, the training data (observed clicks/non-clicks for a bid request) for user response estimation is biased and contains missing data. Only when a bid wins the auction,  are the corresponding labels, i.e., user response $y$ (either click or conversion) and market price $z$, observed. Thus, the probability of a data instance (contains bid request feature, labels, market price) $(\bx, y, z)$ being observed relies on whether the bid $\bidx$ would win or not. Following Chapter \ref{c-auct}, we denote it as $P(\text{win}|\bx, \bidx)$. As before, the generative process of creating observed training data $D = \{(\bx, y, z)\}$ is summarised as:
\begin{align}
\underbrace{q_x(\bx)}_{\text{winning impression}} \equiv \underbrace{P(\text{win}|\bx, \bidx)}_{\text{prob. of winning the auction}} \cdot \underbrace{p_x(\bx)}_{\text{bid request}}, \label{eq:biased-training-data}
\end{align}
where probability $q_x(\bx)$ describes how feature vector $\bx$ is distributed within the training data. As discussed in Chapter \ref{c-auct}, the above equation indicates the relationship (bias) between the p.d.f. (probabilistic density function) of the pre-bid full-volume bid request data (prediction) and the post-bid winning impression data (training); in other words, the predictive models would be trained on $D$, where $\bx \sim q_x(\bx)$, and be finally operated on prediction data $\bx \sim p_x(\bx)$.

In Chapter \ref{c-auct}, we have presented an unbiased estimation of the winning probability $P(\text{win}|\bx, \bidx)$ using a non-parametric survival model. Here we shall introduce a solution of using it for creating bid-aware gradient descent to solve CTR estimation for both logistic regression and factorisation machine as well as neural networks \citep{zhang2016bidaware}.

\begin{figure}[t]
	\centering
	\includegraphics[width=0.8\textwidth]{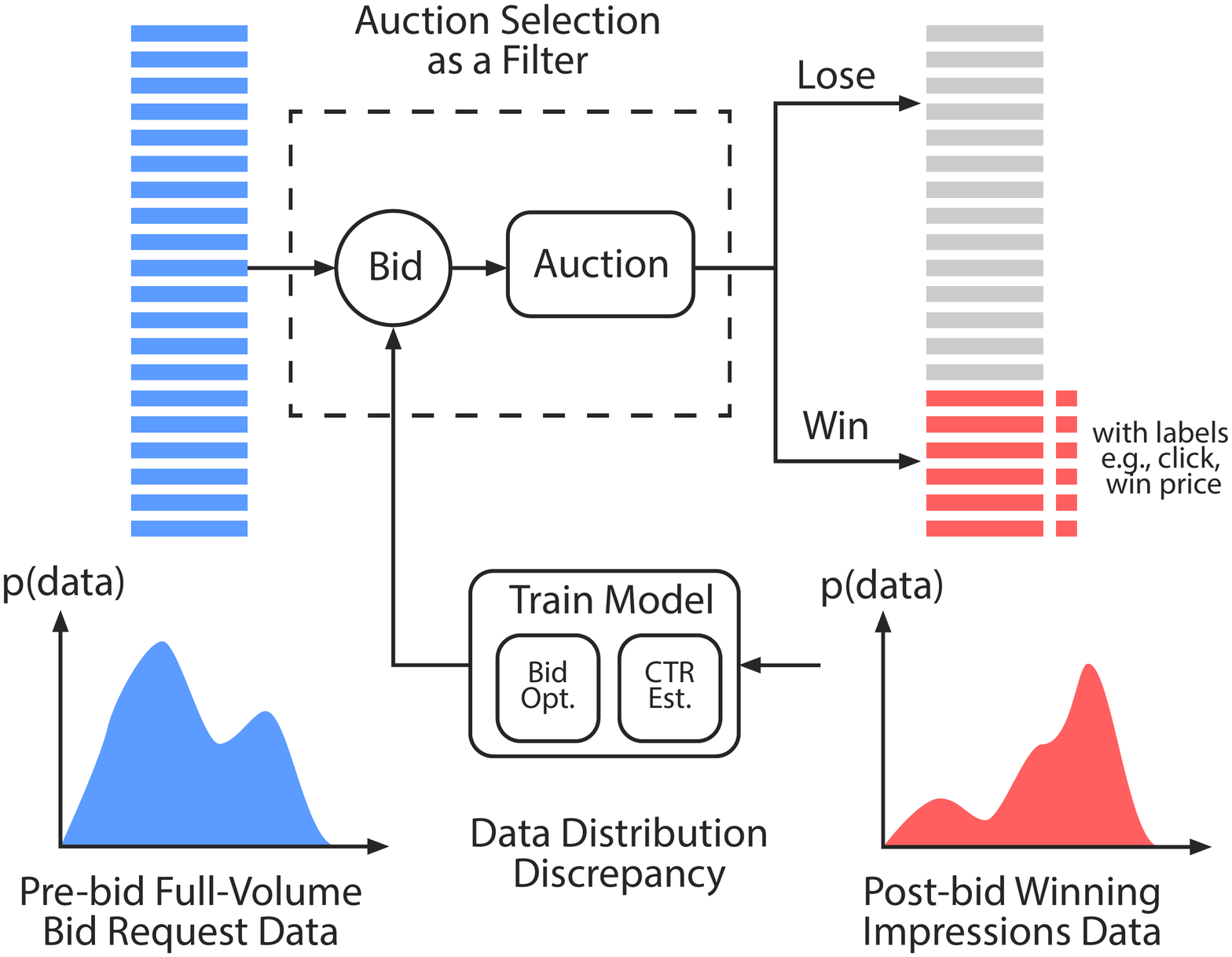}\\
	\caption{From an advertiser perspective, the ad auction selection acts as a dynamic data filter based on bid value, which leads to distribution discrepancy between the post-bid training data (red, right) and pre-bid prediction data (blue, left). Source: \citep{zhang2016bidaware}.}\label{fig:ibias-auction-selection}
\end{figure}

Generally, given a training dataset $D = \{(\bx, y, z)\}$, where the data instance $\bx$ follows the training data distribution $q_x(\bx)$, (the red data distribution in Figure~\ref{fig:ibias-auction-selection}), an \emph{unbiased} supervised learning problem, including previously mentioned logistic regression, matrix factorisation and deep learning, can be formalised into a loss-minimisation problem on prediction data distribution $p_x(\bx)$ (the blue data distribution in Figure~\ref{fig:ibias-auction-selection}):
\begin{align}
\min_{\bt}~~ \mathbb{E}_{\bx \sim p_{x}(\bx)}[\mathcal{L}(y, f_{\bt}(\bx))] + \lambda \Phi(\bt),\label{eq:continous-sl}
\end{align}
where $f_{\bt}(\bx)$ is $\bt$-parametrised prediction model to be learned; $\mathcal{L}(y, f_{\bt}(\bx))$ is the loss function based on the ground truth $y$ and the prediction $f_{\bt}(\bx)$; $\Phi(\bt)$ is the regularisation term that penalises the model complexity; $\lambda$ is the regularisation weight. With Eqs.~(\ref{eq:pq}) and (\ref{eq:win-mp}), one can use importance sampling to reduce the bias of the training data:
\begin{align}
& \mathbb{E}_{\bx \sim p_{x}(\bx)}[\mathcal{L}(y, f_{\bt}(\bx))] = \int_{\bx} p_x(\bx) \mathcal{L}(y, f_{\bt}(\bx)) d\bx \nonumber \\
=& \int_{\bx} q_x(\bx)\frac{\mathcal{L}(y, f_{\bt}(\bx))}{w(\bidx)} d\bx = \mathbb{E}_{\bx \sim q_{x}(\bx)}\Big[\frac{\mathcal{L}(y, f_{\bt}(\bx))}{w(\bidx)}\Big]  \label{eq:importance-sampling}\\
=& \frac{1}{|D|}\sum_{(\bx,y,z) \in D} \frac{\mathcal{L}(y, f_{\bt}(\bx))}{w(\bidx)} = \frac{1}{|D|}\sum_{(\bx,y,z) \in D} \frac{\mathcal{L}(y, f_{\bt}(\bx))}{1- \prod_{b_j < \bidx} \frac{n_j - d_j}{n_j}}. \nonumber
\end{align}

Based on this framework, if the auction winning probability $w(\bidx)$, e.g., Eq.~(\ref{eq:km}), is obtained, one can eliminate the bias for each observed training data instance. Let us look at the case of CTR estimation with logistic regression
\begin{align}
f_{\bt}(\bx) = \frac{1}{1+e^{-\bt^T \bx}}.
\end{align}
With the cross entropy loss between the binary click label $\{0,1\}$ and the predicted probability and L2 regularisation, the framework of Eq.~(\ref{eq:importance-sampling}) is written as
\begin{align}
\min_{\bt} \frac{1}{|D|} \sum_{(\bx,y,z) \in D} \frac{-y \log f_{\bt}(\bx) - (1-y)\log (1-f_{\bt}(\bx))}{1- \prod_{b_j < \bidx} \frac{n_j - d_j}{n_j}} + \frac{\lambda}{2} \|\bt\|_2^2, \label{eq:unbias-lr}
\end{align}
where the winning probability $w(\bidx)=1- \prod_{b_j < \bidx} \frac{n_j - d_j}{n_j}$ is estimated for each observation instance, which is independent from the CTR estimation parameter $\bt$.\footnote{In this section we use $\bt$ to denote the weights of logistic regression instead of $\bs{w}$ in order to avoid the term conflict with the winning probability $w(\bidx)$.} The update rule of $\bt$ is routine using stochastic gradient descent. The derived \emph{Bid-aware Gradient Descent} (BGD) calculation of Eq.~(\ref{eq:unbias-lr}) is\footnote{The term $|D|$ is merged into $\lambda$ for simplicity.}
\begin{align}
\bt \leftarrow (1 - \eta \lambda) \bt + \frac{\eta}{1- \prod_{b_j < \bidx} \frac{n_j - d_j}{n_j}} \Big(y - \frac{1}{1+e^{-\bt^T \bx}}\Big) \bx. \label{eq:ctr-gradient}
\end{align}

From the equation above, one can observe that with a lower winning bid $\bidx$, the probability $1- \prod_{b_j < \bidx} \frac{n_j - d_j}{n_j}$ of seeing the instance in the training set is lower. However, the corresponding gradient from the data instance is higher and vice versa as it is in the denominator.

This is intuitively correct as when a data instance $\bx$ is observed with low probability, e.g., 10\%, one can infer there are 9 more such kind of data instances missed because of auction losing. Thus the training weight of $\bx$ should be multiplied by 10 in order to recover statistics from the full-volume data. By contrast, if the winning bid is extremely high, which leads 100\% auction winning probability, then such data is observed from the true data distribution. Thus there will be no gradient reweighting on this data. Such non-linear relationship  has been well captured in the bid-aware learning update in the gradient updates.






\section{Model comparison}
As discussed in this chapter, there are various user response prediction models.
From the modeling perspective, these models can be generally categorised as linear and non-linear models.

Linear models, including logistic regression \citep{lee2012estimating,mcmahan2013ad} and Bayesian probit regression (with diagonal covariance matrix) \citep{graepel2010web},  directly build the model based on the feature independence assumption. For linear models, the feature interaction patterns are generally captured by building large-scale feature space with combining multi-field features, which could consume much human effort. However, thanks to its high efficiency and high parallelization capability, linear models are able to be fed in much more training data instances (and higher dimensional features) during the same training period, which makes them still highly comparable with the non-linear models in many industrial environments.

Non-linear models, including factorisation machine \citep{rendle2010factorization,juanfield}, tree models \citep{he2014practical} and recently emerged (deep) neural networks models \citep{zhang2016deep,qu2016product}, provide model capacity of automatically learning feature interaction patterns without the need of designing combining features. These non-linear models generally need much more computation resources than the linear ones, and some of them may require multiple stages of model training, as demostrated in \citep{he2014practical}. With the fast development high performance computing (HPC) and the explosion of data volume, non-linear models are more and more applied in commercial platforms for practical user response prediction.

\section{Benchmarking}
There are some real-world benchmark datasets that are publicly avaiable for the research of user response prediction in RTB display advertising.

\begin{description}
	\item[iPinYou] dataset \citep{liao2014ipinyou} comes from the company hosted global RTB algorithm competition during over 10 days in 2013, including 3 seasons. Normally the data of season 2 and 3 are used because of the same data schema, with 24 semi-hashed fields and the user click/conversion responses. Download link:\\ \url{http://data.computational-advertising.org}.
	\item[Criteo 1TB Click Log] is a realistic industry-defining benchmark release by Criteo.
	The dataset contains 24 days of ad click logs, and the data has been cleaned and hashed, including 13 numerical and 26 categorical fields, labeled with the  user click responses. Download link: \url{http://labs.criteo.com/downloads/download-terabyte-click-logs/}.
	\item[Criteo Conversion Log] is another dataset from Criteo focusing on conversion prediction \citep{chapelle2014modeling}.  Download link: \url{http://labs.criteo.com/2013/12/conversion-logs-dataset/}.
	\item[YOYI] runs a major DSP focusing on multi-device display advertising in China.
	YOYI dataset \citep{ran2016cikm} contains 402M impressions, 500K clicks and 428K CNY expense during 8 days in Jan.~2016. 
	Download link: \url{ http://goo.gl/xaao4q}.
	\item[Avazu] releases its 11-day online ad click data for its Avazu Click Prediction  competition on Kaggle. 
	Download link: \url{https://www.kaggle.com/c/avazu-ctr-prediction/data}.
\end{description}

Note that iPinYou and YOYI datasets further contain bidding and auction winning price information, which supports the research of bidding strategy optimisation as will be discussed in Chapter~\ref{c-bid}.

Since there are different settings of user response prediction, thus it could be more feasible to list some benchmarking references \citep{zhang2014real,he2014practical,lee2012estimating,zhang2016deep,qu2016product,ren2016user} than conducting a specific setting of experiments in this survey.

\chapter{Bidding Strategies}
\label{c-bid}
Once it is able to estimate the effectiveness of an ad impression from the previous chapter by estimating user's response rates (e.g., the click-through rate or the conversion rate), the next step is to use it for optimising the bidding. A bidding strategy refers to the logic of deciding a bid price given an ad display opportunity. Good bidding strategy directly leads to effective and accurate RTB ad delivery. Thus designing an optimal bidding strategy is regarded as one of the most important problems in RTB display advertising.

There are two types of approaches towards optimal bidding strategies. A game-theoretical view assumes the rationality of the players (advertisers and publishers) in making decisions \citep{osborne1994course,milgrom2004putting,gummadi2012repeated} and formulates the incentives and the impact of the interplays among the players, whereas a statistical view assumes no or less interplay among the advertisers and publishers and studies decision-making under uncertainty \citep{berger2013statistical}.

This chapter takes the latter approach as assuming advertisers are strategic and rational is questionable in practice \citep{Yuan:2013:RBO:2501040.2501980} and the statistical approach is likely to lead to practically useful solutions, which are indeed the majority solutions for industrial RTB platforms. 
 
\begin{figure}[t]
 \centering
 \includegraphics[width=.7\columnwidth]{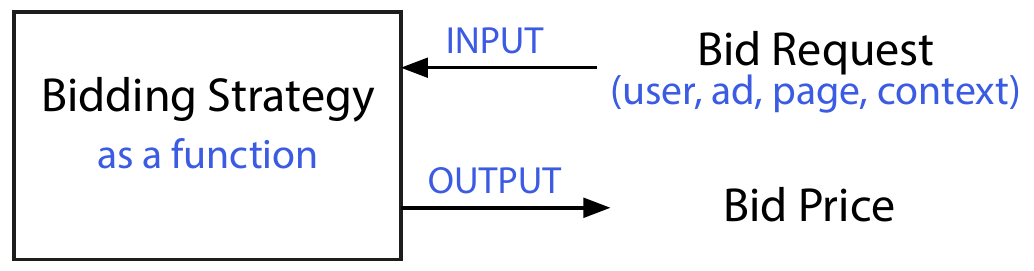}
 \caption{A bidding strategy can be abstracted as a function mapping from the given bid request (in a high dimensional feature space) to a bid price (a non-negative real or integer number).}
 \label{fig:bid-function-abs}
\end{figure}

In the formulation stated in this chapter, the bidding strategy is abstracted as a function which takes in the information of a specific ad display opportunity, i.e., the bid request, and outputs the corresponding bid price for each qualified ad, as illustrated in Figure~\ref{fig:bid-function-abs}. How this bidding function should be designed involves multiple factors, including the auction volume during the campaign's lifetime, the campaign budget and the campaign-specific key performance indicator (KPI), such as the number of clicks or conversions, or advertising revenue.

\section{Bidding problem: RTB vs. sponsored search}
 
Bid optimisation is a well-studied problem in online advertising \citep{feldman2007budget, Hosanagar2008, Ghosh2009, perlich2012bid}. Nonetheless, most research has been so far limited to keyword auction in the context of sponsored search (SS) \citep{edelman2005internet, Animesh2005, Mehta2005AdWords}. Typically, under the scenario of pre-setting the keyword bids (not impression level), the keyword utility, cost and volume are estimated and then an optimisation process is performed to optimise the advertisers' objectives (KPIs) \citep{borgs2007dynamics,zhou2008budget,kitts2004optimal,zhang2014bid}. Given a campaign budget as the cost upper bound, optimising the advertiser performance is defined as a budget optimisation problem \citep{feldman2007budget,muthukrishnan2007stochastic}. Furthermore, \cite{broder2011bid} and \cite{even2009bid} focused on the bid generation and optimisation on broad matched keywords, where query language features are leveraged to infer the bid price of related keywords. \cite{zhang2012joint} proposed to jointly optimise the keyword-level bid and campaign-level budget allocation under a multi-campaign sponsored search account. Some recent work focuses on periodically changing the pre-setting keyword auction price, taking into account the remaining budget and lifetime. For instance, in \citep{amin2012budget,gummadi2011optimal}, Markov decision process was used to perform online decision in tuning the keyword bid price, where the remaining auction volume and budget act as states and the bid price setting as actions.  \cite{kitts2004optimal} proposed to calculate a bid allocation plan during the campaign lifetime, where the bid price on each keyword is set in different discrete time units by considering the market competition and the CTR on different ad positions. However, none of the work evaluates per-impression auction in SS; all the bids are associated with keywords while impression level features are seldom considered, especially for advertisers and their agencies.  Moreover, in SS bid optimisation, search engines play two roles: setting the keyword bids as well as hosting the auctions. The objective function could be diverted to optimise the overall revenue for the search engine \citep{Abrams2006, Radlinski2008, Devanur2009, Zhu2009}, rather than the performance of each individual advertiser's campaigns.
 
The bid optimisation for RTB display advertising is fundamentally different. First, the bids are not determined by pre-defined keywords \citep{Yuan:2013:RBO:2501040.2501980}, but are based on impression level features. Although in general, advertisers (or DSPs) are required to set up their target rules, they need to estimate the value of each ad impression that is being auctioned in real time and return the bid price per auction. Second, in RTB, CPM pricing is generally used. Winning an impression directly results in the cost, despite the fact clicks and conversions can now be directly optimised by advertisers and DSPs. Thus, the dependencies over various effectiveness measures such as eCPC,\footnote{Effective cost per click (eCPC) - The cost of a campaign divided by the total number of clicks delivered.} CPM and the budget constraints need to be studied in a single framework. \cite{Ghosh2009} proposed an algorithm that learns winning bids distribution from full or partial information of auctions in display advertising. The algorithm then makes bidding decisions to achieve the best delivery (number of impressions) within the budget constraint.  In \citep{Chen2011c}, the bid price from each campaign can be adjusted by the publisher side in real time and the target is to maximise the publisher side revenue.

Other relevant problems in RTB have also been studied. \cite{lee2013real} focused on the pacing problem, where the target is to smoothly deliver the campaign budget. From the SSP perspective, the reserve price setting in RTB ad auctions is studied in \citep{yuan2014empirical}. In \citep{lee2012estimating} the sparsity problem of conversion rate estimation is handled by modelling the conversions at different selected hierarchical levels. \cite{dalessandro2012evaluating} investigated the evaluation measures of the display advertising performance and they found the site visit turns out to be a better proxy than the user click. In addition, there is some work on the ad exchange communication problem \citep{chakraborty2010selective,muthukrishnan2009ad}.
 
\section{Concept of quantitative bidding in RTB}

\begin{figure}[t]
	\centering
	\includegraphics[width=.8\columnwidth]{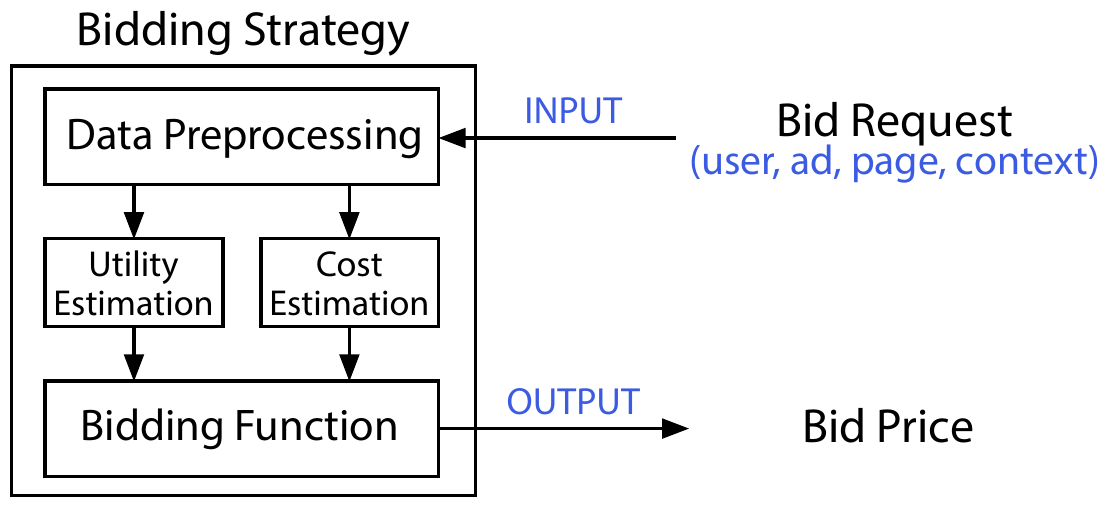}
	\caption{The quantitative bidding: the logic of the bidding function only depends on two (sets of) factors, i.e., the estimated utility and the cost of the ad display opportunity.}
	\label{fig:quantitative-bid}
\end{figure}

In a performance-driven advertising scenario, each ad display opportunity, i.e., bid request, is \emph{quantified} where its \emph{utility}, e.g., the probability of a user clicking on the displayed ad \citep{oentaryo2014predicting} or the expected revenue from this ad impression \citep{ahmed2014scalable,lee2012estimating}, and \emph{cost}, e.g., the cost of winning this ad impression in the auction \citep{cui2011bid}, are carefully estimated. Based on the estimated utility and cost of each bid request, \cite{Zhang:2014:ORB:2623330.2623633} proposed the concept of quantitative bidding. This means that the logic of the bidding function should only depend on two factors: the estimated utility of the ad display opportunity and the estimated cost to win it. All other information can be regarded as independent with the bid price conditioned only by these two factors,\footnote{All the information needed to determine the bid has been reflected in the utility and cost factors, just like the conditional independence in probabilistic graphic models.} as illustrated in Figure~\ref{fig:quantitative-bid}. For example, a sneakers advertiser would like to bid high on users with ages between 15 and 30. This is motivated by the fact that users in such a segment are more likely to be converted to purchase the sneakers after seeing the ads. This is quantified as a higher conversion rate. This is analogous with a high frequency trading strategy in a stock/option market where the trading action is wholly based on the quantified risks and the returns for each asset, regardless of the specific asset attributes or fundamentals \citep{durbin2010all}.

Using the estimated utility and cost, the optimal bidding function to maximise the specific KPI under the target campaign budget and auction volume constraints can be derived.
 
The research frontier and implementations of the utility estimation module and cost estimation module in Figure~\ref{fig:quantitative-bid} have been well investigated in Chapter~\ref{c-ctr} and Section~\ref{sec:landscape}, respectively. In the later sections of this chapter, various bidding strategies and their connections will be discussed.
 
\section{Single-campaign bid optimisation}
 
With the modules of CTR/CVR estimation (Chapter~\ref{c-ctr}) and bid landscape forecasting (Section~\ref{sec:landscape}) ready, bid optimisation techniques seek an optimal bidding strategy to maximise a certain objective (with possible constraints).
 
In this section, the derivation of different bidding strategies under the presumed settings will be discussed.
 
\subsection{Notations and preliminaries}
Define the bidding function as $b(\ctr)$ which returns the bid price given the estimated click-through rate (CTR) $\ctr$ of an ad impression. Various machine learning models for predicting CTR have been introduced in Chapter~\ref{c-ctr}.
 
\vspace{5pt} \noindent \textbf{Winning probability.} From an advertiser's perspective, given the market price distribution $p_z(z)$ and the bid price $b$, the probability of winning the auction is
\begin{align}
w(b) = \int_0^{b} p_z(z) dz. \label{eq:win}
\end{align}
 
If the bid wins the auction, the ad is displayed and the utility and cost of this ad impression are then observed.
 
\vspace{5pt} \noindent \textbf{Utility.} The utility function given the CTR is denoted as $u(\ctr)$. The specific form of $u(\ctr)$ depends on the campaign KPI. For example, if the KPI is click number, then
\begin{align}
u_\text{clk}(\ctr) = \ctr. \label{eq:utility-clk}
\end{align}
 
If the KPI is the campaign's profit, with the advertiser's true value on each click is $v$, then
\begin{align}
u_\text{profit}(\ctr,z) = v\ctr - z. \label{eq:utility-rev}
\end{align}
 
\vspace{5pt} \noindent \textbf{Cost.} If the advertiser wins the auction with a bid $b$, the expected cost is denoted as $c(b)$. In the RTB context there could be first price auctions
\begin{align}
c_1(b) = b, \label{eq:cost-1}
\end{align}
and second price auctions
\begin{align}
c_2(b) = \frac{\int_0^{b}z p_z(z) dz}{\int_0^{b} p_z(z) dz}. \label{eq:cost-2}
\end{align}
 
Different market settings and campaign strategies define different cost functions. If the auction is a first price auction, then $c_1(b)$ should be used. For second price auctions, $c_2(b)$ is reasonably adopted, but \cite{Zhang:2014:ORB:2623330.2623633} still used $c_1(b)$ to conservatively model the upper bound of the second price auctions with possible soft floor prices. Here it is feasible to first use the abstract cost function $c(b)$ in the framework and then specify the implementation of the cost function in specific tasks.
 
\vspace{5pt} \noindent \textbf{Campaign settings.}
Each campaign is set with specific targeting rules, lifetime and budget.
Generally, the targeting rules differentiate the volume, leading to different bid landscapes $p(z)$ and utility (CTR/CVR) distributions. The campaign's targeting rules and lifetime determine the auction volume $T$ it could receive. The campaign's budget $B$ defines the upper bound of its cost during the lifetime.
 
\subsection{Truth-telling bidding}
For non-budget bid optimisation, only $u_\text{profit}(\ctr)$ utility function is meaningful\footnote{If there are no cost-related factors in the utility, one will bid as high as possible to win every impression as there is no budget constraint.}
\begin{align}
U_\text{profit}(b(\cdot)) &= T  \int_{\ctr} \int_{z=0}^{b(\ctr)} u_\text{profit}(\ctr,z) p_z(z) dz \cdot p_r(r)  dr\\
&=T  \int_{\ctr} \int_{z=0}^{b(\ctr)} (v\ctr - z) p_z(z) dz \cdot p_r(r)  dr.
\end{align}
 
Take the gradient of net profit $U_\text{profit}(b(\cdot))$ w.r.t. the bidding function  $b(\ctr)$ and set it to 0,
\begin{align}
\frac{\partial U_\text{profit}(b(\cdot))}{\partial b(\cdot)} = (v\ctr - b(\ctr)) \cdot p_z(b(\ctr)) \cdot p_r(r) = 0,
\end{align}
for all $r$, which derives
\begin{align}
b_{\text{true}}(\ctr) = v \ctr, \label{eq:opt-bid-non-budget}
\end{align}
i.e., the optimal bid price is set as the impression value. Thus the truth-telling bidding is the optimal strategy when there is no budget considered.
 
The truth-telling property of $b_{\text{true}}(\ctr)$ is inherited from the classic second price auctions and the widely adopted VCG auctions in sponsored search \citep{edelman2005internet}, which makes this bidding strategy quite straightforward and widely adopted in industry \citep{lee2012estimating,Chen2011c}. However, the truth-telling bidding strategy is optimal only when the budget and auction volume are not considered.
 
\subsection{Linear bidding}\label{sec:lin-bid}
 
With the campaign lifetime auction volume and budget constraints, the optimal bidding strategies are probably not truth-telling. Extending from the truth-telling bidding strategy, \cite{perlich2012bid} proposed the generalised bidding function with a linear relationship to the predicted CTR for each ad impression being auctioned:
\begin{align}
b_{\text{lin}}(\ctr) = \phi v \ctr, \label{eq:lin-bid}
\end{align}
where $\phi$ is a constant parameter to tune to fit the market competitiveness and the volume.
Specifically, the optimal value of $\phi$ should just maximise the objective with the budget constraint on the training data.
 
\cite{perlich2012bid} proved that the linear bidding function $b_{\text{lin}}(\ctr)$ practically works well in various settings. However, from a scientific perspective, \cite{perlich2012bid} did not provide any theoretic soundness as to why the optimal bidding function should be linear as in Eq.~(\ref{eq:lin-bid}).
 
\subsection{Budget constrained clicks and conversions maximisation}
 
\cite{Zhang:2014:ORB:2623330.2623633,zhang2016optimal} proposed a general bid optimisation framework to incorporate different utility and cost functions.
 
Assuming one wants to find the optimal bidding function $b()$ to maximise the campaign-level KPI $\ctr$ across its auctions over the lifetime, with total bid requests volume $T$ and budget $B$
\begin{align}
\max_{b()} ~~ &T \int_r u(\ctr)  w(b(\ctr)) p_\ctr(\ctr) d\ctr \label{eq:obj} \\
\text{subject to}~~ & T \int_\ctr c(b(\ctr)) w(b(\ctr)) p_\ctr(\ctr) d\ctr = B. \nonumber
\end{align}
 
The Lagrangian of the optimisation problem Eq.~(\ref{eq:obj}) is
\begin{align}
\mathcal{L}(b(\ctr),\lambda) = & \int_{\ctr} u(\ctr) w(b(\ctr))p_\ctr(\ctr)d\ctr - \lambda \int_{\ctr} c(b(\ctr)) w(b(\ctr))p_\ctr(\ctr)d\ctr + \frac{\lambda B}{T}, \label{eq:lagrangian}
\end{align}
where $\lambda$ is the Lagrangian multiplier.
 
\vspace{5pt} \noindent \textbf{Solving $b()$.} Based on calculus of variations, the Euler-Lagrange condition of $b(\ctr)$ is
\begin{align}
\frac{\partial \mathcal{L}(b(\ctr),\lambda)}{\partial b(\ctr)} = 0,
\end{align}
which is
\begin{align}
u(\ctr) p_\ctr(\ctr)\frac{\partial w(b(\ctr))}{\partial b(\ctr)} - \lambda p_\ctr&(\ctr) \Big[ \frac{\partial c(b(\ctr))}{\partial b(\ctr)} w(b(\ctr)) + c(b(\ctr))\frac{\partial w(b(\ctr))}{\partial b(\ctr)} \Big]  = 0\\
\Rightarrow ~~~ \lambda \frac{\partial c(b(\ctr))}{\partial b(\ctr)}w(b(\ctr)) = &\Big[ u(\ctr) -  \lambda c(b(\ctr))\Big] \frac{\partial w(b(\ctr))}{\partial b(\ctr)}.\label{eq:general-condition}
\end{align}
 
Eq.~(\ref{eq:general-condition}) is a general condition of the optimal bidding function, where the specific implementations of winning function $w(b)$, utility function $u(\ctr)$ and cost function $c(b)$ are needed to derive the corresponding form of optimal bidding function.
 
\vspace{5pt} \noindent \textbf{Solving $\lambda$.}
To solve $\lambda$, the bidding function can be explicitly written as $b(r, \lambda)$, and the solution involves to solve
\begin{align}
\frac{\mathcal{L}(b(\ctr, \lambda),\lambda)}{\partial \lambda} &= 0, \\
\Rightarrow T \int_\ctr c(b(\ctr, \lambda)) w(b(\ctr, \lambda)) p_\ctr(\ctr) d\ctr &= B, \label{eq:solve-lambda}
\end{align}
which is to find the $\lambda$ to just exhaust the budget at the same time of running out the auction volume. In practice, Eq.~(\ref{eq:solve-lambda}) has no analytic solution but its numeric solution is very easy to obtain because $c(b)$ and $w(b)$ monotonously increase w.r.t. $b$.
 
Later the analytic solution of $\lambda$ and $b(r, \lambda)$ under some special setting of $w(b)$ and $p_r(r)$ will be shown.
 
\subsubsection{First-price auction}
With cost function $c_1(b)=b$ as in Eq.~(\ref{eq:cost-1}), Eq.~(\ref{eq:general-condition}) is written as
\begin{align}
\lambda \int_0^{b(\ctr)} p_z(z) dz =(u(\ctr) - \lambda b(\ctr)) \cdot   p_z(b(\ctr)) \label{eq:general-condition-first-auction}
\end{align}
which still depends on the detailed form of market price distribution $p_z(z)$ to solve the $b()$. \cite{Zhang:2014:ORB:2623330.2623633} tried an implementation in their paper:
\begin{align}
p_z(z) &= \frac{l}{(l+z)^2},\\
\Rightarrow ~~ w(b) &= \frac{b}{b+l} \label{eq:win-1}.
\end{align}
 
Taking Eq.~(\ref{eq:win-1}) and click utility Eq.~(\ref{eq:utility-clk}) into Eq.~(\ref{eq:general-condition-first-auction}):
\begin{align}
\lambda \frac{b(\ctr)}{b(\ctr) + l} &=(u(\ctr) - \lambda b(\ctr)) \cdot   \frac{b(\ctr)}{(b(\ctr) + l)^2} \\
\Rightarrow ~~ b_{\text{ortb}}(\ctr) &= \sqrt{\frac{u(\ctr) l}{\lambda} + l^2} - l, \label{eq:bid-func-1}
\end{align}
which is the derived optimal bidding function in \citep{Zhang:2014:ORB:2623330.2623633}.
The analytical solution Eq.~(\ref{eq:bid-func-1}) shows that an optimal bidding function should be non-linear. The non-linearity is closely related to the probability of auction winning, but is loosely correlated with the prior distribution of the impression features.
 
If the market price $z$ follows a uniform distribution in $[0,l]$, i.e.,
\begin{align}
p_z(z) = \frac{1}{l},
\end{align}
Eq.~(\ref{eq:general-condition-first-auction}) is rewritten as
\begin{align}
\lambda \frac{b(r)}{l} &= \frac{(u(r) - \lambda b(r))}{l}\\
\Rightarrow b(r) &= \frac{u(r)}{2\lambda}. \label{eq:bid-uniform-mp}
\end{align}
Taking Eq.~(\ref{eq:bid-uniform-mp}) into Eq.~(\ref{eq:solve-lambda}):
\begin{align}
T \int_r \frac{u(r)}{2\lambda}\cdot \frac{u(r)}{2\lambda l} p_r(r) dr = B. \label{eq:sol-tmp}
\end{align}
Taking $u_\text{clk}(r)=r$ as in Eq.~(\ref{eq:utility-clk}) and the special case of uniform CTR distribution $p_r(r) = 1$ into Eq.~(\ref{eq:sol-tmp}):
\begin{align}
T \int_0^1 \frac{r^2}{4\lambda^2 l} dr = B \\
\Rightarrow \lambda = \frac{1}{2}\sqrt{\frac{T}{3Bl}}.
\end{align}
Thus the analytic form of the optimal bidding function is
\begin{align}
b(r) = r \sqrt{\frac{3Bl}{T}}.
\end{align}
 
\subsubsection{Second-price auction}
Taking the definition of winning function Eq.~(\ref{eq:win}) and the second-price cost function Eq.~(\ref{eq:cost-2}) into Eq.~(\ref{eq:general-condition}):
\begin{align}
& \lambda \frac{b(\ctr) p_z(b(\ctr)) w(b(\ctr)) - p_z(b(\ctr)) \int_0^b z p_z(z)dz} {w(b(\ctr))^2} w(b(\ctr)) \nonumber \\
& ~~~~~~~~~~~~~~~~~~~~~~~~~~= (u(\ctr) - \lambda c(b(\ctr))) p_z(b(\ctr)) \\
\Rightarrow ~~~& \lambda (b(\ctr) - c(b(\ctr))) = u(\ctr) - \lambda c(b(\ctr))\\
\Rightarrow ~~~& b_{\text{ortb-lin}}(\ctr) = \frac{u(\ctr)}{\lambda}, \label{eq:opt-bid}
\end{align}
where it can be observed that the optimal bidding function is linear w.r.t. CTR $\ctr$. Just like the linear bidding function discussed in Section~\ref{sec:lin-bid}.
 
The solution of $\lambda$ is obtained by taking Eq.~(\ref{eq:opt-bid}) into the constraint
\begin{align}
T \int_\ctr c(b(\ctr)) w(b(\ctr)) p_\ctr(\ctr) d\ctr = B,
\end{align}
which is rewritten as
\begin{align}
\int_\ctr c\Big(\frac{u(\ctr)}{\lambda}\Big) w\Big(\frac{u(\ctr)}{\lambda}\Big) p_\ctr(\ctr) d\ctr = \frac{B}{T}.\label{eq:budget}
\end{align}
It can be observed that essentially the solution of $\lambda$ makes the equation between the expected cost and the budget. Furthermore, as both $w(\ctr/\lambda)$ and $c(\ctr/\lambda)$ monotonously decrease as $\lambda$ increases, it is quite easy to find a numeric solution of $\lambda$.
 
Similar to what derived in first price auction where the market price $z$ follows a uniform distribution in $[0,l]$, i.e., $p_z(z) = \frac{1}{l}$ and CTR follows a uniform distribution in $[0,1]$ and click utility $u_\text{clk}(r) = r$, one can solve $\lambda$ in Eq.~(\ref{eq:budget}). First, the cost function is
\begin{align}
c(b) = \int_0^b z p(z) dz = \int_0^b z \frac{1}{l} dz = \frac{b^2}{2l}.
\end{align}
Eq.~(\ref{eq:budget}) is rewritten as
\begin{align}
\int_0^1 \frac{r^2}{2 \lambda^2 l} \cdot \frac{r}{\lambda l} dr = \frac{B}{T}\\
\Rightarrow \lambda = \frac{1}{2}\sqrt[3]{\frac{T}{B l^2}}.
\end{align}
Thus the analytic form of the optimal bidding function is
\begin{align}
b(r) = 2 r \sqrt[3]{\frac{B l^2}{T}}.
\end{align}
 
\subsection{Discussions}
 
The implementation of the cost constraint in Eq.~(\ref{eq:obj}) needs careful modelling based on the data. \cite{Zhang:2014:ORB:2623330.2623633} used the cost upper bound, i.e., the bid price, to control the cost and let the total value of cost upper bound be the campaign budget. Here, if one implements the expected cost in second price auction Eq.~(\ref{eq:cost-2}), the cost constraint in Eq.~(\ref{eq:obj}) might not be controlled by the budget. With about 50\% probability, the budget will be exhausted in advance, which is not expected in practice.
 
One can easily find in the first-price bidding function Eq.~(\ref{eq:bid-func-1}), the bid price is jointly determined by utility function $u(r)$, $\lambda$ and market parameter $l$.
Specifically, the bid price is monotonic increasing w.r.t. utility while decreasing w.r.t. $\lambda$.
Moreover, different value settings for parameter $l$ also influence the final bid decision as is shown in \citep{Zhang:2014:ORB:2623330.2623633}.
As is defined in Eq.~(\ref{eq:win-1}), one can tune the parameter $l$ to alter the winning probability so as to fit different market environments.
In fact, one may conclude that the market consideration influences bid price by tuned $l$, the budget constraint controls bid function by $\lambda$, while advertiser's utility expectation $u(r)$ could finally determine the final decision.
 
Let us take a look at the bidding strategy under second-price auctions.
In Eq.~(\ref{eq:opt-bid}), the bid price is mainly determined by utility $u(r)$.
However, the bidding strategy constructs a bridge between utility and budget consideration by $\lambda$.
Move the attention onto Eq.~(\ref{eq:budget}) and it can be found that $\lambda$ is also controlled by this equation which takes both estimated CTR $r$ and winning probability $w(\cdot)$ into consideration.
The latter two factors have low effects on bidding strategy through $\lambda$.
 
From the discussion above, one may find that both bidding strategies under either auction setting are influenced by three factors: advertiser's expected utility, budget constraints and market information.
While under first-price auction setting, the bidding strategy takes utility function and market price together, the strategy upon second-price auctions reflects utility more straightly.
 
For the bidding function under first-price auctions, more attention could be taken on the winning probability estimation problem and consequently to optimise the bidding strategy.
It could be more crucial to take the market information into consideration in the utility function for second-price auctions.
Finally, one may take a step forward to coordinate the optimisation for both CTR estimator and bidding strategy to further push the bidding performance \cite{ren2016user}. 

\section{Multi-campaign statistical arbitrage mining}

\cite{Zhang:2015} studied the problems of arbitrages in real-time bidding based display advertising in a multi-campaign setting.
 
On a display ads trading desk (or DSP), some advertisers would just pay per click or even pay per conversion so as to minimise their risk. From the perspective of the trading desk, it tries to earn the user clicks or conversions in pay-per-view spot market via real-time bidding. It is possible for the trading desk to find some cost-effective cases to earn each click or conversion with a cost lower than the advertiser's predefined cost. In such case, the arbitrage happens: the trading desk earns the difference of cost between the advertiser's predefined cost for each click/conversion and the real cost on it, while advertisers get the user clicks or conversions with no risk. Such click/conversion based transactions act as a complementary role to the mainstream guaranteed delivery and RTB spot market, and is a win-win game if the trading desk would successfully find the arbitrages.
 
In such a scenario, the profit Eq.~(\ref{eq:utility-rev}) is the studied utility for each impression. The bid optimisation framework for a single campaign is
\begin{align}
\max_{b()} ~~ &T \int_r (u(\ctr) - b(\ctr))  w(b(\ctr)) p_\ctr(\ctr) d\ctr \label{eq:sam-obj-single} \\
\text{subject to}~~ & T \int_\ctr b(\ctr) w(b(\ctr)) p_\ctr(\ctr) d\ctr = B. \nonumber
\end{align}
 
Furthermore, such an arbitrage problem can be practically extended to the multiple campaign cases. The trading desk then acts as a \emph{meta-bidder} for multiple campaigns. Each time receiving a bid request, the meta-bidder samples one from $M$ campaigns it serves and then calculates a bid for its ad. If the campaign sampling is a probabilistic process, i.e., sampling the campaign $i$ with a probability $s_i \geq 0$ with $\sum_{i=1}^{M} s_i = 1$. The vector notation of the campaign sampling probabilities is $\bs{s}$.
 
Thus the multi-campaign bid optimisation framework is written as
\begin{align}
\max_{b(), \bs{s}} ~~ &T \sum_{i=1}^{M} s_i \int_r (u(\ctr) - b(\ctr))  w(b(\ctr)) p_\ctr(\ctr) d\ctr \label{eq:sam-obj} \\
\text{subject to}~~ & T \sum_{i=1}^{M} s_i \int_\ctr b(\ctr) w(b(\ctr)) p_\ctr(\ctr) d\ctr = B. \nonumber \\
& \bs{s}^T \bs{1} = 1 \nonumber\\
& \bs{0} \leq \bs{s} \leq \bs{1} \nonumber
\end{align}
 
Given a training set, with the bidding function $b$ and campaign sampling $\bs{s}$, define the meta-bidder profit as $R(b, \bs{s})$ and its cost as $C(b, \bs{s})$:
\begin{align}
R(b, \bs{s}) &= \sum_{t=1}^{T} \sum_{i=1}^{M} s_i \int_r (u(\ctr_i^t) - b(\ctr_i^t))  w(b(\ctr)) s_i^t\\
C(b, \bs{s}) &= \sum_{t=1}^{T} \sum_{i=1}^{M} s_i \int_r b(\ctr_i^t)  w(b(\ctr)) s_i^t
\end{align}
As the training set could change across different times and settings, $R$ and $C$ can be regarded as a random variable, with the expectation $\mathbb{E}[R]$ and the variance $\var[R]$ for profit, and the expectation $\mathbb{E}[C]$ and the variance $\var[C]$ for cost.
 
From a risk management perspective, the meta-bidding needs to control its risk of deficit while optimising its profit. By adding an extra risk control constraint, the meta-bidder optimisation framework is rewritten as
\begin{align}
\max_{b(), \bs{s}} ~~ &\mathbb{E}[R] \label{eq:sam-obj-final} \\
\text{subject to}~~ & \mathbb{E}[C] = B \label{eq:sam-cost} \\
& \var[R] = 1 \label{eq:sam-sampling} \\
& \bs{s}^T \bs{1} = 1 \label{eq:sam-sampling-1}\\
& \bs{0} \leq \bs{s} \leq \bs{1}. \label{eq:sam-sampling-pos}
\end{align}
 
An EM-fashion approach is proposed in \citep{Zhang:2015} to solve the above optimisation problem. Specifically, in the E-step, optimise Eq.~(\ref{eq:sam-obj-final}) with the constraints Eqs.~(\ref{eq:sam-sampling}, \ref{eq:sam-sampling-1}, \ref{eq:sam-sampling-pos}); in M-step, optimise Eq.~(\ref{eq:sam-obj-final}) with the constraint of Eq.~(\ref{eq:sam-cost}). When EM iterations converge, all the constraints are satisfied and the value of objective will get into a local optima.
 
\section{Budget pacing}\label{sec:pacing}
 
Besides the optimal bidding strategy, which can be regarded as a good budget allocation across different display opportunities, advertisers would also pursue a good budget allocation over the time, i.e., to spend the budget smoothly over the time to reach a wider range of audiences \citep{lee2013real}.
 
Generally, there are two types of budget pacing solutions: throttling and bid modification \citep{xu2015smart}. Throttling based methods maintain a pacing rate at each time step, which is the probability of the campaign participating in the incoming auction. Bid modification based methods adaptively adjust the bid price for each incoming auction to achieve the budget pacing target.
 
\cite{lee2013real} provided a straightforward offline throttling based solution to model the budget pacing problem as a linear optimisation problem. Suppose the campaign's daily budget is split into $T$ time slots, and the campaign's daily budget $B$ is allocated across these time slots $[b_1, b_2, \ldots, b_T]$ with $\sum_{t=1}^T b_t = B$. If the incoming bid request $i$ is associated with value $v_i$ and cost $c_i$, and the decision of whether to place a bid for $i$ is denoted as $x_i$, then the linear optimisation problem is
\begin{align}
\max_{\bs{x}}~~ & \sum_{i=1}^{n} v_i x_i\\
\text{subject to}~~ & \sum_{j\in \mathbb{I}_t} c_j x_j \leq b_t~~~~ \forall t \in \{1,2,\ldots, T\},
\end{align}
where $\mathbb{I}_t$ represents the index set of all ad requests coming in at time slot $t$. This solution cannot be applied online as there is no information of the volume, value and cost of the future.
 
Then the authors proposed their online solution of budget pacing:
\begin{align}
\min_{\bs{b}}~~ & \text{-CTR, -AR, eCPC or eCPA}\\
\text{subject to}~~ & \Big|\sum_t^T s(t) - B \Big| \leq \epsilon ~~~ (\text{total spend})\\
& |s(t) - b_t| \leq \delta_t ~~~ (\text{smooth spend})\\
& \text{eCPM} \leq M ~~~ (\text{max CPM})
\end{align}
where $s(t)$ is the actual spend at time slot $t$.
Then, based on CPM stability assumption, the spend $s(t)$ can be factorised as
\begin{align}
s(t) & \propto \text{imps}(t)\\
& \propto \text{reqs(t)} \cdot \frac{\text{bids}(t)}{\text{reqs}(t)} \cdot \frac{\text{imps}(t)}{\text{bids}(t)}\\
& \propto \text{reqs(t)} \cdot \text{pacing\_rate}(t) \cdot \text{win\_rate}(t)
\end{align}
where reqs$(t)$ is the number of received bid requests in time slot $t$; pacing\_rate$(t)$ is the budget pacing rate to be tuned in time slot $t$. Therefore, setting the expected spend $s(t+1)$ as the budget $b_{t+1}$, the pacing rate of the next time slot $t+1$ is
\begin{align}
& \text{pacing\_rate}(t+1)\\
& = \text{pacing\_rate}(t) \cdot \frac{\text{s}(t+1)}{\text{s}(t)} \cdot \frac{\text{reqs}(t)}{\text{reqs}(t+1)} \cdot \frac{\text{win\_rate}(t)}{\text{win\_rate}(t+1)} \\
& = \text{pacing\_rate}(t) \cdot \frac{b_{t+1}}{\text{s}(t)} \cdot \frac{\text{reqs}(t)}{\text{reqs}(t+1)} \cdot \frac{\text{win\_rate}(t)}{\text{win\_rate}(t+1)},
\end{align}
which can be calculated based on real-time performance and the previous round pacing rate.
 
Further throttling based solutions such as \citep{xu2015smart,agarwal2014budget} are normally in the framework of optimisation and online pacing rate control.
 
Bid modification based methods are well investigated in sponsored search \citep{mehta2007adwords,borgs2007dynamics}, where the adjusted bid prices are set on keyword level. In RTB display advertising, feedback control based methods \citep{Chen2011c,karlsson2013applications,zhang2016feedback} are adopted for bid modification in the budget pacing task.
 
In the above work, the feedback controller is embedded in the bidding agent of the DSP. It monitors the real-time KPIs (e.g. CPM, auction winning ratio, CTR etc.) to obtain the error factor to the reference value. Then a feedback control function takes in such an error factor and outputs the control signal, which is used to adjust the bid price for each incoming bid request.
 
For example, \cite{zhang2016feedback} proposed to use proportional-integral-derivative (PID) control function to perform the bid modification to make the campaign achieve the predefined KPIs. As its name implies, a PID controller produces the control signal from a linear combination of the proportional factor, the integral factor and the derivative factor based on the error factor:
\begin{align}
e(t_k) &= x_r - x(t_k) \label{eq:error-factor},  \\
\phi(t_{k+1}) &\leftarrow \lambda_P \underbrace{e(t_k)}_{\text{proportional}} + \lambda_I \underbrace{\sum_{j=1}^{k} e(t_j) \triangle t_j}_{\text{integral}} + \lambda_D \underbrace{\frac{\triangle e(t_k)}{\triangle t_k}}_{\text{derivative}},  \label{eq:pid}
\end{align}
where the error factor $e(t_k)$ is the reference value $x_r$ minus the current controlled variable value $x(t_k)$, the update time interval is given as $\triangle t_j = t_j - t_{j-1}$, the change of error factors is $\triangle e(t_k) = e(t_k) - e(t_{k-1})$, and $\lambda_P$, $\lambda_I$, $\lambda_D$ are the weight parameters for each control factor.
 
Such control signal $\phi(t)$ is then used to adjust the bid price
\begin{align}
b_a(t) = b(t) \exp\{\phi(t)\}, \label{eq:exp-actuator}
\end{align}
where $b(t)$ is the original bid price calculated for the incoming bid request at $t$ and $b_a(t)$ is the adjusted one.
 
Compared with throttling based methods, bid modification based methods are more flexible to achieve various budget pacing targets (bidding zero means no bid). However, such bid control highly depends on the predictable market competition, i.e. bid landscape as discussed in Section~\ref{sec:landscape}, while throttling based methods are more straightforward and normally work well in dynamic marketplaces \citep{xu2015smart}.
 
\section{Benchmarking}
 
Compared to user response prediction datasets, the datasets of bidding strategy optimisation further needs the information about bid price, winning notice and winning prices. To the authors' knowledge, so far there are two RTB datasets containing such information, namely iPinYou and YOYI datasets as mentioned in Chapter~\ref{c-ctr}.
 
Similar to Chapter~\ref{c-ctr}, here list a series of benchmarking literatures about bid optimisation in RTB display advertising \citep{perlich2012bid,zhang2014real,Zhang:2014:ORB:2623330.2623633,Zhang:2015,cai2017real,zhang2017managing}, each of which regards to a specific bid optimisation setting or objective.

\chapter{Dynamic Pricing}
\label{c-dync}

In this chapter, we focus on publishers in the RTB ecosystem. Advertising provides them with major sources of revenue. Therefore, uplifting revenue by using various yield management tools makes one of the key topics on the publisher side. We start with reserve price optimisation and then move to a unified solution that combines various selling channels together from both the future time guaranteed selling and the current time RTB auctions.

\section{Reserve price optimisation}
 
\begin{figure}[t]
	\centering
	\includegraphics[width=\columnwidth]{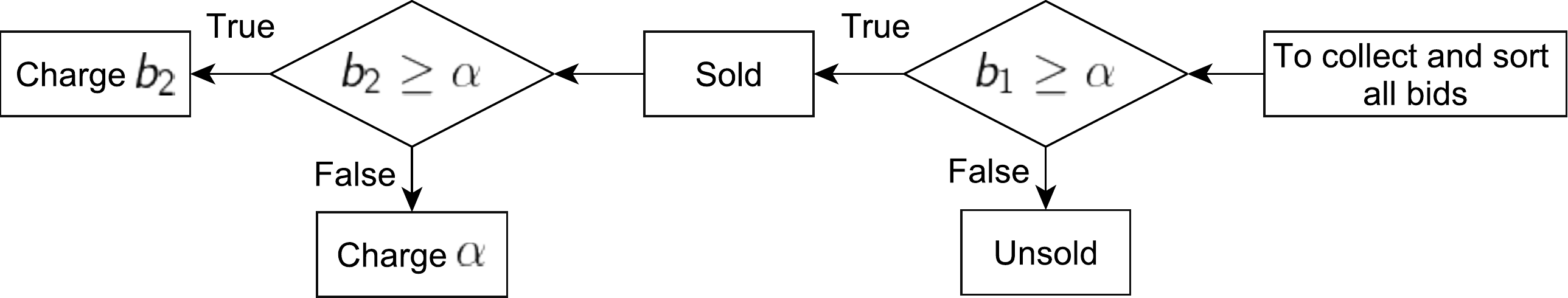}
	\caption{The decision process of second price auctions on the publisher side. The desirable case is $b_1 \geq \alpha > b_2$ where the publisher gains extra payoff of $\alpha - b_2$. The \emph{soft floor prices}, which make the process a lot more complicated, are ignored. Interested readers may refer to~\citep{Yuan:2013:RBO:2501040.2501980} for further discussion.}
	\label{fig-reserve-price-auction}
\end{figure}

An important tool for publisher's yield management is the optimisation of \emph{reserve price}, aka, floor price. A reserve price defines the minimum that a publisher would accept from bidders. It reflects the publisher's private valuation of the inventory, so bids will be discarded if they are below the reserve price. In the second price auction, which is commonly used in RTB, the reserve price could potentially uplift the revenue. Figure~\ref{fig-reserve-price-auction} illustrates how the final price is calculated from bids with a reserve price. Let $b_1, \ldots, b_K$ denote the descending bids and $\alpha$ the reserve price. Then, the desirable case is $b_1 \geq \alpha > b_2$ where the publisher gains extra payoff of $\alpha - b_2$; the neutral case is $b_1 > b_2 \geq \alpha$ where the publisher has no extra gain; and the undesirable case is $ \alpha > b_1 $ where the publisher suffers from a loss of $b_2$.
 
Formally, the problem could be defined as follows. For an auction, denote the final bids as $b_1(t), b_2(t), \cdots, b_K(t)$ where assume $b_1(t) \geq b_2(t) \geq \cdots \geq b_K(t)$. Without a reserve price ($\alpha=0$) the payoff could be denoted as $r(t) = b_2(t)$. Now suppose the publisher sets a non-zero reserve price at each step, denoted by $\alpha(t)$. The payoff function becomes:
\begin{align}
r'(t) = \left\{
    \begin{array}{l l}
        \alpha(t), & b_1(t)\geq \alpha(t) > b_2(t) \\
        b_2(t), & b_2(t) \geq \alpha(t) \\
        0, & \alpha(t) > b_1(t)
    \end{array}
\right.
\end{align}
The overall income is $R(T) = \sum_{t=1}^T r'(t)$. It is feasible to assume there is zero payoff when the reserve price is too high. In practice, publishers usually redirect these impressions to managed campaigns or other ad marketplaces for reselling to reduce the risk of over-optimisation.
 
This optimisation problem has been previously studied in the context of sponsored search~\citep{Edelman2006, Even-Dar2008, Ostrovsky2009, Xiao2009}. However, the problem in the RTB context is different and unique. Firstly, the optimal auction theory requires knowledge of  the distribution of the advertisers' private yet true assessments of the impression before calculating the optimal reserve price~\citep{Edelman2006}. In RTB, it becomes a lot harder to learn the distribution. As previously mentioned, an advertiser is required to submit a bid for each impression using his own algorithm, which is never disclosed to publishers and could rely heavily on privately-owned user interest segments. Besides, various practical constraints such as the budget, campaign life time, and even irrationality divert advertisers from bidding at private values. This difference makes the private value based algorithm inefficient in practice. Secondly, unlike sponsored search, an advertiser does not have the keyword constraint and faces an almost unlimited supply of impressions in RTB. Setting up an aggressive reserve price would easily move the advertisers away from those placements and force them to look for something cheaper.
 
In the RTB context, the reserve price optimisation problem has been studied in \citep{yuan2014empirical}. The authors present the analysis on bids to reject the Log-normal distribution hypothesis, propose a set of heuristic rules to effectively explore the optimal reserve prices, and look at the actions from the buy side to reject the attrition hypothesis. We will first briefly introduce the Optimal Auction Theory, then describe their work as follows.
 
\subsection{Optimal auction theory}
\label{sec-optimal-auction-theory}
Regardless of the existence of reserve price, bidders are encouraged to bid their private values in the second price auctions \citep{Myerson1981, matthews1995technical}. Note that this dominant strategy does not hold in modern sponsored search where quality scores are generally used \citep{edelman2005internet} in ad ranking. Without quality scores, the strategy of bidding at the private value forms part of the Nash equilibrium of the system, meaning as time elapses advertisers have no incentive to change their bids, given that all other factors remain the same. In this non-cooperative game \citep{osborne1994course}, the winner could, but would not, lower his bid to let other competitors win because losing the auction is not beneficial in either short-term or long-term (lowering the bid while still winning has no effect since the winner always pays the second highest bid).
 
Suppose the publisher knows the bidders' private value distribution. The optimal auction theory mathematically defines the optimal reserve price~\citep{Xiao2009,Myerson1981}. Suppose there are $K$ bidders and they are risk-neutral and symmetric, i.e., having identical value distributions. Each bidder $k \in K$ has private information on the value of an impression, drawn from distribution $F_k(x)$, where $F_k(x)$ denotes the probability that the advertiser's private evaluation value is less than or equal to a certain number $x$. Usually it is assumed Log-normal~\citep{Ostrovsky2009} or Uniform distribution~\citep{Myerson1981}. Assuming private values are independently distributed, the distribution over value vector is
\begin{align*}
F(\cdot) = F_1(\cdot) \times \cdots F_K(\cdot),
\end{align*}
and then the optimal reserve price is given as (c.f., ~\citep{osborne1994course}
for details):
\begin{align}\label{eq-reserve-price-game}
\alpha = \frac{1 - F(\alpha)}{F'(\alpha)} + v_P,
\end{align}
where $F'(\alpha)$ is the density function, the first order derivative of $F(\alpha)$ and $v_P$ is the publisher's private value. In practice, $v_P$ could be obtained from a guaranteed contract with a flat CPM, or from another ad network where the average revenue is known.
 
The theory was evaluated in~\citep{Ostrovsky2009} and showed mixed results, as shown in Table~\ref{tbl-ostrovsky-2009}. The results showed that the reserve price optimisation could lead to substantial revenue increase in some cases. Also, this proved for the first time that the Optimal Auction Design theory is applicable in practice.
 
\begin{table}
    \caption {Experiment results from~\citep{Ostrovsky2009}}
    \label{tbl-ostrovsky-2009}
    \begin{tabular}{|p{5cm}|r|r|r|}
    \hline
     Group & \parbox{2cm}{\raggedleft Estimated impact on revenue} & t-statistic & p-value \\ \hline
    Keywords with fewer than 10 searches per day & -2.19\% & -2.36 & 0.0183 \\ \hline
    Keywords with at least 10 searches per day & 3.30\% & 2.32 & 0.0201 \\ \hline
    Optimal reserve price $< .2$ & -9.19\% & -11.1 & <0.0001 \\ \hline
    Optimal reserve price $>= .2$ & 3.80\% & 5.41 & <0.0001 \\ \hline
    Average \# of bidders $< 5.5$ & 10.06\% & 7.29 & <0.0001 \\ \hline
    Average \# of bidders $>= 5.5$ & 2.54\% & 3.59 & <0.0003 \\ \hline
    \end{tabular}
\end{table}
 
\subsubsection{Drawbacks in RTB practice}
 
\cite{yuan2014empirical} evaluated the Optimal Auction Theory in the RTB practice. The authors implemented it following the Log-normal distribution assumption of bidders' private values. They also adopted the symmetric assumption, i.e., there is only one distribution for all bidders. Under these assumptions the optimality of the auction under GSP is proved in~\citep{Edelman2006}. The estimation of Log-normal's mean and standard deviation was obtained using the training dataset (impression-level logs from 14 Dec 2012 to 18 Jan 2013).
 
A few drawbacks were reported mostly due to the difficulty of learning bidders' private values, i.e., $F(x)$. Firstly, a bidder could have a complex private value distribution which does not follow Log-normal. In RTB an advertiser computes a bid for each individual impression based on the contextual~\citep{Broder:2007:SAC:1277741.1277837} and behavioural~\citep{yan2009much} data. The data is fed into their private models which are never disclosed to publishers or other advertisers. This is very different from sponsored search where search engines run bidding algorithms for advertisers and host auctions as a publisher at the same time. Also, in sponsored search the auctions are based on keywords, so the population of the bidders are relatively more stable, whereas in RTB, the auctions are at the impression level and the advertisers are given more flexibility on what to bid and how much to bid.
 
\begin{figure}
	\centering
	\includegraphics[width=\columnwidth]{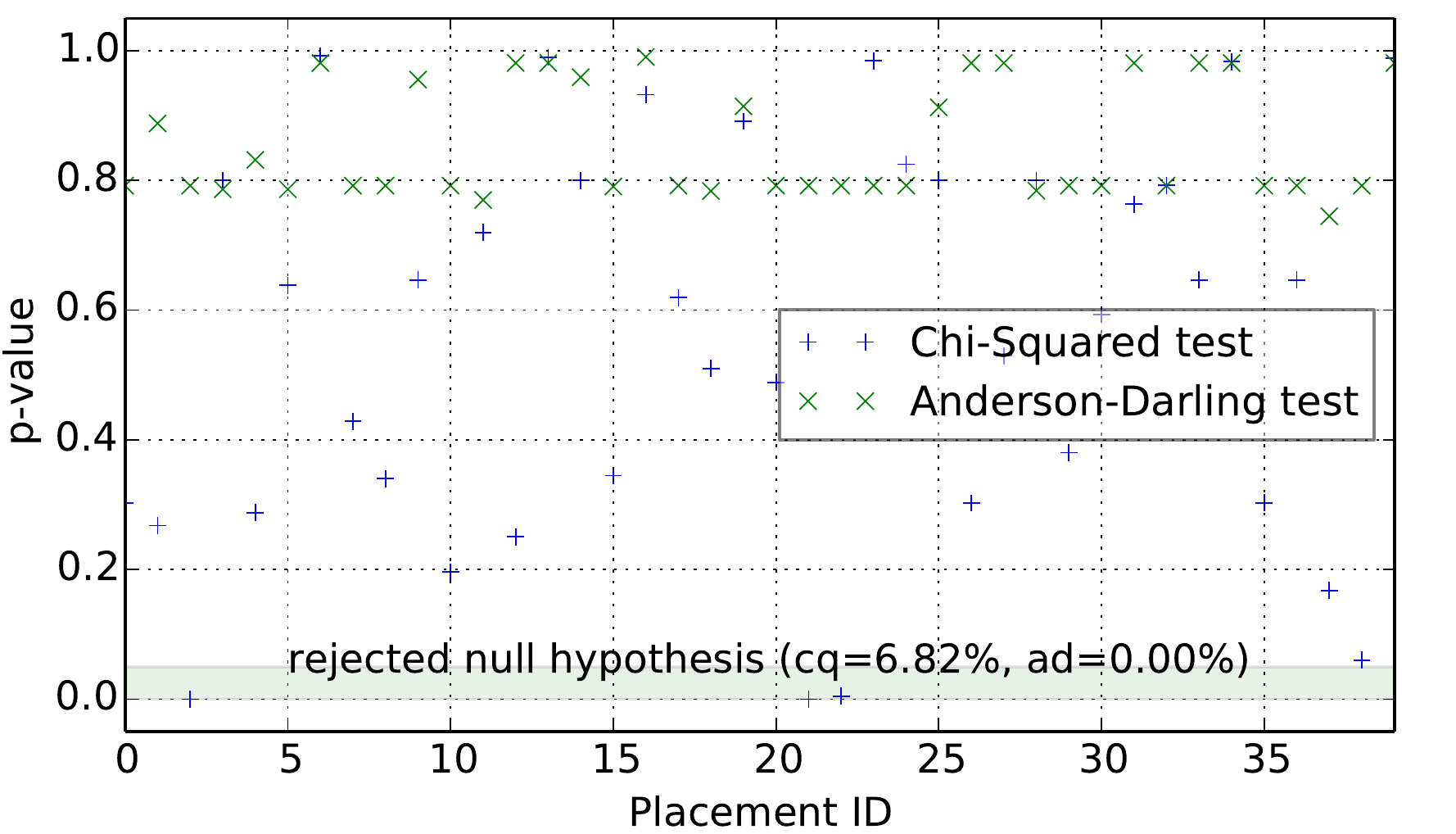}
	\caption{
	Tests in \citep{yuan2014empirical} showed only bids from 3 out of 44 placements (6.82\%) accept the Uniform distribution hypothesis. The Uniform distribution is tested by Chi-Squared (CQ) test and the Log-normal distribution is tested by Anderson-Darling (AD) test. }
	\label{fig-reserve-price-distribution-test-placement}
\end{figure}

\begin{figure}
	\centering
	\includegraphics[width=\columnwidth]{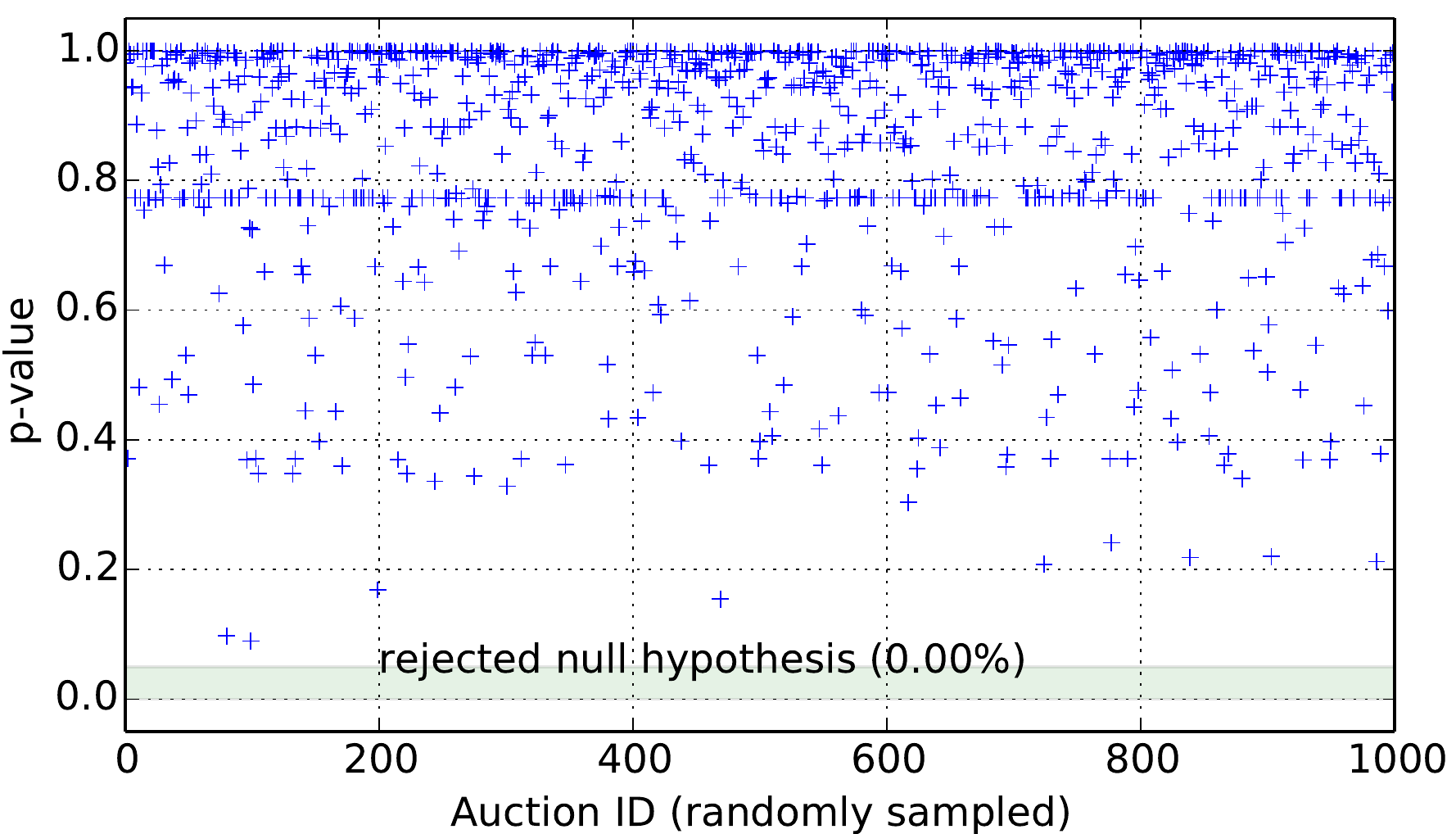}
		\caption{
	Tests in \citep{yuan2014empirical} showed only bids from less than 0.1\% of all auctions accept the Log-normal distribution hypothesis. The plot shows a random sample of 1000 auctions. Only the Log-normal distribution is tested by Anderson-Darling test. }
	\label{fig-reserve-price-distribution-test-auction}
\end{figure}

Following previous works in Game Theory, where researchers use an exploration period to collect and fit the private value distributions~\citep{Edelman2006, Ostrovsky2009}, Uniform distribution at placement level and Log-normal distribution at both placement and impression level were tested in~\citep{yuan2014empirical}. Although these distributions are widely adopted in research literature~\citep{Myerson1981,Ostrovsky2009}, only a small portion of tests returned positive results as shown in Figures~\ref{fig-reserve-price-distribution-test-placement} and~\ref{fig-reserve-price-distribution-test-auction}.
 
Secondly, it is assumed that advertisers bid at their private values in the second price auction. However, in practice, an advertiser may not know clearly his private valuation of an impression. Instead, he wants to achieve the best possible performance. Also in different stages (prospecting, optimisation, and retargeting, etc.) of an advertising campaign, the bidding strategy could change. This makes the bidding activity vary greatly across the limited flight time of a campaign.
 
\begin{figure}[t]
	\centering
	\includegraphics[width=\columnwidth]{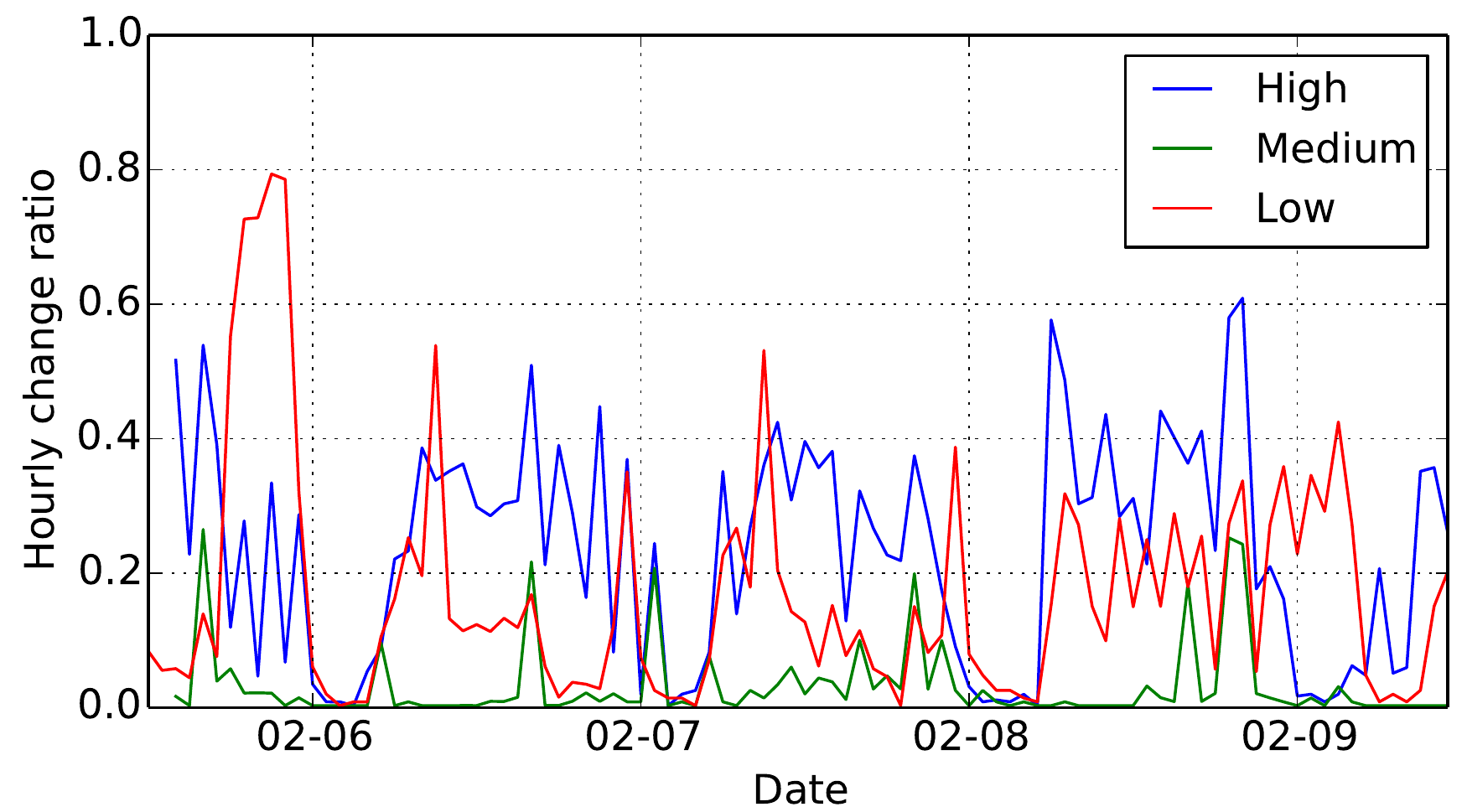}
	\caption{The change of winners for placements with different levels of competition in 4 days, reported in \citep{yuan2014empirical}.
	The fact that a bidder does not always win could add difficulty to reserve price detection if undisclosed. The result also implies the change rate does not necessary relate to the competition level.}
	\label{fig-reserve-price-winner-change}
\end{figure}

Thirdly, there are other practical constraints including accessibility of auction details, noise introduced by the frequent change of auction winners, c.f. Figure~\ref{fig-reserve-price-winner-change}. These constraints need careful consideration when implementing the theory in RTB practice.
 
\subsection{Game tree based heuristics}
 
\begin{figure}[t]
	\centering
	\includegraphics[width=\columnwidth]{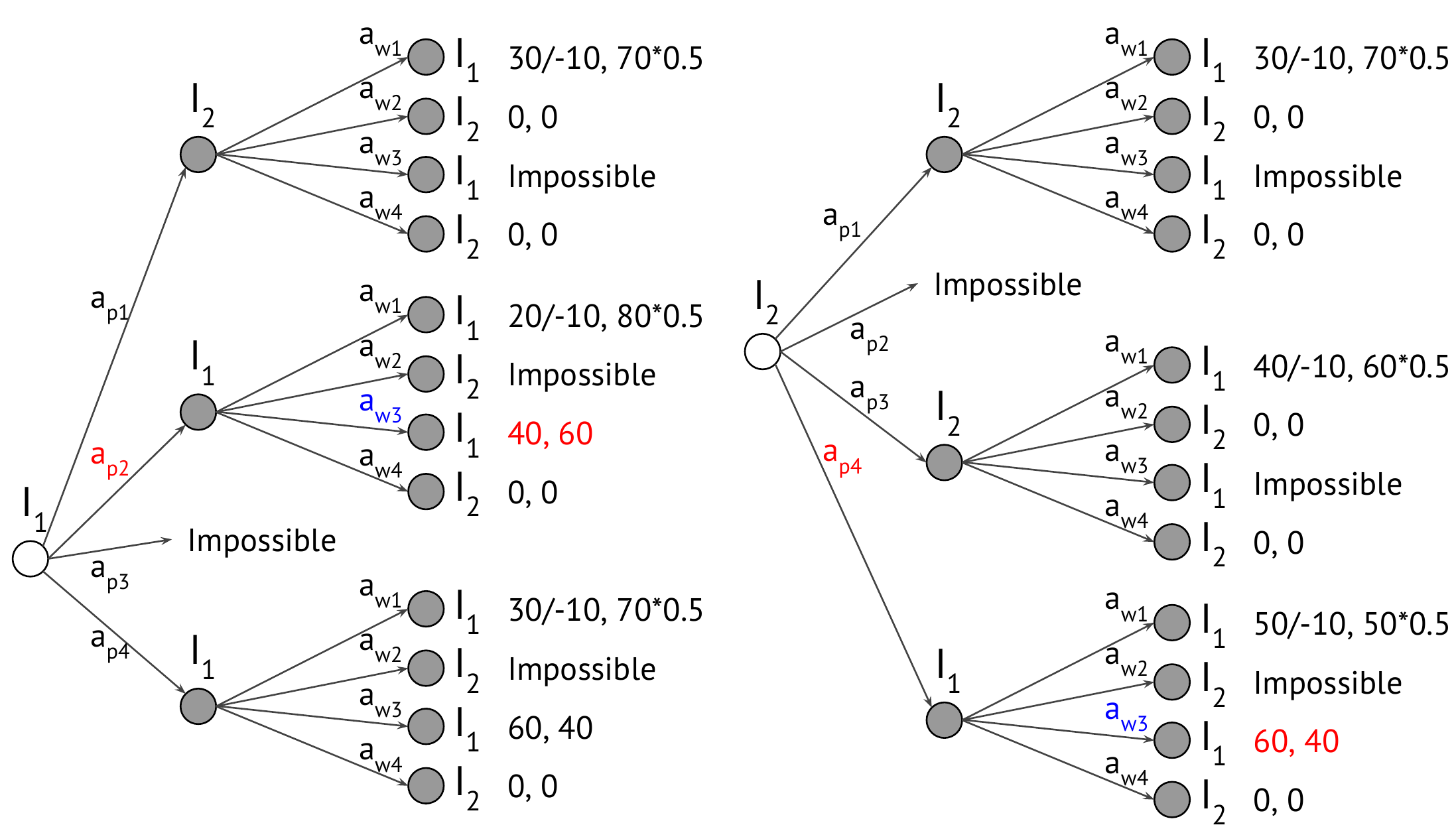}
	\caption{The game between the winner and the publisher in the reserve price problem. \cite{yuan2014empirical} analysed the game between the winner and the publisher in the reserve price problem.
		At the leaf nodes the result information set as well as the payoffs of (winner, publisher) are presented. Note for the action $a_{ w1 }$ the payoff of the winner could be negative if he has already been bidding the maximal affordable price. It is assumed that these cases happen at a chance of $50\%$ due to no utilisation of historical data. Thus, the payoff of the publisher is discounted by $0.5$ in these cases.}
	\label{fig-reserve-price-game-tree}
\end{figure}

\cite{yuan2014empirical} proposed a set of heuristics based on the game tree analysis. They dropped the \emph{repeated} nature of auctions and assume the seller only considered the current auction thus did not learn the private values from history. The extensive form representation of the modified game is described as follows. The game tree is given in Figure~\ref{fig-reserve-price-game-tree}.
 
\begin{itemize}
	\item Player: the winner of auctions (advertisers) $w$ and the publisher $p$.
 
	\item The information set $\mathbb{I}$ before acting is the same for the winner and the publisher. It has two decision nodes:\\
	$I_1$, the winning bid $b$ is equal to or higher than the current reserve price $\alpha$;\\
	$I_2$, the winning bid is lower than the reserve price.
	
	\item The action set of the winner $A_w$:\\
	$a_{w1}$, to increase $b$ to higher than $\alpha$;\\
	$a_{w2}$, to increase $b$ to lower than $\alpha$;\\
	$a_{w3}$, to decrease or hold $b$ to higher than $\alpha$;\\
	$a_{w4}$, to decrease or hold $b$ to lower than $\alpha$.
	
	\item The action set of the publisher $A_p$:\\
	$a_{p1}$, to increase or hold $\alpha$ to higher than $b$;\\
	$a_{p2}$, to increase or hold $\alpha$ to lower than $b$;\\
	$a_{p3}$, to decrease $\alpha$ to higher than $b$;\\
	$a_{p4}$, to decrease $\alpha$ to lower than $b$.
	
	\item The sequence of move: first the publisher, then the winner.	
\end{itemize}

Based on the analysis, the authors claim the following set of heuristics to be the dominant strategy for the publisher.
\begin{align}
  s^\ast_w ( I ) = \left\{
	\begin{array}{l l}
		a_{ w3 }, & \textrm{if} \; I = I_1 \\
		a_{ w1 }, & \textrm{if} \; I = I_2
	\end{array}
\right.
\end{align}
 
The heuristics indicate that the bid price should be gradually reduced but increased again when lost the auction.
 
\subsection{Exploration with a regret minimiser}
 
\cite{cesa2013regret} took a more theoretical approach to the problem. In their work they made similar assumptions and abstracted the problem as follows:
 
\begin{quote}
  A seller is faced with repeated auctions, where each
auction has a (different) set of bidders, and each bidder draws
bids from some fixed unknown distribution which is the same
for all bidders. It is important to remark that we need not
assume that the bidders indeed bid their private value. Our
assumption on the bidders' behaviour, a priori, implies that
if they bid using the same strategy, their bid distribution
is identical. The sell price is the second-highest bid, and the
seller's goal is to maximize the revenue by only relying on
information regarding revenues on past auctions.
\end{quote}
 
The authors proposed an online algorithm that optimise the seller's reserve price and showed that after $T$ steps ($T$ repetitions of the auction) the algorithm has a regret of only $\mathcal{O}(\sqrt{T})$. The work was inspired by \cite{kleinberg2003value} who discretised the range of reserve prices to $\Theta(T^{1/3})$ price bins, and uses some efficient multi-armed bandit algorithm over the bins \citep{auer2002finite}.
 
The proposed algorithm works in stages where each stage contains a few time steps. For stage 1, the algorithm does exploration by setting the reserve price $\alpha$ to 0. Suppose this is played for $T_1$ steps and the revenues $R_1(0), \dots, R_{T_1}(0)$ are observed, thus the empirical distribution of the second highest price is
$$
\hat{F}_{2,1}(x) = \frac{1}{T_1} \left | \left\{ t=1,\dots, T_1 : R_t(0) \leq x \right\} \right |
$$
and the initial estimation on the reserve price is
$$
\hat{\mu}_1(\alpha) = \mathbb{E}\left [B^{(2)} \right] + \int _0^\alpha \hat{F}_{2,1}(t)dt - \alpha \beta^{-1} (\hat{F}_{2,1}(\alpha)).
$$
 
For every following step $t$ in stage $i$, play $\alpha_t=\hat{\alpha}_i$ and observe the revenues $R_1(\hat{\alpha}_i, \dots, R_{T_i}(\hat{\alpha}_i$, where $\hat{\alpha}_i$ is computed as follows
$$
\hat{\alpha}^{\ast}_{i-1} = \arg \max \hat{\mu}_{i-1}(\alpha)
$$
with constraints
$$
\alpha \in [\hat{\alpha}_{i-1}, 1],
$$
$$
\hat{F}_{2, i-1}(\alpha) < 1-a,
$$
where $a$ is the approximation parameter and $a \in (0, 1]$. Let
$$
P_i=\left\{ \alpha \in [\hat{\alpha}_{i-1}, 1] :
  \hat{\mu}_{i-1}(\alpha) \geq \hat{\mu}_{i-1}(\hat{\alpha}^{\ast}_{i-1}
  -2C_{\delta, i-1}  \hat{\alpha}^{\ast}_{i-1}  - 2C_{\delta, i-1} \alpha
\right\},
$$
where $\delta \in (0, 1]$ is the confidence level and the confidence interval is defined as
$$
C_{\delta, i}(\alpha)=\alpha\sqrt{\frac{2}{1-\hat{F}_{2, i}(\alpha)T_i} \ln \frac{6S}{\delta}},
$$
where $S = S(T)$ is either the total number of stages or an upper bound thereof. Then set
$$
\hat{\alpha}_i = \min P_i \cap \left\{ \alpha : \hat{F}_{2, i-1}(\alpha) \leq 1-a \right \}.
$$
 
At the end of every time step, the empirical distribution is updated as
$$
\hat{F}_{2, i}(x) = \frac{1}{T_i} | { t=1, \dots, T_i : R_t(\hat{\alpha}) \leq x} |
$$
and the estimated reserve price is updated as
$$
\hat{\mu}_i(\alpha) =
	\mathbb{E} [ B^{(2)} ] +
	\int _0^{\hat{\alpha}_i} F_2(t)dt +
	\int_{\hat{\alpha}_i}^{\alpha} \hat{F}_{2, i}(t)dt -
	\alpha \beta^{-1}(\hat{F}_{2, i}(\alpha)).
$$
 
The proof is omitted here. Interested readers may refer to \citep{cesa2013regret} for more details.
 
\section{Programmatic direct}

As discussed in Chapter \ref{c-intro}, there are two major ways of selling impressions in display advertising. They are either sold in RTB spot through auction mechanisms or in advance via guaranteed contracts. The former has achieved a significant automation via real-time bidding (RTB); however, the latter is still mainly done over-the-counter through direct sales.
 
Guaranteed inventories stand for guaranteed contracts sold by top tier websites. Generally, they are: highly viewable because of good position and size; rich in the first-party data (publishers' user interest database) for behaviour targeting; flexible in format, size, device, etc.; audited content for brand safety. Therefore, it is not surprising that guaranteed inventories are normally sold in bulk at high prices in advance than those sold on the spot market.

\begin{figure}[t]
	\scriptsize
	\centering
	\includegraphics[width=0.45\linewidth]{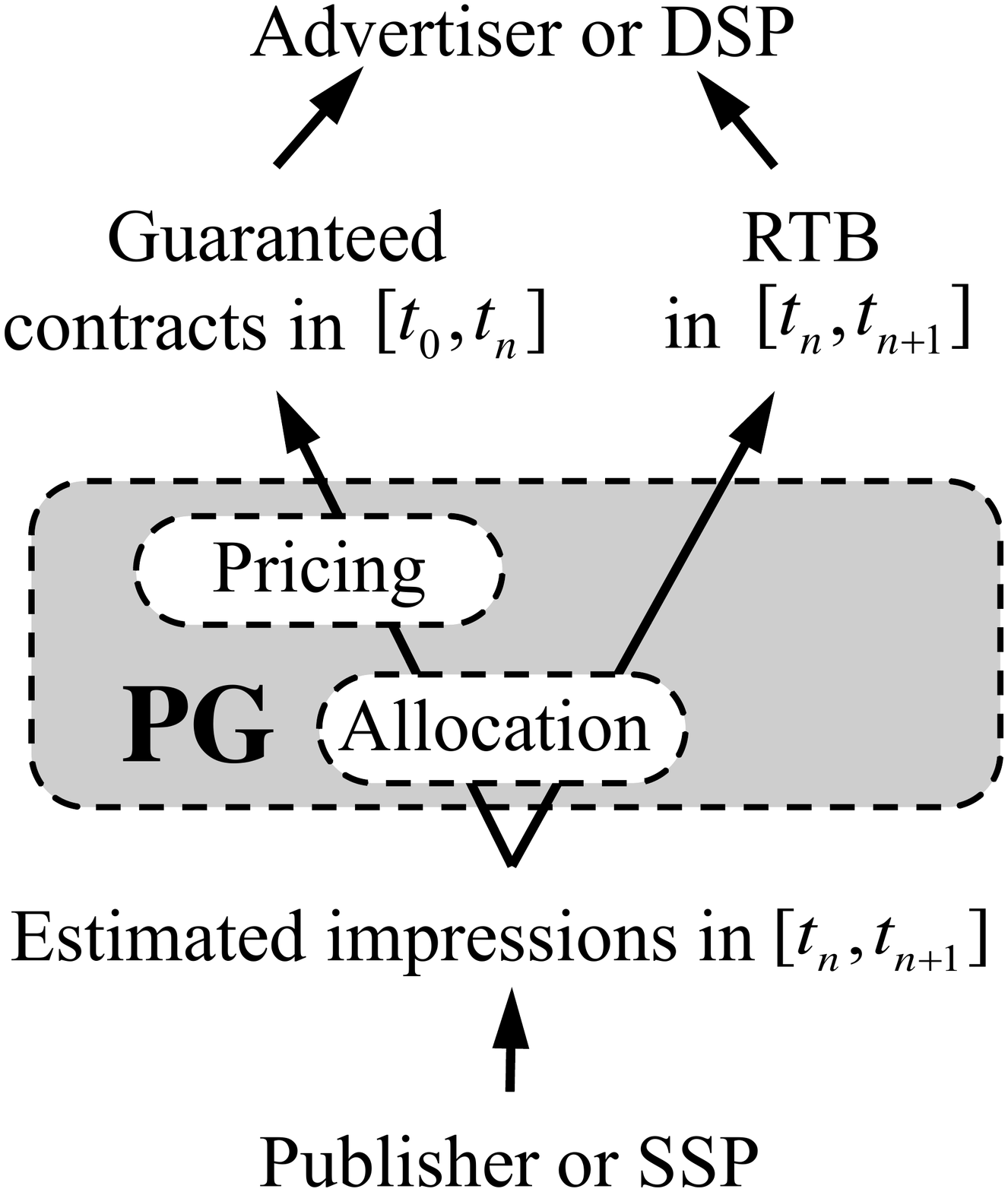}
	\vspace{-10pt}
	\caption{A systematic view of programmatic guarantee (PG) in display advertising reported by \cite{Chen:2014:DPM:2648584.2648585}: $[t_0, t_n]$ is the time period to sell the guaranteed impressions that will be created in future period $[t_n, t_{n+1}]$.}
	\label{fig:shift_of_selling}
\end{figure}

Programmatic guarantee (PG), sometimes called programmatic reserve/premium, is a new concept that has gained much attention recently. Notable examples of some early services on the market are \texttt{iSOCKET.com}, \texttt{BuySellAds.com} and \texttt{ShinyAds.com}. It is essentially an allocation and pricing engine for publishers or supply-side platforms (SSPs) that brings the automation into the selling of guaranteed inventories apart from RTB. Figure~\ref{fig:shift_of_selling} illustrates how PG works for a publisher (or SSP) in display advertising. For a specific ad slot (or user tag\footnote{Group of ad slots which target specific types of users.}), the estimated total impressions in a future period can be evaluated and allocated algorithmically at the present time between the guaranteed market and the spot market. Impressions in the former are sold in advance via guaranteed contracts until the delivery date while those in the latter are auctioned off in RTB. Unlike the traditional way of selling guaranteed contracts, there is no negotiation process between publisher and advertiser. The guaranteed price (i.e., the fixed per impression price) will be listed in ad exchanges dynamically like the posted stock price in financial exchanges. Advertisers or demand-side platforms (DSPs) can see a guaranteed price at a time, monitor the price changes over time and purchase the needed impressions directly at the corresponding guaranteed prices a few days, weeks or months earlier before the delivery date.
 
\cite{Chen:2014:DPM:2648584.2648585} proposed a mathematical model that allocates and prices the future impressions between real-time auctions and guaranteed contracts. Similar problems have been studied in many other industries. Examples include retailers selling fashion and seasonal goods and airline companies selling flight tickets~\citep{Talluri_2004}. However, in display advertising, impressions are with uncertain salvage values because they can be auctioned off in real-time on the delivery date. The combination with RTB requires a novel solution.

 Under conventional economic assumptions, it shows that the two ways can be seamlessly combined programmatically and the publisher's revenue can be maximized via price discrimination and optimal allocation. In the model, advertisers are assumed to be risk-averse, and they would be willing to purchase guaranteed impressions if the total costs are less than their private values. Also an advertiser's purchase behaviour can be affected by both the guaranteed price and the time interval between the purchase time and the impression delivery date. The dynamic programming solution suggests an optimal percentage of future impressions to sell in advance and provides an explicit formula to calculate at what prices to sell. It is found that the optimal guaranteed prices are dynamic and are non-decreasing over time.
They also showed that the model adopts different strategies in allocation and pricing according to the level of competition. In a less competitive market, lower prices of the guaranteed contracts will encourage the purchase in advance and the revenue gain is mainly contributed by the increased competition in future RTB. In a highly competitive market, advertisers are more willing to purchase the guaranteed contracts and thus higher prices are expected. The revenue gain is largely contributed by the guaranteed selling.

\section{Ad options and first look contracts}

In theory, RTB auction has many desirable economic properties. However, it suffers a number of limitations including the uncertainty in payment prices for advertisers, the volatility in the publisher's revenue, and the weak loyalty between advertisers and publishers. \emph{Options} contracts, as a concept, have been introduced recently into online advertising from finance \citep{Black_1973} to solve the non-guaranteed delivery problem as well as provide advertisers with greater flexibility \citep{chen2014lattice,chen2013multi}.  In practice, the option contract has been implemented as a ``First Look'' tactic that is widely offered by
publishers who offer prioritised access to selected advertisers within an open real-time bidding (RTB) market environment~\citep{Yuan:2013:RBO:2501040.2501980}. Instead of the winning
impression going to the highest bid in RTB, ``First Look''
affords first the right of refusal for an impression within an exchange
based on a pre-negotiated floor or fixed price. If the buyer requests it,
they are guaranteed to win the impression. This privilege is typically
granted in return for a commitment. Formally, an ad option is a contract in which an advertiser can have a right but not
obligation to purchase future impressions or clicks from a specific ad
slot or keyword at a pre-specified price. The pre-negotiated price is
usually called the strike price in finance. In display advertising,
the price can be charged as either a CPM or CPC depending on the underlying ad format. The
corresponding winning payment price of impressions or clicks from
real-time auctions is called the underlying price. The publisher or
search engine grants this right in exchange for a certain amount of
upfront fee, called the option price. The option is more
flexible than guaranteed contracts as on the delivery date, if the
advertiser thinks that the spot market is more beneficial, he can join
online auctions as a bidder and his cost of not using an ad option is only the option price.

\begin{figure}[tp]
	\centering
	\includegraphics[width=1\linewidth]{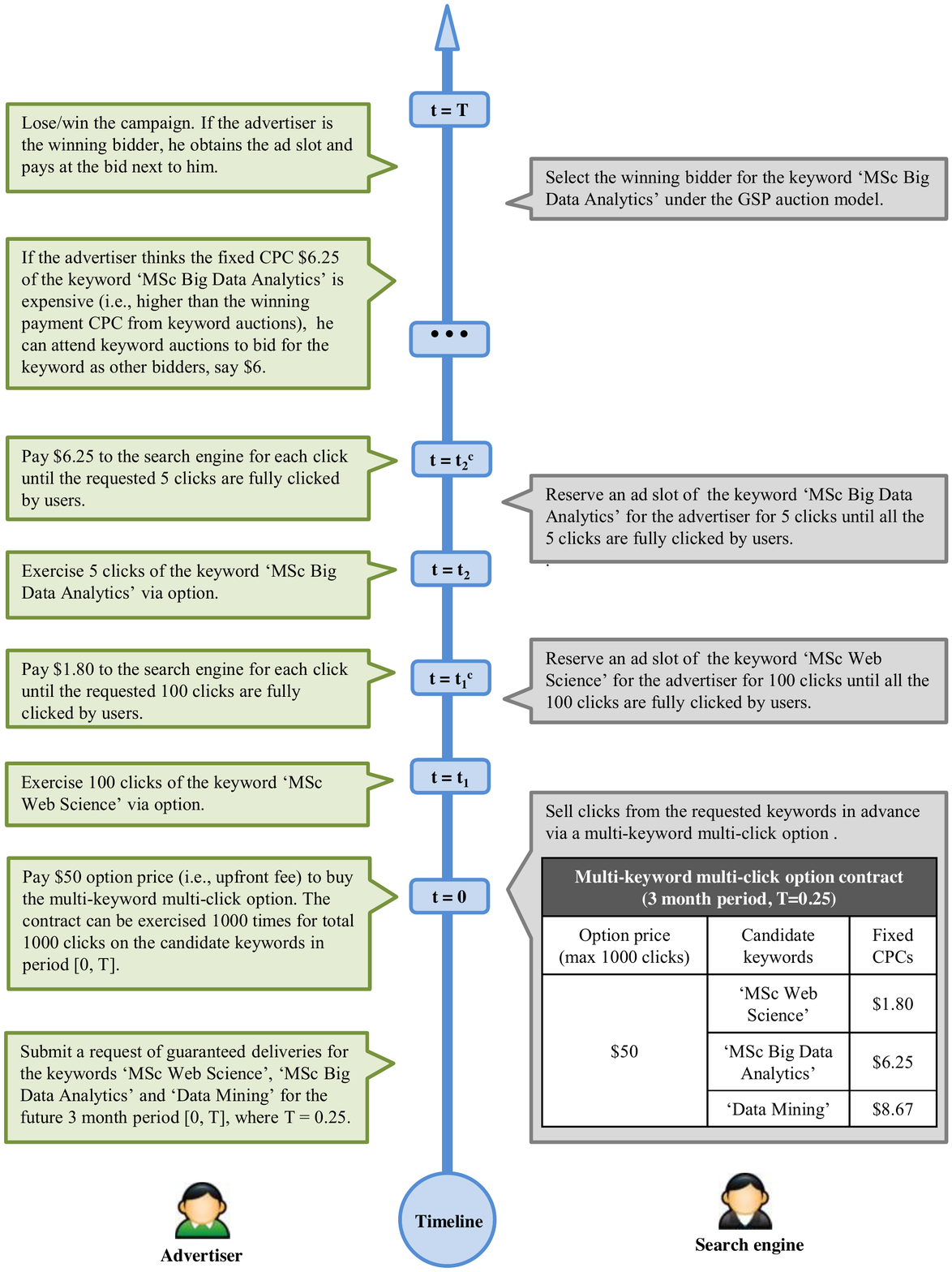}
	\caption{\cite{chen2013multi} designed the schematic view of buying, selling and exercising a multi-keyword multi-click ad option.}
	\label{fig:option_structure}
\end{figure}

\cite{chen2013multi} illustrated such an idea. Suppose that a computer science department creates a new master degree programme on \lq{}Web Science and Big Data Analytics\rq{} and is interested in an advertising campaign based around relevant search terms such as \lq{}MSc Web Science\rq{}, \lq{}MSc Big Data Analytics\rq{} and \lq{}Data Mining\rq{}, etc. Similarly, in display advertising, webpages and underlying user interests are classified into predefined categories and therefore can be equally used as targeting categories (keywords). The campaign is to start immediately and last for three months and the goal is to generate at least 1000 clicks on the ad which directs users to the homepage of this new master programme. The department (i.e., advertiser) does not know how the clicks will be distributed among the candidate keywords, nor how much the campaign will cost if based on keyword auctions. However, with the ad option, the advertiser can submit a request to the search engine to lock-in the advertising cost. The request consists of the candidate keywords, the overall number of clicks needed, and the duration of the contract. The search engine responds with a price table for the option, as shown in Figure~\ref{fig:option_structure}. It contains the option price and the fixed CPC for each keyword. The CPCs are fixed yet different across the candidate keywords. The contract is entered into when the advertiser pays the option price.

During the contract period $[0, T]$, where $T$ represents the contract expiration date (in terms of year format and is three months in this example), the advertiser has the right, at any time, to exercise portions of the contract, for example, to buy a requested number of clicks for a specific keyword. This right expires after time $T$ or when the total number of clicks have been purchased, whichever is sooner.  For example, at time $t_1 \leq T$ the advertiser may exercise the right for 100 clicks on the keyword \lq{}MSc Web Science\rq{}. After receiving the exercise request, the search engine immediately reserves an ad slot for the keyword for the advertiser until the ad is clicked 100 times. In our current design, the search engine decides which rank position the ad should be displayed as long as the required number of clicks is fulfilled - it is assumed there are adequate search impressions within the period. It is also possible to generalise the study in this section and define a rank specific option where all the parameters (CPCs, option prices etc.) become rank specific. The advertiser can switch among the candidate keywords and also monitor the keyword auction market. If, for example, the CPC for the keyword \lq{}MSc Web Science\rq{} drops below the fixed CPC, then the advertiser may choose to participate in the auction rather than exercise the option for the keyword. If later in the campaign, the spot price for the keyword `MSc Web Science' exceeds the fixed CPC, the advertiser can then exercise the option.
 
Figure~\ref{fig:option_structure} illustrates the flexibility of the proposed ad option. Specifically, (i) the advertiser does not have to use the option and can participate in keyword auctions as well, (ii) the advertiser can exercise the option at any time during the contract period, (iii) the advertiser can exercise the option up to the maximum number of clicks, (iv) the advertiser can request any number of clicks in each exercise provided the accumulated number of exercised clicks does not exceed the maximum number, and (v) the advertiser can switch among keywords at each exercise with no additional cost. Of course, this flexibility complicates the pricing of the option, which is discussed next.

One of the key issues for ad options contracts is their pricing and valuation. \cite{wang2012selling}~and~\cite{chen2013multi} proposed ad options
between buying and non-buying the future impressions and consider the situation where the underlying price follows a geometric Brownian motion (GBM)~\citep{Samuelson_1965_2}.  \cite{chen2013multi} investigated a special option for sponsored search whereby an advertiser can target a set of keywords for a certain number of total clicks in the future. Each candidate keyword can be specified with a strike price and the option buyer can exercise the option multiple times at any time prior to or on the contract expiration date. According to~\citep{yuan2014empirical}, there is only a very
small number of ad slots whose CPM or CPC satisfies the
GBM assumption. \cite{chen2014lattice} addressed the issue and provided a more general pricing framework, based on lattice methods. It used a stochastic volatility (SV) model to describe the underlying price movement for cases where the GBM assumption is not valid. Based on the SV model, a censored binomial lattice is constructed for option pricing. Lattice methods can also be used for pricing display ad options with the GBM underling. Several binomial and trinomial lattice methods were examined to price a display ad option and deduce the close-form solutions to examine their convergence performance.
 

\chapter{Attribution Models}
\label{c-attri}
Online advertising provides feasibility to track users' interaction on the displayed ads such as the clicks. However, a user's final conversion (e.g. item purchase) is usually contributed by multiple ad events, namely touchpoints. Thus theoretically the credit of such a conversion should be properly allocated over these touchpoints.

\begin{figure}
  \centering
  \includegraphics[width=0.9\columnwidth]{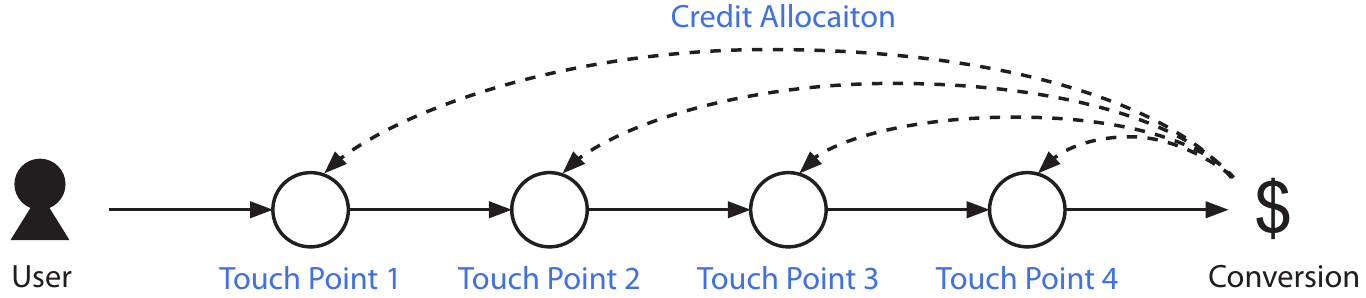}
  \caption{An illustration of conversion attribution over multiple touchpoints.}\label{fig:conv-attr}
\end{figure}

As illustrated in Figure~\ref{fig:conv-attr}, conversion attribution is the problem of assigning credit to one or more channels for driving the user to the desirable actions such as making a purchase. It is important to have a ``right'' attribution model in order to reallocate budgets on different channels and campaigns \citep{geyik2014multi}. However, such a problem is theoretically and practically difficult to solve since there is no ``ground-truth'' data indicating how the credit should be perfectly allocated across different channels. In this chapter, we present a series of models proposed for the conversion attribution problem.
 
Designing a good conversion attribution is important for RTB display advertising although it is not unique in this area. 
Because of the high flexibility of impression-level decision making in RTB, correctly or reasonably allocating the conversion credit across the multiple touchpoints would effectively avoid cheating and gaming the ad systems.

\section{Heuristic models}\label{sec:heuristics}
Basically, heuristic models are based on human-created rules possibly according to their business experiences. Such models are simple, straightforward, and widely adopted in industry.
 
A list of heuristic models discussed by Google Analytics \citep{kee2012attribution} are provided in Table~\ref{tab:heuristic-models}.
 
\begin{description}
	\item [Last touch:] the last touchpoint earns 100\% credit. Last Touch Attribution (LTA) model is the most widely adopted conversion attribution mechanism in display advertising \citep{dalessandro2012causally}. It is quite straightforward and easy to implement. The advertiser only needs to check the last touchpoint before the user conversion and count such a conversion on it. However, LTA obviously encourages DSPs to focus the campaign budget on the late touchpoints, such as retargeted users, i.e., the users who have already showed their interest on the advertised products. As such, less budget will be put on the new users, which indeed hurts the advertisers' benefit in the long term.
 
	\item [First touch:] the first touchpoint earns 100\% credit. This model drives the DSPs to run any performance-driven campaign just like branding campaigns because the target is to cover as many new users as possible.
 
	\item [Linear touch:] the conversion credit is evenly allocated across all the touchpoints. This model might be useful if campaigns are designed to maintain users' awareness throughout the campaigns' lifetime.
 
	\item [Position based:] the first and last touchpoints are highlighted during the customer journey, which directly (but might not effectively) encourage branding and performance-driven campaigns.
 
	\item [Time decay:] the fresh touchpoints are regarded as more contributive than the old ones. This model might be helpful if the business (e.g. promotion, sales) is operated in a short period.
 
	\item [Customised:] the advertiser can generally create their own rules to allocate the conversion credit across the touchpoints.
\end{description}
 
\begin{table}[t]
\centering
\caption{Several heuristic attribution models on touchpoints in Figure~\ref{fig:conv-attr}.}\label{tab:heuristic-models}
\begin{tabular}{|c|rrrr|}\hline
Model & \multicolumn{4}{c|}{Attribution}\\ \hline
touchpoint & 1 & 2 & 3 & 4\\ \hline
Last touch & 0\% & 0\% & 0\% & 100\% \\
First touch & 100\% & 0\% & 0\% & 0\% \\
Linear touch & 25\% & 25\% & 25\% & 25\% \\
Time decay & 10\% & 20\% & 30\% & 40\% \\
Position based & 40\% & 10\% & 10\% & 40\% \\
Customised & 5\% & 25\% & 15\% & 55\% \\
\hline
\end{tabular}
\end{table}
 
Although the above heuristic models are still popular, they are usually from the advertisers' intuition or personal experience, instead of the data. It is not hard to see that the heuristic models are far from optimal. One may easily game the systems, particularly the last-touch attribution and first-touch attribution models. For instance, the so-called ``cookie bombing'' strategies of some advertisers or DSPs are to game the ad systems by advocating quantity over quality. One can deliver the ads as cheaply as possible, regardless of targeting, content or frequency, and drive the number of reached users in order to just get credited by either the last-touch attribution or the first-touch attribution models. We next introduce multi-touch and data-driven attribution models.
 
\section{Shapley value}
 
We provide a quick introduction to Shapley value \citep{shapley1952value}, a concept from cooperative game theory which aims to fairly allocate the utility of the game across the coalition of players.
 
In the online advertising scenario, Shapely value $V_k$ of the $k$-th touchpoint of the customer journey is
 
\begin{align}
V_k &= \sum_{S \subseteq C\backslash k} w_{S,k} \cdot (\mathbb{E}[y|S \cup k] - \mathbb{E}[y|S]), \label{eq:shapely}\\
w_{S,k} &= \frac{|S|!(|C| - |S| - 1)!}{|C|!} \label{eq:shapely-w},
\end{align}
 
where $C$ is the set of the whole channels; $S$ is an arbitrary subset of $C \backslash  k$, including the empty set $\varnothing$; $y$ denotes the achieved utility of the coalition game. The calculation is straightforward: the Shapely value of $k$ is the weighted average of the improvement of the expected $y$ and the weight is the probability of a $|C|$-length sequence starting with $S,k$.
 
The factors $\mathbb{E}[y|S \cup k]$ and $\mathbb{E}[y|S]$ in Eq.~(\ref{eq:shapely}) are calculated from the data statistics (simply by counting). However, the weight terms come purely from permutation and combination without considering the data distribution.
 
In the next section, we shall introduce a series of full data-driven probabilistic models.
 
\section{Data-driven probabilistic models}
 
\cite{Shao:2011:DMA:2020408.2020453} first presented the concept of data-driven multi-touch attribution. In their paper, two models are implemented, namely bagged logistic regression and a simple probabilistic model.
 
\subsection{Bagged logistic regression}
 
The idea of bagged logistic regression is to predict whether the user is going to convert given the current ad touch events. Without the consideration of repeated touches from the same channel, the input data of the logistic regression is the vector of the user's ad touch events $\bs{x} = [x_1, x_2, \ldots, x_n]$, where $x_i$ is the binary value indicating whether the user has been touched by the channel $i$. The corresponding $y$ is the binary value of whether the user made the final conversion.
 
\cite{Shao:2011:DMA:2020408.2020453} proposed to use bagging process to train the logistic regression with sampled data instances and sampled channels so that the averaged weight of each channel is more robust and reliable. The averaged weight of each dimension is regarded as the importance of the channel used for the conversion credit and budget allocation.
 
\subsection{A simple probabilistic model}
 
\cite{Shao:2011:DMA:2020408.2020453} also proposed a simple probabilistic model which combines the first- and second-order channel conditional conversion probabilities.
 
For a given dataset, the first-order conditional conversion probability of channel $i$ is calculated as
\begin{align}
P(y=1|x_i) = \frac{N_{\text{positive}}(x_i)}{N_{\text{positive}}(x_i) + N_{\text{negative}}(x_i)},
\end{align}
where $N_{\text{positive}}(x_i)$ and $N_{\text{negative}}(x_i)$ are the number of users ever exposed to channel $i$ with and without final conversions, respectively. Similarly, the second-order conditional conversion probability of a channel pair $i,j$ is calculated as
\begin{align}
P(y=1|x_i,x_j) = \frac{N_{\text{positive}}(x_i,x_j)}{N_{\text{positive}}(x_i,x_j) + N_{\text{negative}}(x_i,x_j)},
\end{align}
where $N_{\text{positive}}(x_i,x_j)$ and $N_{\text{negative}}(x_i,x_j)$ are the number of users exposed to channels $i$ and $j$ with and without conversions, respectively.
 
Then the contribution of channel $i$ based on the dataset is summarised as
\begin{equation}
\label{eq:shao-attr}
\begin{aligned}
V(x_i) &= P(y|x_i) + \frac{1}{2N_{j\neq i}} \sum_{j\neq i} \Big( P(y|x_i, x_j) - P(y|x_i) - P(y|x_j) \Big)\\
&= \frac{1}{2} P(y|x_i) + \frac{1}{2N_{j\neq i}} \sum_{j\neq i} \Big( P(y|x_i, x_j) - P(y|x_j) \Big),
\end{aligned}
\end{equation}
where $N_{j\neq i}$ denotes the number of channels other than $i$.
 
From the above equation we can see that the proposed probabilistic model is indeed a simplification to the Shapley value model: (i) it only considers the first- and second-order channel combinations; (ii) the weight of the first order conditional probability $P(y|x_i)$ is arbitrarilyset as $1/2$, which is neither consistent with Eq.~(\ref{eq:shapely-w}) nor calculated by the data.
 
\subsection{An extension to the probabilistic model}
 
Extending from both Shapley value \citep{shapley1952value} and the probabilistic model \citep{Shao:2011:DMA:2020408.2020453}, \cite{dalessandro2012causally} proposed a causal framework for multi-touch attribution. The importance of channel $i$ is calculated as
 
\begin{equation}
\label{eq:perlich-attr}
V(x_i) = \sum_{S \subseteq C\backslash i} w_{S,i} ( P(y|S, x_i) - P(y|S) ),
\end{equation}
 
where $C$ is the set of the whole channels; $S$ is an arbitrary subset of $C \backslash  k$, including the empty set $\varnothing$, just like Eq.~(\ref{eq:shapely-w}). But the calculation of probability $w_{S,i}$ is totally based on the data observation instead of the permutation and combination in Eq.~(\ref{eq:shapely-w}) without considering the data distribution.
 
\section{Other models}
 
There are more relevant papers about conversion attribution.
\cite{abhishek2012media} developed a hidden Markov model (HMM) to tackle the attribution problem based on the concept of a conversion funnel.
\cite{anderl2014mapping} proposed a Markov graph to model the first- and high-order Markov walks in the customer journey.
\cite{wooff2015time} suggested an asymmetric ``bathtub shape'' time-weighted attribution model for online retailer advertising.
\cite{zhang2014multi} assumed the time-decay attribution patterns (similar with the one discussed in Section~\ref{sec:heuristics}) and proposed to leverage Cox time survival model to calculate the credit allocation.
Besides the previously mentioned causal probabilistic inference model \citep{dalessandro2012causally}, there are further causal inference research for conversion attribution \citep{barajas2015online,barajas2015estimating}.
\cite{xu2014path} proposed a mutually exciting point process to model the path to the purchase in online advertising.
People also focused on the path to the purchase \citep{xu2014path}.
The incremental utility given a channel is also formally studied in \citep{sinha2014estimating} with an econometric model and in \citep{xu2016lift} with a boosted tree Machine Learning model.
 
After the discussion of various multi-touch attribution models, we explain how they can be leveraged for budget allocation \citep{geyik2014multi} and bidding \citep{xu2016lift}, yielding better advertising performance.
 
 
 
\section{Applications of attribution models}
 
With a multi-touch attribution model, from the micro perspective, the credit of a particular conversion can be assigned over multiple previous touchpoints, which motivates MTA-based bidding strategies in each channel \citep{xu2016lift}. From the macro perspective, it is feasible for the advertiser to make a sensible budget allocation over different channels to optimise the overall advertising performance \citep{geyik2014multi}.
 
\subsection{Lift-based bidding}
 
The traditional bidding strategies discussed in Chapter~\ref{c-bid} are called \emph{value-based} bidding strategies as the bidding is based on the estimated value (i.e. utility) of the potential impression. For example, let $\theta$ be the conversion rate of the ad impression and $r$ be the value of the conversion, then the truth-telling bidding strategy will bid
\begin{align}
b_{\text{value}} = r \times \theta. \label{eq:value-bid}
\end{align}
 
Recently, \cite{xu2016lift} proposed a concept of \emph{lift-based} bidding strategies, where the lift of conversion rate of the user after showing the ad is estimated as $\Delta \theta$ and the corresponding bid price is
\begin{align}
b_{\text{lift}} = r \times \Delta \theta. \label{eq:lift-bid}
\end{align}
The lift CVR indeed corresponds to the conversion credit assigned to such touchpoints from a multi-touch attribution model:
\begin{align}
P(\text{attribution}|\text{conversion}) = \frac{\Delta \theta}{\theta}\\
b_{\text{lift}} = r \times \theta \times P(\text{attribution}|\text{conversion}) \label{eq:lift-bid-conv-attr}.
\end{align}
By contrast, $P(\text{attribution}|\text{conversion})=1$ in Eq.~(\ref{eq:lift-bid-conv-attr}) for last-touch attribution.
 
The basic assumption of lift-based bidding strategies is the user could make the conversion in any context, even if there is no ad exposure at all. As such, in any context with any previous touches, denoted as $\mathcal{H}$, there is an underlying conversion rate $\theta = P(\text{conv}|\mathcal{H})$ for each user, and an ad impression $h$ is possible to lift the user's conversion rate $\Delta \theta = P(\text{conv}|\mathcal{H},h) - P(\text{conv}|\mathcal{H})$.
 
\cite{xu2016lift} proposed to leverage gradient boosting decision trees (GBDT) to estimate $P(\text{conv}|\mathcal{H})$ for each case and then calculated the CVR lift. They further proved that such a lift-based bidding strategy (\ref{eq:lift-bid}) yields more conversions than the value-based bidding strategy (\ref{eq:value-bid}). This is intuitive: the value-based bidding strategy focuses the campaign budget on high-CVR users but such users are already likely to convert and further ad exposures do not improvement the CVR much; the lift-based bidding strategy allocate the budget on the impressions according to the impression contributions on the CVR, which improves the expected conversion number of the campaign.
 
Unfortunately, the experiment of Xu et al. showed that although such a lift-based bidding strategy indeed brought more conversions to the advertiser, more conversions are assigned to the competitive value-based bidding strategies than the lift-based one because of the last-touch attribution mechanism. Only when multi-touch attribution mechanism is adopted for all the campaigns of the advertiser or even in the whole RTB marketplace, can the lift-based bidding strategies be widely used, which will push the market to a higher efficient one.
 
\subsection{Budget allocation}
 
\cite{geyik2014multi} proposed a framework of performance-driven campaign budget allocation across different channels with a certain attribution model as input. Suppose there are $n$ channels of a campaign $X={x_1, x_2, \ldots, x_n}$, the maximum spending capability of each channel $x_i$ is $S_i$, the campaign global budget is $B$, and the ROI of each channel is $R_i$, then the budget allocation problem is formulated as
\begin{align}
\max_{B_1,\ldots,B_n} ~~& \sum_{i=1}^n R_i B_i\\
\text{subject to} ~~& 0 \leq B_j \leq S_j ~~ \forall j\in \{1,\ldots,n\}\\
& \sum_{i=1}^{n} B_i \leq B,
\end{align}
where $B_i$ is the budget allocated to the channel $x_i$, which are to be optimised.
 
The attribution model reflects in the ROI calculation of each channel $R_i$:
\begin{align}
R_i = \frac{\sum_a P(x_i|a) r_a}{\text{Cost in $x_i$}},
\end{align}
where $a$ represents an observed conversion, $r_a$ is the monetised value of $a$, $P(x_i|a)$ is the conversion credit assigned to channel $x_i$ according to the attribution model.
 
For the multi-touch attribution model, \cite{geyik2014multi} adopted the one proposed by \cite{Shao:2011:DMA:2020408.2020453}
\begin{align}
P(x_i|a) = \frac{V(x_i)}{\sum_{j=1}^n V(x_j)},
\end{align}
where $V(x_j)$ is calculated as in Eq.~(\ref{eq:shao-attr}).
In their 12-day online experiment, where each channel is the line item (sub-campaign) of the campaign, the budget allocation bucket with MTA consistently outperformed the one with LTA.

\chapter{Fraud Detection}
\label{c-fraud}
As reported by Interactive Advertising Bureau's (IAB) in 2015, ad fraud is costing the U.S. marketing and media industry an estimated \$8.2 billion each year \citep{iab2015what}. The report contributes \$4.6 billion, or 56\%, of the cost to ``invalid traffic'', of which 70\% is performance based, e.g., CPC and CPA, and 30\% is CPM based. These are already staggering numbers comparing with the annual spend of \$59.6 billion in U.S., however, because ad fraud is hard to detect and tools to protect advertisers are immature, the actual numbers could be much higher.
 
Although researchers have devoted a great effort in auction mechanism design to avoid manipulation, as discussed in Chapter~\ref{c-auct}, there are indeed opportunities to game the system, especially when participants do not aim to win auctions, or the seller does not have genuine impressions to sell. Ad fraud has existed since the beginning of sponsored search, mainly in the form of click fraud. In recent years it has been gaining traction, as RTB is now being widely adopted \citep{fulgoni2016fraud}. The distributed structure of RTB ad exchanges makes it easier to commit and conceal fraud. In this chapter, we first review different types of ad fraud, and then introduce countermeasures, focusing on impression fraud which is getting more and more prevalent in RTB \citep{stone2011understanding,crussell2014madfraud}.
 
\section{Ad fraud types}
 
Ad fraud types have been explained well in \citep{daswani2008online,stone2011understanding}. In general, we could follow the definition by \cite{google2016ad}:
\begin{quote}
"Invalid traffic including both impressions, clicks, and conversions which are not to be the result of the genuine user interests."
\end{quote}
 
There are generally three types of ad fraud, which corresponds to the three types of commonly used pricing models:
 
\begin{itemize}
	\item Impression fraud, where the fraudster generates fake bid requests, sells them in ad exchanges, and gets paid when advertisers buy them to get impressions;
	
	\item Click fraud, where the fraudster generates fake clicks after loading an ad; and
	
	\item Conversion fraud, where the fraudster completes some actions, e.g., filling out a form, downloading and installing an app, after loading an ad.
\end{itemize}
 
Note that different types of fraud often appear together. For example, click fraud usually comes with impression fraud, as described by \cite{daswani2008online}, to achieve a reasonable CTR in analytical reports. In this chapter we discuss the fraud issues mainly in the context of RTB, however, note that they usually have a very close relationship and similar techniques with ones in traditional advertising channels, e.g., click fraud in Sponsored Search.
 
\section{Ad fraud sources}
 
Ad fraud is generated from a variety of sources. Due to its profit potential, many parties, especially the supply side, are attracted to the business, creating complex structures to take advantage of the distributed online advertising ecosystem. In this section we describe a few sources, including pay-per-view networks, botnets, and competitor's attack. If one could know where the fraud is from and how it is created, he/she may be in a better position to detect and filter it from the normal traffic.
 
\subsection{Pay-per-view networks}
 
\begin{figure}
	\centering
	\includegraphics[width=\columnwidth]{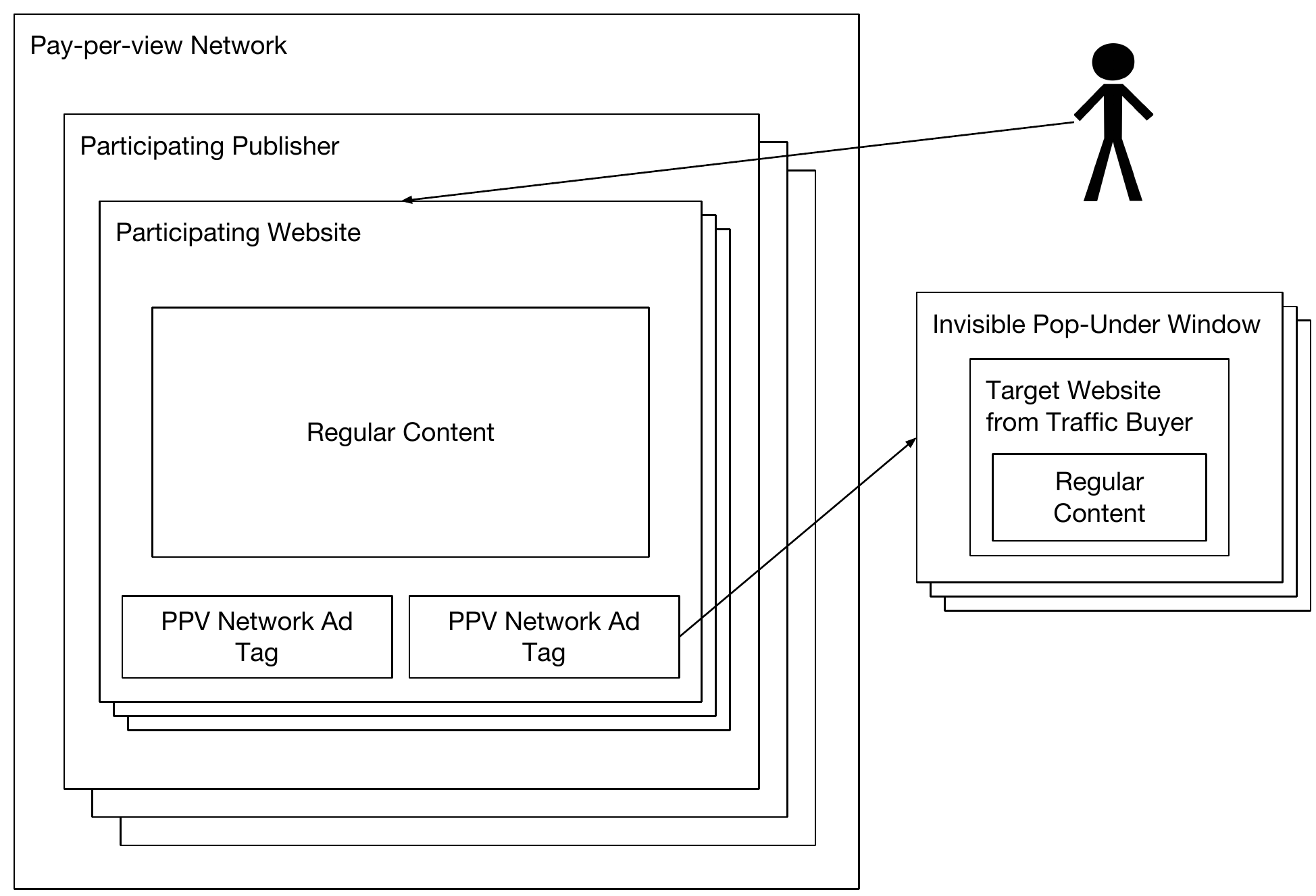}
	\caption{How pay-per-view (PPV) networks generate traffic.}
	\label{fig-ad-fraud-ppv-network}
\end{figure}

Pay-per-view (PPV) networks have been comprehensively studied in \citep{springborn2013impression}. Authors set up honeypot websites and purchased traffic from public available traffic-generation service providers. These providers usually offer a specified volume of traffic at a target website over a specified time period. Many of them support advanced features like geography targeting, mobile traffic, or click events. In their study, authors report that most of the traffic comes from the PPV networks, which pays legitimate publishers for implementing their tags. These tags are used to create Pop-Under windows which load the target website. Pop-Under windows are below the current browser window and usually have a size of 0x0 pixels, thus cannot be easily discovered and closed by ordinary Internet users. This process is illustrated in Figure \ref{fig-ad-fraud-ppv-network}.
 
The characteristics of the PPV network traffic are reported by \cite{springborn2013impression}:	Most of the purchased traffic does not follow the normal diurnal cycle and there is little interaction from purchased traffic.  Also it has been shown that there are large number of incomplete loads from the PPV traffic, where the best case is approximately 60\%. Many different IP addresses are used especially for larger purchases. Some PPV networks are able to provide thousands of unique IPs with little overlap with the public IP blacklists. In addition, there is a great diversity of User Agents, but almost half of the views have a height or width of 0, which is consistent with the idea of using Pop-Under windows to generate fake impressions.
 
These characteristics then shed light on building a detection and filtering system. Based on these discoveries, a few countermeasures are proposed in the paper. Advertisers could employ these methods to get a better control of ad spend.
 
\begin{itemize}
	\item Viewport size check: valid impressions will not be displayed in a 0x0 viewport, which is invisible to users;
 
	\item A referer blacklist, which checks if the traffic is from the PPV networks;
 
	\item	A publisher blacklist, which avoids buying traffic from publishers who participate in the PPV networks.
\end{itemize}
 
\subsection{Botnets}
 
\cite{feily2009survey} provided a good survey on botnet and its detection. Botnets are usually built with compromised end users' computers. These computers are installed with one or multiple software packages, which run autonomously and automatically. Taking over personal computers helps botnets avoid detection. It diversifies the IP addresses and geographic locations, masking the loads of traffic they send across the Internet. Computers could get infected by accessing a hacked Wi-Fi network, web browser or operating system vulnerabilities, worms, installing software with \emph{Trojan horses} or backdoors. Once infected, they join a network and listen to and execute commands issued by the botnet owner, or the botnet master.
 
In history, botnets have been used to steal credentials, to extort money with the threat of deploying Distributed Denial of Service (DDoS), and to send spams. Recently, botnets have been used to conduct ad fraud more often and have been reported to steal billions of dollars from advertisers every year \citep{stone2011understanding}.
 
Once a computer is under the control of a botnet, it can generate Internet traffic as the botnet master commands. The traffic is then sold to publishers who believe they could make more money by reselling the traffic to ad networks or ad exchanges they are part of and do not get caught, or directed to target websites with ad tags set up by botnet master himself.
 
There are mainly two ways to generate traffic on an infected computer:
 
\begin{itemize}
	\item Hijack the original network traffic and inject / replace ad code \citep{thomas2015ad};
	
	\item Open browser windows which are invisible to end users to load target website \citep{stone2011understanding}.
\end{itemize}
 
Note that in the second case, the botnet softwares are capable of generating clicks, too.
 
There are a few ways of detecting a botnet as discussed by \cite{vratonjic2010isps}.
 
\begin{itemize}
	\item Signature based detection, which extracts software / network package signature from known botnet activities;
	
	\item Anomaly based detection, which attempts to detect botnets based on several network traffic anomalies such as high network latency, high volumes of traffic, traffic on unusual ports, and unusual system behaviour that could indicate presence of malicious bots in the network;
	
	\item DNS based detection, which focuses on analysing DNS traffic which is generated by communication of bots and the controller;
	
	\item Mining based detection, which uses Machine Learning techniques to cluster or classify botnet traffic.
\end{itemize}
 
\begin{figure}
	\centering
	\includegraphics[width=\columnwidth]{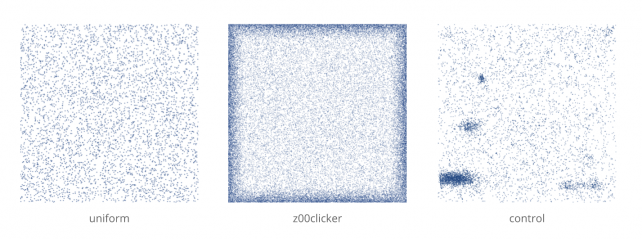}
	\caption{\cite{alex2015inside} reported the unique click patterns from a botnet \emph{z00clicker}.}
	\label{fig-ad-fraud-z00clicker}
\end{figure}

An example of the last approach is Google's analysis on \emph{z00clicker}, as reported in \citep{alex2015inside} and shown in Figure \ref{fig-ad-fraud-z00clicker}. The click pattern of the known botnet is obviously from the one generated by ordinary users. By mining and classifying click patterns of impressions, one may be able to identify undiscovered botnets, too.

\subsection{Competitors' attack}
 
Advertisers spend their budget in online advertising to buy impressions and clicks. This gives an advertiser's competitors a chance to attack by intentionally loading and clicking its ads, especially in the Sponsored Search scenario, as described by \cite{daswani2008online}. These fraudulent clicks, which will be marked invalid if identified, usually have the intention to drain the competitor's advertising budget. Once the budget is completely drained, not only will the attacker's ads be shown exclusively to target users, but he could also pay less due to the nature of the Second Price Auction. 
 
Similarly, a competitor could choose to load advertiser's ads intentionally and repeatedly, but without clicking on them. The process is similar to impression fraud in the PPV networks and by Botnets discussed above. In Sponsored Search, this attack usually results in a low CTR of ads which in turn heavily affects the Quality Score. With a low Quality Score the advertiser will suffer from suboptimal ranking positions, and/or will have to pay more to secure an advertising slot. In the RTB context, this attack could aim to deplete the advertiser's budget, then the competitor could buy the specific inventory at a lower price. 

\subsection{Other sources}
 
There are other sources of ad fraud, or cheating behaviours, in the context of RTB \citep{stone2011understanding}:
 
\begin{itemize}
	\item Hired spammers;
	
	\item Keyword stuffing, where a publisher stuffs the webpage with irrelevant keywords (usually invisible to users) in the hope of retrieving high value ads;
	
	\item Impression stuffing, where a publisher "stacks" many banners together or make them invisible to get a large number of impressions per page view;
	
	\item Coercion, where the publisher explicitly ask users to click on ads to support the website, or obfuscates ads with regular content;
	
	\item Forced browser action, where the attacker forces the user's browser to load additional webpages (e.g., Pop-up ads) or click on ads.
\end{itemize}
 
\section{Ad fraud detection with co-visit networks}
 
Ad fraud detection is usually an unsupervised learning problem and it is difficult to capture the ground-truth. \cite{stitelman2013using} proposed a method to identify malicious website clusters (which generate fraud impressions) by looking at the co-visit network. The co-visit network is defined on a bi-partite graph $G=\left\langle B,W,E \right\rangle$ of browsers (users) $B$ and websites $W$ that the browsers are observed visiting. $E$ is the set of edges between the browsers and the websites they are observed at over a pre-specified period of time (e.g., one day or one week). After normalising the number of browsers on each website and introducing a threshold on the co-visitation rate, the co-visit network could be established on the projection:
$$
G_W^n = \left\langle V_W \subseteq W,
E = {(x,y): x,y \in W,[\Gamma_G(x) \cap \Gamma_G(y)] / \Gamma_G(x) \geq n} \right \rangle.
$$
 
\begin{figure}
	\centering
	\includegraphics[width=\columnwidth]{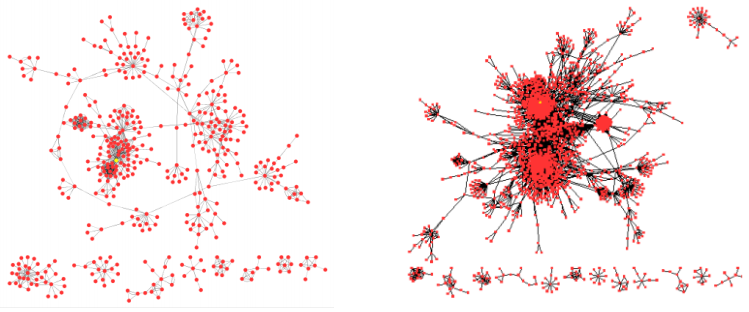}
	\caption{The co-visit networks of Dec 2010 (left) and Dec 2011 (right) reported by  \cite{stitelman2013using}.}
	\label{fig-ad-fraud-co-visit-networks}
\end{figure}

In the network, an edge will be created between two website nodes if there are at least $n\times 100 \%$ of users on website $x$ also visited website $y$. By setting the threshold to $50\%$ the authors report two networks from Dec. 2010 and Dec. 2011 respectively, as shown in Figure \ref{fig-ad-fraud-co-visit-networks}.
 
\begin{figure}
	\centering
	\includegraphics[width=\columnwidth]{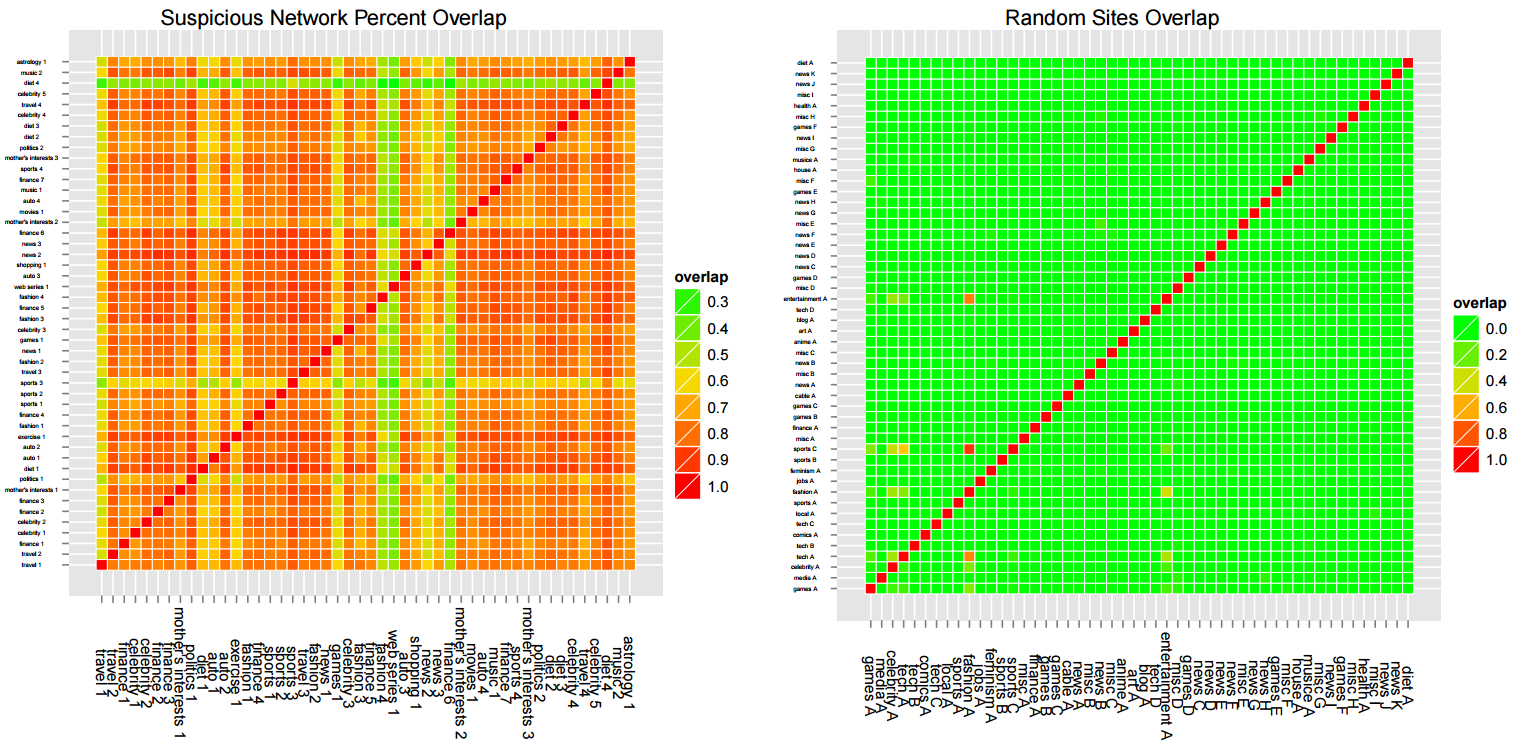}
	\caption{The percentage of user overlap for a network of suspicious websites on the left and for a random group of websites on the right reported by \cite{stitelman2013using}.}
	\label{fig-ad-fraud-clusters-comparison}
\end{figure}

The authors claim the clustered websites are usually malicious because their users do not show the same behaviour as in random samples. In other words, users typically do not share the tastes in choosing websites, unless the websites are extremely popular. This difference is shown in Figure \ref{fig-ad-fraud-clusters-comparison}.
 
\subsection{Feature engineering}
 
Ad fraud detection has a highly targeted goal which is very different from other topics in online advertising. Therefore, it is important to develop a unique and comprehensive feature engineering workflow to capture the characteristics of traffic. In \citep{oentaryo2014detecting}, the authors reported the results of Fraud Detection in Mobile Advertising (FDMA) 2012 Competition. The competition was the first of its kind which used datasets from real-world advertising companies. The challenge of the competition was considered a classification problem with a publisher dataset and a click dataset. Participants were asked to determine the \emph{status} of a publisher among \emph{OK}, \emph{Observation}, and \emph{Fraud}. The organisers of the competition used \emph{Average Precision} metrics \citep{zhu2004recall} to evaluate models, which favours algorithms capable of ranking the few useful items ahead of the rest.
 
The summary from competition teams reveals a lot of insights of important features in this task, for example,
 
\begin{itemize}
	\item Total and average number of clicks and standard deviation of different time intervals;
	
	\item Total, distinct, and average number of Referer URLs and standard deviation;
	
	\item Total, distinct, and average number of Device User Agents and standard deviation;
	
	\item Total, distinct, and average number of IPs and standard deviation;
	
	\item Country, city, or finer grain geo-location of users;	
\end{itemize}
 
Second order features could be created by combination. Temporal features could be added, too. For example, total number of clicks with \texttt{Browser=Chrome} and \texttt{day-part=Morning}.
 
\begin{figure}[t]
	\centering
	\includegraphics[width=\columnwidth]{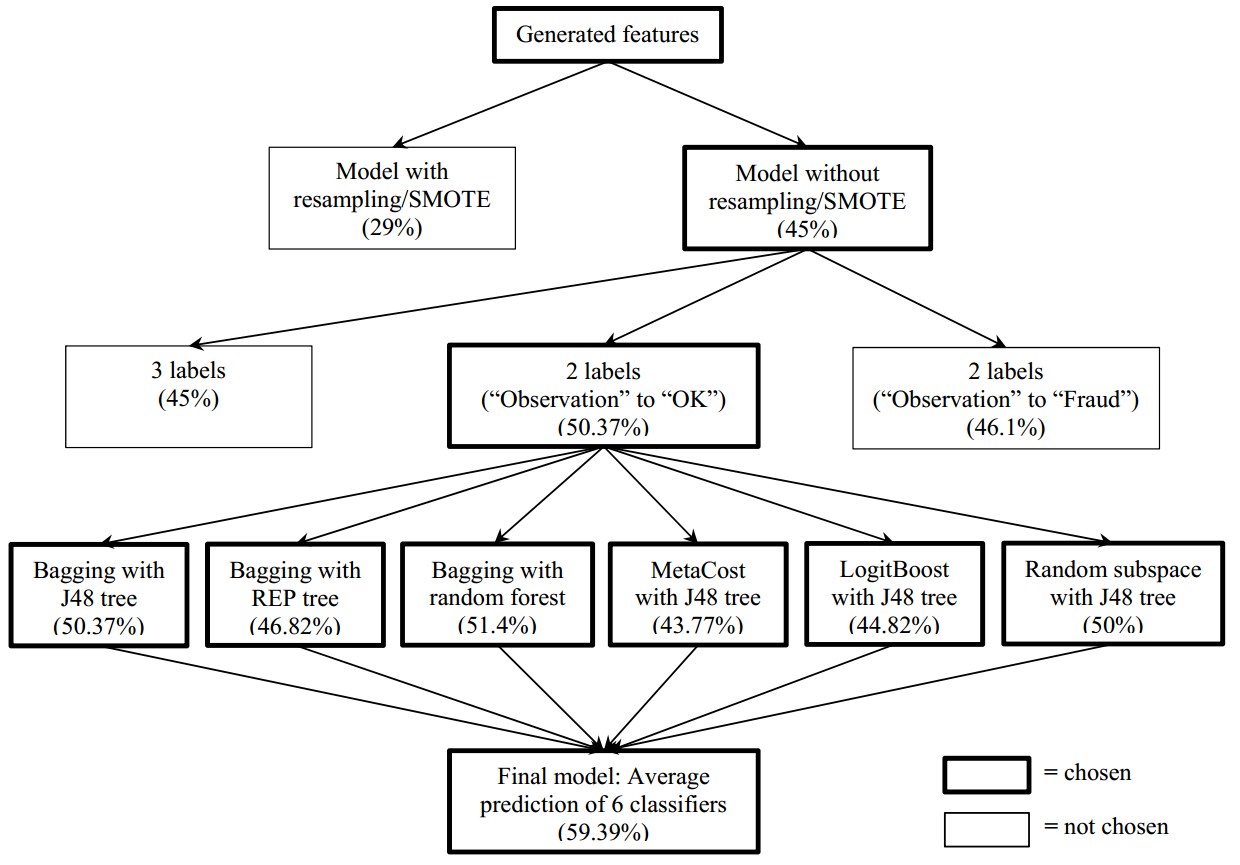}
	\caption{An ensemble model for Ad fraud detection in FDMA 2012 competition \citep{oentaryo2014detecting}.}
	\label{fig-ad-fraud-ensembles}
\end{figure}

With these features, the competition participants have built classification models. Most of them chose to use ensemble models. For example, the second winning team have reported the following structure as shown in Figure~\ref{fig-ad-fraud-ensembles}.
 
\begin{figure}[t]
	\centering
	\includegraphics[width=0.9\columnwidth]{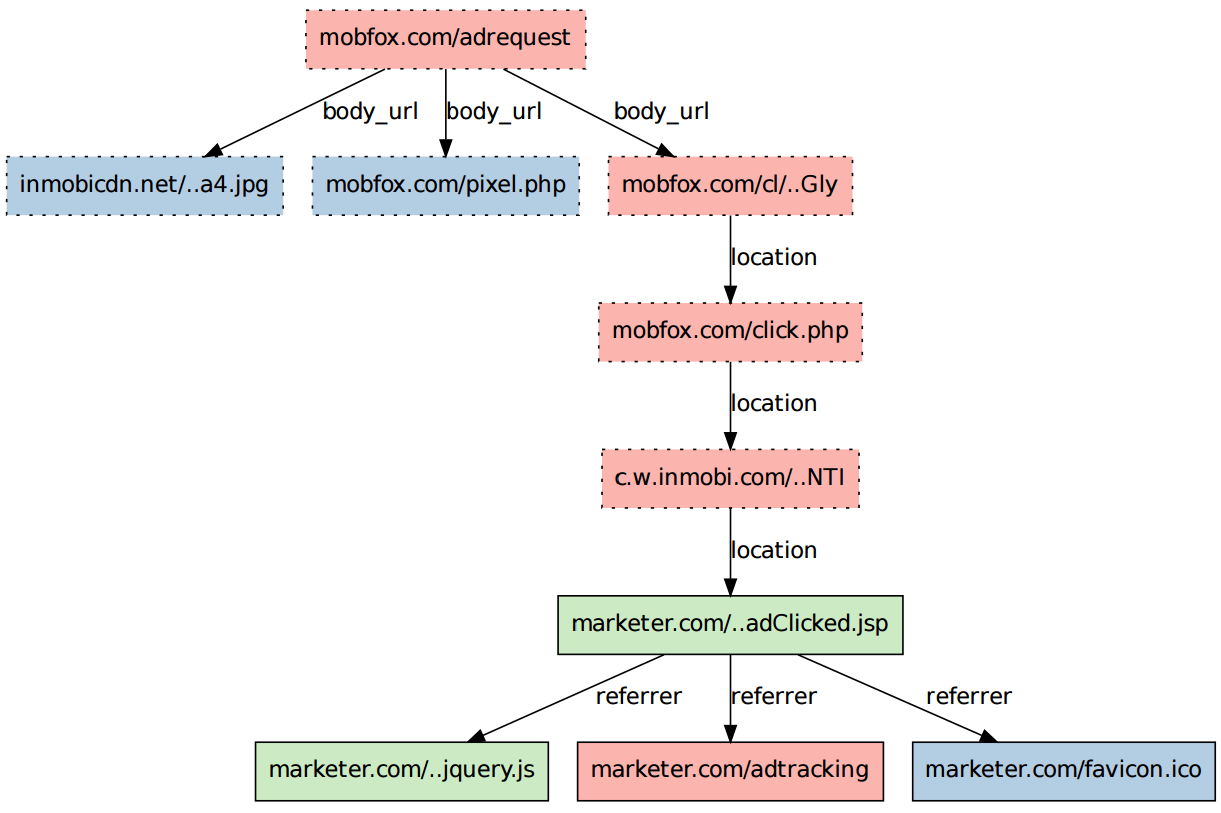}
	\caption{Example ad request tree with click illustrated by \cite{crussell2014madfraud}. Nodes in blue are images and nodes in green are static web content. Nodes with a dotted outline are for requests with a known ad provider hostname.}
	\label{fig-ad-fraud-ad-request-tree}
\end{figure}

Given different context, the detector could have access to more features. For example, in \citep{crussell2014madfraud} the authors report the feature importance for ad fraud detection in Android apps. Nine out of the top ten most important features were derived from the query parameter of URLs, such as the number of enumeration parameters, the number of low and high entropy parameters, and the total number of parameters. Also, the authors were able to construct the ad request tree, as shown in Figure \ref{fig-ad-fraud-ad-request-tree}. Such trees proved valuable in providing the complete query parameters, depth and width as additional features.
 
\section{Viewability methods}
 
In order to reduce advertisers' unnecessary cost on the traffic from trivially design robots and non-intentional traffic from true users, viewability methods are designed to add into the ad impression counting mechanism.
 
\cite{zhang2015empirical} investigated users' short-term memory after viewing the ad creatives (the text, picture, or video that the advertiser wants to show to users) in different settings of displayed pixel percentage and exposure time. The goal of the study was to find how the displayed pixel percentage and the exposure time influence the users' ad recall, and which impression viewability measurement best matches the users' remembered ad.
 
\begin{figure}
\centering
\includegraphics[width=0.8\columnwidth]{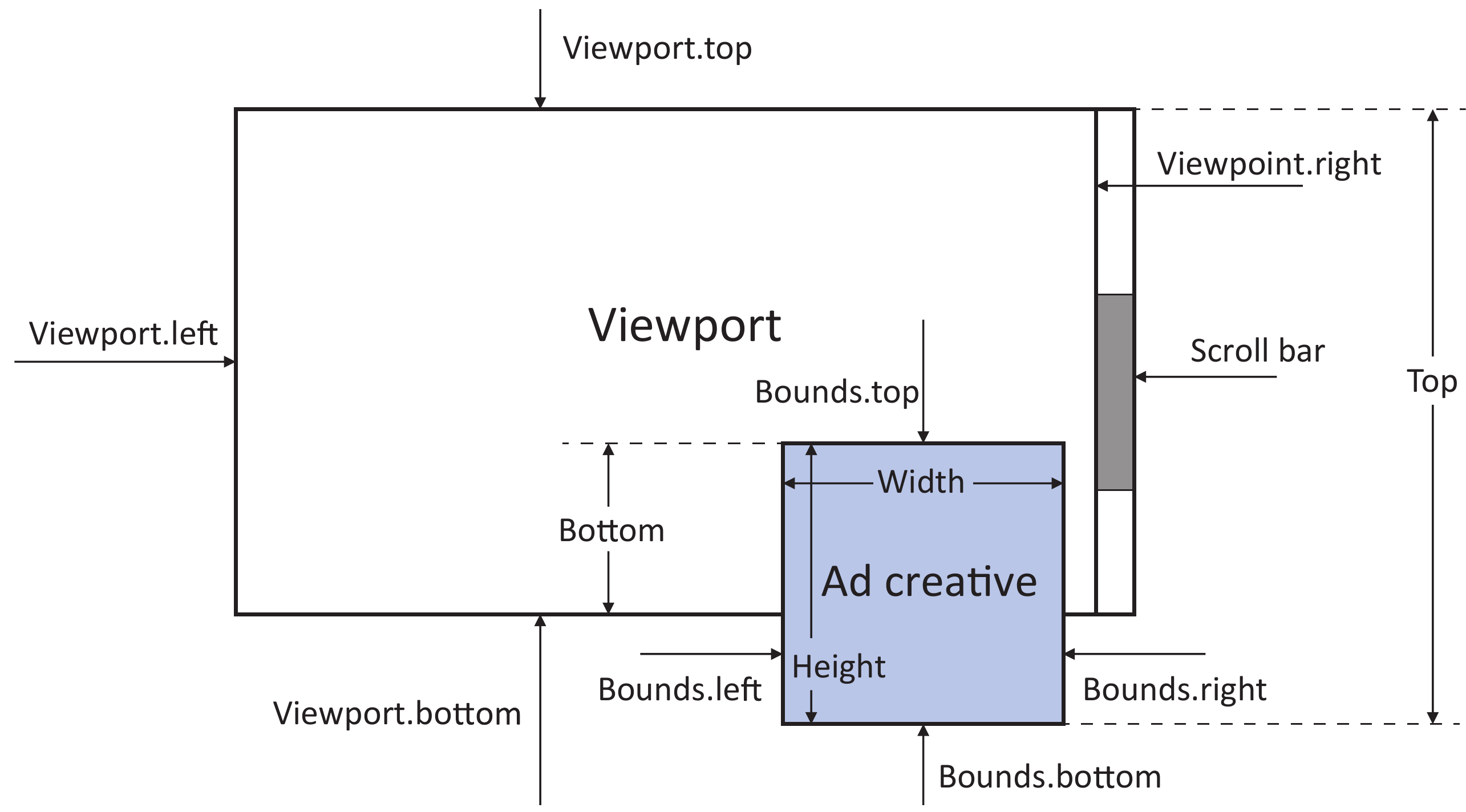}
\caption{Advanced pixel percentage tracking diagram from \cite{zhang2015empirical}. }\label{fig:adImp}
\end{figure}
 
 
The displayed pixel percentage for rectangle ad creative in the viewport can be calculated by the displayed height percentage times the displayed width percentage. Therefore, the bounds of browser's viewport and each ad creative were tracked by \cite{zhang2015empirical}.
 
Figure \ref{fig:adImp} shows the relationship of the variables. In webpage coordinates, the upper left point is the origin point. The lower place means the higher y-axis value and the right place means the higher x-axis value. Specifically, the four ratios are calculated as follows.

{\small
\begin{align*}
\texttt{Top}& = \texttt{min(1, (bounds.bottom - viewport.top) / height)}\\
\texttt{Bottom}& = \texttt{min(1, (viewport.bottom - bounds.top) / height)}\\
\texttt{Left}& = \texttt{min(1, (bounds.right - viewport.left) / width)}\\
\texttt{Right}& = \texttt{min(1, (viewport.right - bounds.left) / width)}\\
\texttt{Pixel\%} & = \texttt{Top} \times \texttt{Bottom} \times \texttt{Left} \times \texttt{Right}
\end{align*}
}
 
In Figure~\ref{fig:adImp}, \texttt{Top}=\texttt{Left}=\texttt{Right}=1, \texttt{Bottom}=0.6, thus the pixel percentage is 60\%. Given an impression measurement with the pixel percentage threshold 50\%, the measurement will count this ad impression. Note that when any of the four factors is negative value, the entire ad creative is outside of the viewport, thus the pixel percentage is calculated as zero.

 
The exposure time is associated with a pixel percentage threshold. For example, if the pixel percentage is 50\%, only after half pixels have been shown in the viewport does the tracking system start to count the exposure time. If one does not want any pixel percentage threshold, just set it as 0\%. If the measured exposure time has surpassed the predefined threshold, e.g., 2 seconds, then the measurement counts this ad impression.
 
Specifically, \cite{zhang2015empirical} used the tick counts based methods to calculate the exposure time.
For example, for the measurement of 50\% pixel percentage and 2 seconds exposure time, the tick counter will start to track the time once the pixel percentage meets 50\%. Then tick counter calls the pixel percentage tracking algorithm every 0.1 second for 20 times.
Every time the pixel percentage tracking algorithm checks whether the current pixel percentage is no less than 50\%. If it returns false, the tick counter will restart the counting. If the tick counter counts up to 20, the exposure time and pixel percentage thresholds are both reached, thus the measurement counts this ad impression.
 
The user study on 20 participants were conducted in the experiment of \citep{zhang2015empirical}. The empirically optimal threshold of display percentage is 75\% and that of exposure time is 2 seconds.
 
Besides the research study, there exists similar industrial criteria to filter out the useless ad traffic. In 2013, Google announced that the advertisers were charged only for the viewed ad impressions, where an ad was considered as viewed only if the pixel percentage was no less than 50\% and the exposure time was no less than 1 second \citep{google2013view}.
 
\section{Other methods}
 
There are a few other methods to fight ad fraud \citep{stone2011understanding}. \emph{Bluff} ads, or \emph{honeypot} ads are the ones sent by ad networks or exchanges to publishers. The ads contain irrelevant display information (either texts or images)  and act as a \emph{litmus} test for the legitimacy of the individual clicking on the ads \citep{haddadi2010fighting}. Fraudsters could be identified if they register a high CTR on bluff ads.
 
One can also check website popularity and ranking: the number of impressions a publisher is generating for their webpage can be checked against known, trusted website rankings such as Alexa or Compete. If the publisher has much more traffic than their page ranking would suggest, this would be indicative of fraudulent activity \citep{naor1998secure}.

\chapter{The Future of RTB}
\label{c-future}
 
In RTB based display advertising, there are fruitful research opportunities that can be extended from the present study. 
In a broader scope, RTB display advertising has become a significant battlefield
for big data. As the advertising transactions are aggregated
across websites in real time, the RTB display advertising
industry has a unique opportunity to understand the internet
traffic, user behaviours, and online transactions. 
As reported, a middle range Demand Side Platform (DSP) currently would process more than 30 billion
ad impressions daily, while the New York Stock Exchange trades around 12
billion shares daily. It is fair to say that
the transaction volume from display advertising has already
surpassed that of the financial market.
Yet, both markets bear a certain similarity. While the financial market uses a double
auction to create bid and ask quotes, RTB (Real-time
Bidding) display advertising adopts the second-price auction
to gather bid quotes from advertisers once an impression
is being generated. As more targeted algorithms are being proposed, advertising
optimisation becomes more resembling to that of
the financial market trading and tends to be driven by the
marketing profit and Return-On-Investment (ROI). That is,
there is an explicit and measurable campaign goal of acquiring
new users and obtaining sales from the acquired users. Thus, one of the next challenges is to properly bid for an ad impression to drive the profit and ROI. This challenge becomes essential
to performance-driven campaigns.

For a DSP, better models are to be explored in order to improve user response estimation \citep{zhang2016deep,qu2016product,shioji2017neural}. In particular, deep learning methods have started to attract more attention since 2016 and have already demonstrated their potential in performance. For bid optimisation, a more natural treatment would be dynamic decision making instead of a static one. We would expect deep reinforcement learning techniques \citep{mnih2015human} to be explored for modelling the bidding decision process. For instance, the bidding for a given ad campaign would repeatedly
happen during its lifespan before the budget runs out. As
such, each bid is strategically correlated by the constrained
budget and the overall effectiveness of the campaign (e.g.,
the rewards from generated clicks), which is only observed
after the campaign has completed. Thus, an optimal bidding strategy should sequentially and dynamically allocate the budget across all the available impressions on the basis of both the immediate and future rewards \citep{cai2017real}.

As the industry becomes more mature, the RTB infrastructure would be able to take more means for trading ad placements. For instance, \emph{header bidding} has recently emerged as a new way of conducting direct auctions, by getting rid of the inefficiencies of a "waterfall" sequence of selling their ad inventories. A futures exchange could be established in order to reduce the uncertainty and risk of real-time inventory buying \citep{chen2014lattice,wang2012selling,Chen:2014:DPM:2648584.2648585}.

 
 
To sum up, RTB display advertising is an important and challenging playground for the most advanced interdisciplinary research and practice of information retrieval, data science, machine learning and economics. We have presented a diverse set of current research topics and their solutions in this monograph. We expect further deep and advanced research and industrial innovation on RTB to emerge, both algorithmically and systematically.

\appendix
\chapter{RTB Glossary}
\label{c-rtb-glossary}
A summary of the terminology used in the monograph, which is also presented in \citep{zhang2016uclthesis}.

\begin{description}
\item[Ad Exchange] A marketplace which connects the media sellers (publishers) and buyers (advertisers) via network message parsing with a predefined protocol, and select the buyers for each sold media inventory (ad impression) by auctions.
\item[Ad Inventory] A notion of the advertising volume regarded as the virtual assets owned by the publisher. The unit of ad inventory is an ad display opportunity, i.e., an ad impression.
\item[Ad Slot] A region of the page to place the ad creative.
\item[AWR, Auction Winning Ratio] From the micro perspective, AWR means the probability of winning a specific ad auction with a specific bid value; from the macro perspective, AWR means the impression number divided by the participated auction number from a certain volume during a certain period.
\item[Bid, Bid Value, Bid Price] The amount of the money the advertiser wants to pay for the ad display opportunity being auctioned.
\item[Bid Optimisation] The designing of the bidding function such that the consequent advertising performance, measured by some KPIs, is optimised as mush as possible.
\item[Bidding Agent] A functional module of performing bid calculation for each received bid request and a qualified ad in DSP.
\item[Bidding Function] The function abstracted from the bidding strategy inputs a bid request and possibly some environment information and outputs the bid price.
\item[Bidding Strategy] The bidding logic which inputs a bid request and possibly some environment information and outputs the bid price.
\item[Budget] The total amount of money available for advertising cost during a campaign lifetime.
\item[Campaign] A series of ads sharing the same advertising target and making up an integrated marketing communication.
\item[Channel] A particular way of deliver the ads. For example, sponsored search on Google AdWords, feed ads on Facebook and RTB display ads via Google AdX.
\item[Click] A click on the ad creative from a page, which directs the user to the landing page of the ad.
\item[Conversion] An event showing a user has become a customer of the advertiser. The conversion event can be defined by various of actions, such as a successful page landing, a registration on the advertiser's website, an email subscription, making a deposit, a product purchase etc.
\item[CPA, Cost per Action or Cost per Acquisition] A predefined amount of money the advertiser pays the ad agent (DSP in RTB display advertiser, search engine in sponsored search) when a specified action has been observed on the delivered ad impression. The action can be defined by various of actions, such as a successful page landing, a registration on the advertiser's website, an email subscription, making a deposit, a product purchase etc.
\item[CPC, Cost per Click] A predefined amount of money the advertiser pays the ad agent (DSP in RTB display advertiser, search engine in sponsored search) when a user click has been observed on the delivered ad impression.
\item[CPM, Cost per Mille] A predefined amount of money the advertiser pays the ad agent (DSP in RTB display advertiser, search engine in sponsored search) for each delivered ad impression, often counted by one thousand of the same cases of ad impressions.
\item[Creative] The content of a specific ad, often in the format of images for display advertising and text for sponsored search. Javascript based creatives are also allowed in some ad exchanges to enable interactive creatives. The hyperlink on the creative points to the landing page that the advertiser wants the user to browse.
\item[CTR, Click-Through Rate] From the micro perspective, CTR means the probability of a specific user in a specific context clicking a specific ad; from the macro perspective, CTR means the click number divided by the impression number from a certain volume during a certain period.
\item[CVR, Conversion Rate] From the micro perspective, CVR means the probability of the user conversion is observed after showing the ad impression; from the macro perspective, CVR means the conversion number divided by the impression number from a certain volume during a certain period.
\item[DMP, Data Management Platform] The platform which collects, analyses and trades user behaviour information. DSPs are its major clients.
\item[DSP, Demand-Side Platform] The platform which serves advertisers to manage their campaigns and submits real-time bidding responses for each bid request to the ad exchange via computer algorithms.
\item[eCPA, Effective Cost per Action (or Acquisition)] The average cost for acquiring an action, also called efficient cost per action or expected cost per action in some references.
\item[eCPC, Effective Cost per Click] The average cost for acquiring a click, also called efficient cost per click or expected cost per click in some references.
\item[First-Price Auction] The auction where the winner, i.e., the participator with the highest bid value, pays her bid value.
\item[Floor Price] The lowest acceptable price that the publisher would sell the ad impression to any advertisers in the auction.
\item[Impression] An ad display in front of the user.
\item[KPI, Key Performance Indicator] A certain quantitative measurement of advertising performance, such as impression number, click number, conversion number, CPM, eCPC, eCPA, AWR, CTR etc.
\item[Market Price] A different name of winning price, defined on a specific bid request, which means the lowest bid value to win the auction of this bid request, i.e., the highest bid value from other competitors of this auction.
\item[ROI, Return on Investment] The ratio of the profit (revenue minus cost) gained from advertising over the advertising cost.
\item[RTB, Real-Time Bidding] A display ads trading mechanism where the ad inventory is traded on impression level via an instant ad auction with the bid values returned from the advertisers calculated in real time, e.g., less than 100ms.
\item[Second-Price Auction] The auction where the winner, i.e., the participant with the highest bid value, pays the second highest bid.
\item[Soft Floor Price] The soft floor price is set such that if the winner's bid is lower than the soft floor price but higher than the hard floor price, the winner pays his/her bid as the cost.
\item[SSP, Supply-Side Platform] The platform which serves publishers to manage the ad inventory of the sites. Upon each page loading, the SSP sends the ad request for each of the RTB ad slot to the ad exchange. Once the ad exchange returned the ID or code of the winning ad, SSP calls the corresponding ad server for the ad creative.
\item[Spot Market] In programmatic buying advertising, the RTB spot market means the transactions of ad inventory can be delivered in real time.
\item[Trading Desk] RTB trading desks (TDs) are automated systems in RTB ecosystem, mainly on the demand side. Typically they have a user-friendly interface, which allows for planning, configuring, running, optimizing, and reporting of display ad campaigns.
\item[User Segmentation] The subsets of users divided by users' demographical information, e.g., age, gender, location and occupation, or interest categories or tags. Normally, user segmentation is provided by DMP or ad exchange to help advertisers perform demographical or behavioural targeting. The bidding strategy can also highly leverage such information to perform effective bidding.
\item[Winning Price] A different name of market price, defined on a specific bid request, which means the lowest bid value to win the auction of this bid request.
\end{description}

\backmatter

\bibliographystyle{apalike} 
\bibliography{rtb}

\begin{thebibliography}{}

\bibitem[Abhishek et~al., 2012]{abhishek2012media}
Abhishek, V., Fader, P., and Hosanagar, K. (2012).
\newblock Media exposure through the funnel: A model of multi-stage
  attribution.
\newblock {\em Available at SSRN 2158421}.

\bibitem[Abrams, 2006]{Abrams2006}
Abrams, Z. (2006).
\newblock Revenue maximization when bidders have budgets.
\newblock In {\em Proceedings of the Seventeenth Annual ACM-SIAM Symposium on
  Discrete Algorithm}, pages 1074--1082. Society for Industrial and Applied
  Mathematics.

\bibitem[Acar et~al., 2014]{acar2014web}
Acar, G., Eubank, C., Englehardt, S., Juarez, M., Narayanan, A., and Diaz, C.
  (2014).
\newblock The web never forgets: Persistent tracking mechanisms in the wild.
\newblock In {\em Proceedings of the 2014 ACM SIGSAC Conference on Computer and
  Communications Security}, pages 674--689. ACM.

\bibitem[Agarwal et~al., 2014]{agarwal2014budget}
Agarwal, D., Ghosh, S., Wei, K., and You, S. (2014).
\newblock Budget pacing for targeted online advertisements at linkedin.
\newblock In {\em Proceedings of the 20th ACM SIGKDD International Conference
  on Knowledge Discovery and Data Mining}, pages 1613--1619. ACM.

\bibitem[Agrawal et~al., 2009]{agrawal2009diversifying}
Agrawal, R., Gollapudi, S., Halverson, A., and Ieong, S. (2009).
\newblock Diversifying search results.
\newblock In {\em Proceedings of the Second ACM International Conference on Web
  Search and Data Mining}, pages 5--14. ACM.

\bibitem[Ahmed et~al., 2014]{ahmed2014scalable}
Ahmed, A., Das, A., and Smola, A.~J. (2014).
\newblock Scalable hierarchical multitask learning algorithms for conversion
  optimization in display advertising.
\newblock In {\em Proceedings of the Seventh ACM International Conference on
  Web Search and Data Mining}, pages 153--162. ACM.

\bibitem[Ahmed et~al., 2011]{ahmed2011scalable}
Ahmed, A., Low, Y., Aly, M., Josifovski, V., and Smola, A.~J. (2011).
\newblock Scalable distributed inference of dynamic user interests for
  behavioral targeting.
\newblock In {\em Proceedings of the 17th ACM SIGKDD International Conference
  on Knowledge Discovery and Data Mining}, pages 114--122. ACM.

\bibitem[Alrwais et~al., 2012]{Alrwais:2012:DGC:2420950.2420954}
Alrwais, S.~A., Gerber, A., Dunn, C.~W., Spatscheck, O., Gupta, M., and
  Osterweil, E. (2012).
\newblock Dissecting ghost clicks: Ad fraud via misdirected human clicks.
\newblock In {\em Proceedings of the 28th Annual Computer Security Applications
  Conference}, ACSAC '12, pages 21--30, New York, NY, USA. ACM.

\bibitem[Amin et~al., 2012]{amin2012budget}
Amin, K., Kearns, M., Key, P.~B., and Schwaighofer, A. (2012).
\newblock Budget optimization for sponsored search: Censored learning in
  {MDP}s.
\newblock In {\em Proceedings of the 28th Conference on Uncertainty in
  Artificial Intelligence}.

\bibitem[Anagnostopoulos et~al.,
  2007]{Anagnostopoulos:2007:JCA:1321440.1321488}
Anagnostopoulos, A., Broder, A.~Z., Gabrilovich, E., Josifovski, V., and
  Riedel, L. (2007).
\newblock Just-in-time contextual advertising.
\newblock In {\em Proceedings of the Sixteenth ACM Conference on Conference on
  Information and Knowledge Management}, pages 331--340. ACM.

\bibitem[Anderl et~al., 2014]{anderl2014mapping}
Anderl, E., Becker, I., Wangenheim, F.~V., and Schumann, J.~H. (2014).
\newblock Mapping the customer journey: A graph-based framework for online
  attribution modeling.
\newblock {\em Available at SSRN 2343077}.

\bibitem[Animesh et~al., 2005]{Animesh2005}
Animesh, A., Ramachandran, V., and Viswanathan, S. (2005).
\newblock Online advertisers bidding strategies for search, experience, and
  credence goods: An empirical investigation.
\newblock In {\em Second Workshop on Sponsored Search Auctions}.

\bibitem[Auer et~al., 2002]{auer2002finite}
Auer, P., Cesa-Bianchi, N., and Fischer, P. (2002).
\newblock Finite-time analysis of the multiarmed bandit problem.
\newblock {\em Machine learning}, 47(2-3):235--256.

\bibitem[Baeza-Yates et~al., 1999]{baeza1999modern}
Baeza-Yates, R., Ribeiro-Neto, B., et~al. (1999).
\newblock {\em Modern information retrieval}, volume 463.
\newblock ACM press New York.

\bibitem[Bahdanau et~al., 2014]{bahdanau2014neural}
Bahdanau, D., Cho, K., and Bengio, Y. (2014).
\newblock Neural machine translation by jointly learning to align and
  translate.
\newblock {\em arXiv preprint arXiv:1409.0473}.

\bibitem[Balseiro et~al., 2015]{balseiro2015repeated}
Balseiro, S.~R., Besbes, O., and Weintraub, G.~Y. (2015).
\newblock Repeated auctions with budgets in ad exchanges: Approximations and
  design.
\newblock {\em Management Science}, 61(4):864--884.

\bibitem[Balseiro and Candogan, 2015]{balseiro2015optimal}
Balseiro, S.~R. and Candogan, O. (2015).
\newblock Optimal contracts for intermediaries in online advertising.
\newblock {\em Available at SSRN 2546609}.

\bibitem[Barajas et~al., 2015]{barajas2015estimating}
Barajas, J., Akella, R., Flores, A., and Holtan, M. (2015).
\newblock Estimating ad impact on clicker conversions for causal attribution: A
  potential outcomes approach.
\newblock In {\em Proceedings of the 2015 SIAM International Conference on Data
  Mining}, pages 640--648. SIAM.

\bibitem[Barajas~Zamora, 2015]{barajas2015online}
Barajas~Zamora, J. (2015).
\newblock {\em Online display advertising causal attribution and evaluation}.
\newblock PhD thesis, UC Santa Cruz: Electrical Engineering.

\bibitem[Barth, 2011]{barth2011web}
Barth, A. (2011).
\newblock The web origin concept.
\newblock Technical report, Internet Engineering Task Force.

\bibitem[BBC, 2013]{google2013view}
BBC (2013).
\newblock Google to charge advertisers viewed for seen ads.
\newblock \url{http://www.bbc.com/news/business-25356956}.
\newblock Accessed: 2016-07.

\bibitem[Bengio et~al., 2007]{bengio2007greedy}
Bengio, Y., Lamblin, P., Popovici, D., Larochelle, H., et~al. (2007).
\newblock Greedy layer-wise training of deep networks.
\newblock {\em Advances in Neural Information Processing Systems}, 19:153.

\bibitem[Berger, 1985]{berger2013statistical}
Berger, J.~O. (1985).
\newblock {\em Statistical decision theory and Bayesian analysis}.
\newblock Springer Science \& Business Media.

\bibitem[Black and Scholes, 1973]{Black_1973}
Black, F. and Scholes, M. (1973).
\newblock The pricing of options and corporate liabilities.
\newblock {\em Journal of Political Economy}, 81(3):637--654.

\bibitem[Boda et~al., 2011]{boda2011user}
Boda, K., F{\"o}ldes, {\'A}.~M., Guly{\'a}s, G.~G., and Imre, S. (2011).
\newblock User tracking on the web via cross-browser fingerprinting.
\newblock In {\em Nordic Conference on Secure IT Systems}, pages 31--46.
  Springer.

\bibitem[Borgs et~al., 2007]{borgs2007dynamics}
Borgs, C., Chayes, J., Immorlica, N., Jain, K., Etesami, O., and Mahdian, M.
  (2007).
\newblock Dynamics of bid optimization in online advertisement auctions.
\newblock In {\em Proceedings of the 16th International Conference on World
  Wide Web}, pages 531--540. ACM.

\bibitem[Breiman, 1996]{breiman1996bagging}
Breiman, L. (1996).
\newblock Bagging predictors.
\newblock {\em Machine learning}, 24(2):123--140.

\bibitem[Breiman, 2001]{breiman2001random}
Breiman, L. (2001).
\newblock Random forests.
\newblock {\em Machine learning}, 45(1):5--32.

\bibitem[Breiman et~al., 1984]{breiman1984classification}
Breiman, L., Friedman, J.~H., Stone, C.~J., and Olshen, R.~A. (1984).
\newblock {\em Classification and regression trees}.
\newblock CRC press.

\bibitem[Broder et~al., 2007]{Broder:2007:SAC:1277741.1277837}
Broder, A., Fontoura, M., Josifovski, V., and Riedel, L. (2007).
\newblock A semantic approach to contextual advertising.
\newblock In {\em Proceedings of the 30th Annual International ACM SIGIR
  Conference on Research and Development in Information Retrieval}, pages
  559--566. ACM.

\bibitem[Broder et~al., 2011]{broder2011bid}
Broder, A., Gabrilovich, E., Josifovski, V., Mavromatis, G., and Smola, A.
  (2011).
\newblock Bid generation for advanced match in sponsored search.
\newblock In {\em Proceedings of the Fourth ACM International Conference on Web
  Search and Data Mining}, pages 515--524. ACM.

\bibitem[Cai et~al., 2017]{cai2017real}
Cai, H., Ren, K., Zhang, W., Malialis, K., and Wang, J. (2017).
\newblock Real-time bidding by reinforcement learning in display advertising.
\newblock In {\em The Tenth ACM International Conference on Web Search and Data
  Mining}. ACM.

\bibitem[Cesa-Bianchi et~al., 2013]{cesa2013regret}
Cesa-Bianchi, N., Gentile, C., and Mansour, Y. (2013).
\newblock Regret minimization for reserve prices in second-price auctions.
\newblock In {\em Proceedings of the Twenty-Fourth Annual ACM-SIAM Symposium on
  Discrete Algorithms}, pages 1190--1204. Society for Industrial and Applied
  Mathematics.

\bibitem[Chakraborty et~al., 2010]{chakraborty2010selective}
Chakraborty, T., Even-Dar, E., Guha, S., Mansour, Y., and Muthukrishnan, S.
  (2010).
\newblock Selective call out and real time bidding.
\newblock In {\em International Workshop on Internet and Network Economics},
  pages 145--157. Springer.

\bibitem[Chapelle, 2014]{chapelle2014modeling}
Chapelle, O. (2014).
\newblock Modeling delayed feedback in display advertising.
\newblock In {\em Proceedings of the 20th ACM SIGKDD International Conference
  on Knowledge Discovery and Data Mining}, pages 1097--1105. ACM.

\bibitem[Chapelle, 2015]{chapelle2015offline}
Chapelle, O. (2015).
\newblock Offline evaluation of response prediction in online advertising
  auctions.
\newblock In {\em Proceedings of the 24th International Conference on World
  Wide Web}, pages 919--922. ACM.

\bibitem[Chapelle et~al., 2014]{chapelle2014simple}
Chapelle, O., Manavoglu, E., and Rosales, R. (2014).
\newblock Simple and scalable response prediction for display advertising.
\newblock {\em ACM Transactions on Intelligent Systems and Technology (TIST)},
  5(4):61.

\bibitem[Chen and Wang, 2015]{chen2014lattice}
Chen, B. and Wang, J. (2015).
\newblock A lattice framework for pricing display advertisement options with
  the stochastic volatility underlying model.
\newblock {\em Electronic Commerce Research and Applications}, 14(6):465--479.

\bibitem[Chen et~al., 2015]{chen2013multi}
Chen, B., Wang, J., Cox, I.~J., and Kankanhalli, M.~S. (2015).
\newblock Multi-keyword multi-click advertisement option contracts for
  sponsored search.
\newblock {\em ACM Transactions on Intelligent Systems and Technology (TIST)},
  7(1):5.

\bibitem[Chen et~al., 2014]{Chen:2014:DPM:2648584.2648585}
Chen, B., Yuan, S., and Wang, J. (2014).
\newblock A dynamic pricing model for unifying programmatic guarantee and
  real-time bidding in display advertising.
\newblock In {\em Proceedings of the Eighth International Workshop on Data
  Mining for Online Advertising}, pages 1:1--1:9. ACM.

\bibitem[Chen et~al., 2011a]{chen2011feature}
Chen, T., Zheng, Z., Lu, Q., Zhang, W., and Yu, Y. (2011a).
\newblock Feature-based matrix factorization.
\newblock {\em arXiv preprint arXiv:1109.2271}.

\bibitem[Chen et~al., 2011b]{Chen2011c}
Chen, Y., Berkhin, P., Anderson, B., and Devanur, N.~R. (2011b).
\newblock Real-time bidding algorithms for performance-based display ad
  allocation.
\newblock In {\em Proceedings of the 17th ACM SIGKDD International Conference
  on Knowledge Discovery and Data Mining}, pages 1307--1315. ACM.

\bibitem[Clarke et~al., 2008]{clarke2008novelty}
Clarke, C. L.~A., Kolla, M., Cormack, G.~V., Vechtomova, O., Ashkan, A.,
  B{\"u}ttcher, S., and MacKinnon, I. (2008).
\newblock Novelty and diversity in information retrieval evaluation.
\newblock In {\em Proceedings of the 31st Annual International ACM SIGIR
  Conference on Research and Development in Information Retrieval}, pages
  659--666. ACM.

\bibitem[Collobert and Weston, 2008]{collobert2008unified}
Collobert, R. and Weston, J. (2008).
\newblock A unified architecture for natural language processing: Deep neural
  networks with multitask learning.
\newblock In {\em Proceedings of the 25th International Conference on Machine
  Learning}, pages 160--167. ACM.

\bibitem[Crussell et~al., 2014]{crussell2014madfraud}
Crussell, J., Stevens, R., and Chen, H. (2014).
\newblock Madfraud: Investigating ad fraud in android applications.
\newblock In {\em Proceedings of the 12th Annual International Conference on
  Mobile Systems, Applications, and Services}, pages 123--134. ACM.

\bibitem[Cui et~al., 2011]{cui2011bid}
Cui, Y., Zhang, R., Li, W., and Mao, J. (2011).
\newblock Bid landscape forecasting in online ad exchange marketplace.
\newblock In {\em Proceedings of the 17th ACM SIGKDD International Conference
  on Knowledge Discovery and Data Mining}, pages 265--273. ACM.

\bibitem[Dabrowska, 1987]{dabrowska1987non}
Dabrowska, D.~M. (1987).
\newblock Non-parametric regression with censored survival time data.
\newblock {\em Scandinavian Journal of Statistics}, pages 181--197.

\bibitem[Dai et~al., 2007]{dai2007transferring}
Dai, W., Xue, G.-R., Yang, Q., and Yu, Y. (2007).
\newblock Transferring naive bayes classifiers for text classification.
\newblock In {\em AAAI}, volume~7, pages 540--545.

\bibitem[Dalessandro et~al., 2014]{dalessandro2014scalable}
Dalessandro, B., Chen, D., Raeder, T., Perlich, C., Han~Williams, M., and
  Provost, F. (2014).
\newblock Scalable hands-free transfer learning for online advertising.
\newblock In {\em Proceedings of the 20th ACM SIGKDD International Conference
  on Knowledge Discovery and Data Mining}, pages 1573--1582. ACM.

\bibitem[Dalessandro et~al., 2015]{dalessandro2012evaluating}
Dalessandro, B., Hook, R., Perlich, C., and Provost, F. (2015).
\newblock Evaluating and optimizing online advertising: Forget the click, but
  there are good proxies.
\newblock {\em Big Data}, 3(2):90--102.

\bibitem[Dalessandro et~al., 2012]{dalessandro2012causally}
Dalessandro, B., Perlich, C., Stitelman, O., and Provost, F. (2012).
\newblock Causally motivated attribution for online advertising.
\newblock In {\em Proceedings of the Sixth International Workshop on Data
  Mining for Online Advertising and Internet Economy}, page~7. ACM.

\bibitem[Daswani et~al., 2008]{daswani2008online}
Daswani, N., Mysen, C., Rao, V., Weis, S., Gharachorloo, K., and Ghosemajumder,
  S. (2008).
\newblock Online advertising fraud.
\newblock {\em Crimeware: Understanding New Attacks and Defenses}, 40(2):1--28.

\bibitem[Deng et~al., 2013]{deng2013deep}
Deng, L., He, X., and Gao, J. (2013).
\newblock Deep stacking networks for information retrieval.
\newblock In {\em 2013 IEEE International Conference on Acoustics, Speech and
  Signal Processing}, pages 3153--3157. IEEE.

\bibitem[Devanur and Hayes, 2009]{Devanur2009}
Devanur, N.~R. and Hayes, T.~P. (2009).
\newblock The adwords problem: online keyword matching with budgeted bidders
  under random permutations.
\newblock In {\em Proceedings of the 10th ACM conference on Electronic
  commerce}, pages 71--78. ACM.

\bibitem[Dreller, 2010]{Dreller2010Brief}
Dreller, J. (2010).
\newblock A brief history of paid search advertising.
\newblock \url{https://goo.gl/Rls5GQ} (Accessed: 13-Jun-2017).

\bibitem[Durbin, 2010]{durbin2010all}
Durbin, M. (2010).
\newblock {\em All about high-frequency trading}.
\newblock McGraw Hill Professional.

\bibitem[Eckersley, 2010]{eckersley2010unique}
Eckersley, P. (2010).
\newblock How unique is your web browser?
\newblock In {\em International Symposium on Privacy Enhancing Technologies
  Symposium}, pages 1--18. Springer.

\bibitem[Edelman and Ostrovsky, 2007]{edelman2007strategic}
Edelman, B. and Ostrovsky, M. (2007).
\newblock Strategic bidder behavior in sponsored search auctions.
\newblock {\em Decision support systems}, 43(1):192--198.

\bibitem[Edelman et~al., 2005]{edelman2005internet}
Edelman, B., Ostrovsky, M., and Schwarz, M. (2005).
\newblock Internet advertising and the generalized second price auction:
  Selling billions of dollars worth of keywords.
\newblock Technical report, National Bureau of Economic Research.

\bibitem[Edelman and Schwarz, 2006]{Edelman2006}
Edelman, B. and Schwarz, M. (2006).
\newblock {Optimal auction design in a multi-unit environment: The case of
  sponsored search auctions}.
\newblock {\em Unpublished manuscript, Harvard Business School}.

\bibitem[Even~Dar et~al., 2008]{Even-Dar2008}
Even~Dar, E., Feldman, J., Mansour, Y., and Muthukrishnan, S. (2008).
\newblock Position auctions with bidder-specific minimum prices.
\newblock {\em Lecture Notes in Computer Science (including subseries Lecture
  Notes in Artificial Intelligence and Lecture Notes in Bioinformatics)}, 5385
  LNCS:577--584.

\bibitem[Even~Dar et~al., 2009]{even2009bid}
Even~Dar, E., Mirrokni, V.~S., Muthukrishnan, S., Mansour, Y., and Nadav, U.
  (2009).
\newblock Bid optimization for broad match ad auctions.
\newblock In {\em Proceedings of the 18th International Conference on World
  Wide Web}, pages 231--240. ACM.

\bibitem[Feily et~al., 2009]{feily2009survey}
Feily, M., Shahrestani, A., and Ramadass, S. (2009).
\newblock A survey of botnet and botnet detection.
\newblock In {\em 2009 Third International Conference on Emerging Security
  Information, Systems and Technologies}, pages 268--273. IEEE.

\bibitem[Feldman et~al., 2007]{feldman2007budget}
Feldman, J., Muthukrishnan, S., Pal, M., and Stein, C. (2007).
\newblock Budget optimization in search-based advertising auctions.
\newblock In {\em Proceedings of the 8th ACM conference on Electronic
  Commerce}, pages 40--49. ACM.

\bibitem[Friedman, 2001]{friedman2001greedy}
Friedman, J.~H. (2001).
\newblock Greedy function approximation: a gradient boosting machine.
\newblock {\em Annals of Statistics}, pages 1189--1232.

\bibitem[Friedman, 2002]{friedman2002stochastic}
Friedman, J.~H. (2002).
\newblock Stochastic gradient boosting.
\newblock {\em Computational Statistics \& Data Analysis}, 38(4):367--378.

\bibitem[Friedman et~al., 2001]{friedman2001elements}
Friedman, J.~H., Hastie, T., and Tibshirani, R. (2001).
\newblock {\em The elements of statistical learning}, volume~1.
\newblock Springer series in statistics Springer, Berlin.

\bibitem[Fulgoni, 2016]{fulgoni2016fraud}
Fulgoni, G.~M. (2016).
\newblock Fraud in digital advertising: A multibillion-dollar black hole.
\newblock {\em Journal of Advertising Research}, pages JAR--2016.

\bibitem[Geyik et~al., 2014]{geyik2014multi}
Geyik, S.~C., Saxena, A., and Dasdan, A. (2014).
\newblock Multi-touch attribution based budget allocation in online
  advertising.
\newblock In {\em Proceedings of 20th ACM SIGKDD Conference on Knowledge
  Discovery and Data Mining}, pages 1--9. ACM.

\bibitem[Ghosh et~al., 2009]{Ghosh2009}
Ghosh, A., Rubinstein, B.~I., Vassilvitskii, S., and Zinkevich, M. (2009).
\newblock Adaptive bidding for display advertising.
\newblock In {\em Proceedings of the 18th International Conference on World
  Wide Web}, pages 251--260. ACM.

\bibitem[Gomer et~al., 2013]{gomer2013network}
Gomer, R., Rodrigues, E.~M., Milic-Frayling, N., and Schraefel, M.~C. (2013).
\newblock Network analysis of third party tracking: User exposure to tracking
  cookies through search.
\newblock In {\em Web Intelligence (WI) and Intelligent Agent Technologies
  (IAT), 2013 IEEE/WIC/ACM International Joint Conferences on}, volume~1, pages
  549--556. IEEE.

\bibitem[Google, 2016]{google2016ad}
Google (2016).
\newblock Ad traffic quality resource center.
\newblock \url{http://www.google.com/ads/adtrafficquality/index.html}.
\newblock Accessed: 13-Jun-2017.

\bibitem[Gorla, 2016]{gorla2016bi}
Gorla, J. (2016).
\newblock {\em A bi-directional unified Model for information retrieval}.
\newblock PhD thesis, University College London.

\bibitem[Gorla et~al., 2013]{gorla2013probabilistic}
Gorla, J., Lathia, N., Robertson, S.~E., and Wang, J. (2013).
\newblock Probabilistic group recommendation via information matching.
\newblock In {\em Proceedings of the 22nd International Conference on World
  Wide Web}, pages 495--504. ACM.

\bibitem[Graepel et~al., 2010]{graepel2010web}
Graepel, T., Candela, J.~Q., Borchert, T., and Herbrich, R. (2010).
\newblock Web-scale bayesian click-through rate prediction for sponsored search
  advertising in microsoft's bing search engine.
\newblock In {\em Proceedings of the 27th International Conference on Machine
  Learning}, pages 13--20.

\bibitem[Graham, 2010]{Graham2010Brief}
Graham, R. (2010).
\newblock A brief history of digital ad buying and selling.
\newblock \url{https://goo.gl/BZbuWY} (Accessed: 13-Jun-2017).

\bibitem[Greene, 2005]{greene2005censored}
Greene, W.~H. (2005).
\newblock Censored data and truncated distributions.
\newblock {\em Available at SSRN 825845}.

\bibitem[Gummadi et~al., 2011]{gummadi2011optimal}
Gummadi, R., Key, P.~B., and Proutiere, A. (2011).
\newblock Optimal bidding strategies in dynamic auctions with budget
  constraints.
\newblock In {\em Communication, Control, and Computing (Allerton), 2011 49th
  Annual Allerton Conference on}, pages 588--588. IEEE.

\bibitem[Gummadi et~al., 2012]{gummadi2012repeated}
Gummadi, R., Key, P.~B., and Proutiere, A. (2012).
\newblock Repeated auctions under budget constraints: Optimal bidding
  strategies and equilibria.
\newblock In {\em Eighth Workshop on Ad Auctions}.

\bibitem[Haddadi, 2010]{haddadi2010fighting}
Haddadi, H. (2010).
\newblock Fighting online click-fraud using bluff ads.
\newblock {\em ACM SIGCOMM Computer Communication Review}, 40(2):21--25.

\bibitem[He et~al., 2015]{he2015delving}
He, K., Zhang, X., Ren, S., and Sun, J. (2015).
\newblock Delving deep into rectifiers: Surpassing human-level performance on
  imagenet classification.
\newblock In {\em Proceedings of the IEEE International Conference on Computer
  Vision}, pages 1026--1034.

\bibitem[He et~al., 2014]{he2014practical}
He, X., Pan, J., Jin, O., Xu, T., Liu, B., Xu, T., Shi, Y., Atallah, A.,
  Herbrich, R., Bowers, S., et~al. (2014).
\newblock Practical lessons from predicting clicks on ads at facebook.
\newblock In {\em Proceedings of the Eighth International Workshop on Data
  Mining for Online Advertising}, pages 1--9. ACM.

\bibitem[Hinton et~al., 2012]{hinton2012deep}
Hinton, G., Deng, L., Yu, D., Dahl, G.~E., Mohamed, A.-r., Jaitly, N., Senior,
  A., Vanhoucke, V., Nguyen, P., Sainath, T.~N., and Kingsbury, B. (2012).
\newblock Deep neural networks for acoustic modeling in speech recognition: The
  shared views of four research groups.
\newblock {\em IEEE Signal Processing Magazine}, 29(6):82--97.

\bibitem[Hosanagar and Cherepanov, 2008]{Hosanagar2008}
Hosanagar, K. and Cherepanov, V. (2008).
\newblock Optimal bidding in stochastic budget constrained slot auctions.
\newblock In {\em Proceedings of the 9th ACM conference on Electronic
  Commerce}, pages 20--20. ACM.

\bibitem[{Interactive Advertising Bureau}, 2015]{iab2015what}
{Interactive Advertising Bureau} (2015).
\newblock What is an untrustworthy supply chain costing the us digital
  advertising industry?

\bibitem[Jansen, 2007]{Jansen2007Sponsored}
Jansen, B.~J. (2007).
\newblock {Sponsored Search: Is Money a Motivator for Providing Relevant
  Results}.
\newblock {\em IEEE Computer}.

\bibitem[Joachims, 2002]{joachims2002optimizing}
Joachims, T. (2002).
\newblock Optimizing search engines using clickthrough data.
\newblock In {\em Proceedings of the eighth ACM SIGKDD International Conference
  on Knowledge Discovery and Data Mining}, pages 133--142. ACM.

\bibitem[Johnson, 1999]{johnson1999survival}
Johnson, N.~L. (1999).
\newblock {\em Survival models and data analysis}, volume~74.
\newblock John Wiley \& Sons.

\bibitem[Jones et~al., 2000]{jones2000probabilistic}
Jones, K.~S., Walker, S., and Robertson, S.~E. (2000).
\newblock A probabilistic model of information retrieval: development and
  comparative experiments: Part 2.
\newblock {\em Information Processing \& Management}, 36(6):809--840.

\bibitem[Juan et~al., 2016]{juanfield}
Juan, Y., Zhuang, Y., Chin, W.-S., and Lin, C.-J. (2016).
\newblock Field-aware factorization machines for {CTR} prediction.
\newblock In {\em Proceedings of the 9th ACM Conference on Recommender
  Systems}.

\bibitem[Kan et~al., 2016]{ran2016cikm}
Kan, R., Zhang, W., Rong, Y., Zhang, H., Yu, Y., and Wang, J. (2016).
\newblock User response learning for directly optimizing campaign performance
  in display advertising.
\newblock In {\em Proceedings of the ACM International Conference on
  Information \& Knowledge Management}. ACM.

\bibitem[Kantrowitz, 2015]{alex2015inside}
Kantrowitz, A. (2015).
\newblock Inside google's secret war against ad fraud.
\newblock {\em Ad Age}.

\bibitem[Kaplan and Meier, 1958]{kaplan1958nonparametric}
Kaplan, E.~L. and Meier, P. (1958).
\newblock Nonparametric estimation from incomplete observations.
\newblock {\em Journal of the American Statistical Association},
  53(282):457--481.

\bibitem[Karlsson and Zhang, 2013]{karlsson2013applications}
Karlsson, N. and Zhang, J. (2013).
\newblock Applications of feedback control in online advertising.
\newblock In {\em American Control Conference (ACC), 2013}, pages 6008--6013.
  IEEE.

\bibitem[Karp, 2008]{Karp2008Google}
Karp, S. (2008).
\newblock Google adwords: A brief history of online advertising innovation.
\newblock \url{https://goo.gl/bh1HVW} (Accessed: 13-Jun-2017).

\bibitem[Kee, 2012]{kee2012attribution}
Kee, B. (2012).
\newblock Attribution playbook - google analytics.
\newblock
  \url{http://services.google.com/fh/files/misc/attribution_playbook.pdf}.
\newblock Accessed: 13-Jun-2017.

\bibitem[Kenny and Marshall, 2001]{Kenny2011Contextual}
Kenny, D. and Marshall, J.~F. (2001).
\newblock Contextual marketing: The real business of the internet.
\newblock \url{http://hbswk.hbs.edu/archive/2124.html} (Accessed: 13-Jun-2017).

\bibitem[Kitts and Leblanc, 2004]{kitts2004optimal}
Kitts, B. and Leblanc, B. (2004).
\newblock Optimal bidding on keyword auctions.
\newblock {\em Electronic Markets}, 14(3):186--201.

\bibitem[Kleinberg and Leighton, 2003]{kleinberg2003value}
Kleinberg, R. and Leighton, T. (2003).
\newblock The value of knowing a demand curve: Bounds on regret for online
  posted-price auctions.
\newblock In {\em Proceedings of 44th Annual IEEE Symposium on Foundations of
  Computer Science.}, pages 594--605. IEEE.

\bibitem[Koren et~al., 2009]{koren2009matrix}
Koren, Y., Bell, R., Volinsky, C., et~al. (2009).
\newblock Matrix factorization techniques for recommender systems.
\newblock {\em Computer}, 42(8):30--37.

\bibitem[Lang et~al., 2012]{lang2012handling}
Lang, K.~J., Moseley, B., and Vassilvitskii, S. (2012).
\newblock Handling forecast errors while bidding for display advertising.
\newblock In {\em Proceedings of the 21st International Conference on World
  Wide Web}, pages 371--380. ACM.

\bibitem[LeCun and Bengio, 1995]{lecun1995convolutional}
LeCun, Y. and Bengio, Y. (1995).
\newblock Convolutional networks for images, speech, and time series.
\newblock {\em The Handbook of Brain Theory and Neural Networks},
  3361(10):1995.

\bibitem[LeCun et~al., 1998]{lecun1998gradient}
LeCun, Y., Bottou, L., Bengio, Y., and Haffner, P. (1998).
\newblock Gradient-based learning applied to document recognition.
\newblock {\em Proceedings of the IEEE}, 86(11):2278--2324.

\bibitem[Lee et~al., 2013]{lee2013real}
Lee, K.-C., Jalali, A., and Dasdan, A. (2013).
\newblock Real time bid optimization with smooth budget delivery in online
  advertising.
\newblock In {\em Proceedings of the Seventh International Workshop on Data
  Mining for Online Advertising}, page~1. ACM.

\bibitem[Lee et~al., 2012]{lee2012estimating}
Lee, K.-c., Orten, B., Dasdan, A., and Li, W. (2012).
\newblock Estimating conversion rate in display advertising from past
  erformance data.
\newblock In {\em Proceedings of the 18th ACM SIGKDD International Conference
  on Knowledge Discovery and Data Mining}, pages 768--776. ACM.

\bibitem[Li et~al., 2009]{li2009transfer}
Li, B., Yang, Q., and Xue, X. (2009).
\newblock Transfer learning for collaborative filtering via a rating-matrix
  generative model.
\newblock In {\em Proceedings of the 26th Annual International Conference on
  Machine Learning}, pages 617--624. ACM.

\bibitem[Li and Lu, 2016]{lideep}
Li, H. and Lu, Z. (2016).
\newblock Deep learning for information retrieval.
\newblock In {\em Proceedings of the 39th International ACM SIGIR Conference on
  Research and Development in Information Retrieval}, pages 1203--1206. ACM.

\bibitem[Liao et~al., 2014]{liao2014ipinyou}
Liao, H., Peng, L., Liu, Z., and Shen, X. (2014).
\newblock {iPinYou} global rtb bidding algorithm competition dataset.
\newblock In {\em Proceedings of the Eighth International Workshop on Data
  Mining for Online Advertising}, pages 1--6. ACM.

\bibitem[Liao et~al., 2005]{liao2005logistic}
Liao, X., Xue, Y., and Carin, L. (2005).
\newblock Logistic regression with an auxiliary data source.
\newblock In {\em Proceedings of the 22nd International Conference on Machine
  Learning}, pages 505--512. ACM.

\bibitem[Liu, 2009]{liu2009learning}
Liu, T.-Y. (2009).
\newblock Learning to rank for information retrieval.
\newblock {\em Foundations and Trends in Information Retrieval}, 3(3):225--331.

\bibitem[Loni et~al., 2014]{loni2014cross}
Loni, B., Shi, Y., Larson, M., and Hanjalic, A. (2014).
\newblock Cross-domain collaborative filtering with factorization machines.
\newblock In {\em European Conference on Information Retrieval}, pages
  656--661. Springer.

\bibitem[Mangalampalli et~al., 2011]{mangalampalli2011feature}
Mangalampalli, A., Ratnaparkhi, A., Hatch, A.~O., Bagherjeiran, A., Parekh, R.,
  and Pudi, V. (2011).
\newblock A feature-pair-based associative classification approach to
  look-alike modeling for conversion-oriented user-targeting in tail campaigns.
\newblock In {\em Proceedings of the 20th international conference companion on
  World wide web}, pages 85--86. ACM.

\bibitem[Matthews, 1995]{matthews1995technical}
Matthews, S.~A. (1995).
\newblock {\em A technical primer on auction theory I: Independent private
  values}, volume 1096.
\newblock Center for Mathematical Studies in Economics and Management Science,
  Northwestern University.

\bibitem[McAfee, 2011]{mcafee2011design}
McAfee, R.~P. (2011).
\newblock The design of advertising exchanges.
\newblock {\em Review of Industrial Organization}, 39(3):169--185.

\bibitem[McMahan et~al., 2013]{mcmahan2013ad}
McMahan, H.~B., Holt, G., Sculley, D., Young, M., Ebner, D., Grady, J., Nie,
  L., Phillips, T., Davydov, E., Golovin, D., et~al. (2013).
\newblock Ad click prediction: a view from the trenches.
\newblock In {\em Proceedings of the 19th ACM SIGKDD International Conference
  on Knowledge Discovery and Data Mining}, pages 1222--1230. ACM.

\bibitem[Mehta et~al., 2007a]{Mehta2005AdWords}
Mehta, A., Saberi, A., Vazirani, U., and Vazirani, V. (2007a).
\newblock Adwords and generalized online matching.
\newblock {\em Journal of the ACM (JACM)}, 54(5):22.

\bibitem[Mehta et~al., 2007b]{mehta2007adwords}
Mehta, A., Saberi, A., Vazirani, U., and Vazirani, V. (2007b).
\newblock Adwords and generalized online matching.
\newblock {\em Journal of the ACM (JACM)}, 54(5):22.

\bibitem[Menezes and Monteiro, 2005]{menezes2005introduction}
Menezes, F.~M. and Monteiro, P.~K. (2005).
\newblock {\em An introduction to auction theory}.
\newblock OUP Oxford.

\bibitem[Menon et~al., 2011]{menon2011response}
Menon, A.~K., Chitrapura, K.-P., Garg, S., Agarwal, D., and Kota, N. (2011).
\newblock Response prediction using collaborative filtering with hierarchies
  and side-information.
\newblock In {\em Proceedings of the 17th ACM SIGKDD International Conference
  on Knowledge Discovery and Data Mining}, pages 141--149. ACM.

\bibitem[Milgrom, 2004]{milgrom2004putting}
Milgrom, P.~R. (2004).
\newblock {\em Putting auction theory to work}.
\newblock Cambridge University Press.

\bibitem[Mnih et~al., 2015]{mnih2015human}
Mnih, V., Kavukcuoglu, K., Silver, D., Rusu, A.~A., Veness, J., Bellemare,
  M.~G., Graves, A., Riedmiller, M., Fidjeland, A.~K., Ostrovski, G., Petersen,
  S., Beattie, C., Sadik, A., Antonoglou, I., King, H., Kumaran, D., Wierstra,
  D., Legg, S., and Hassabis, D. (2015).
\newblock Human-level control through deep reinforcement learning.
\newblock {\em Nature}, 518(7540):529--533.

\bibitem[Muthukrishnan, 2009]{muthukrishnan2009ad}
Muthukrishnan, S. (2009).
\newblock Ad exchanges: Research issues.
\newblock In {\em International Workshop on Internet and Network Economics},
  pages 1--12. Springer.

\bibitem[Muthukrishnan et~al., 2010]{muthukrishnan2007stochastic}
Muthukrishnan, S., P{\'a}l, M., and Svitkina, Z. (2010).
\newblock Stochastic models for budget optimization in search-based
  advertising.
\newblock {\em Algorithmica}, 58(4):1022--1044.

\bibitem[Myerson, 1981]{Myerson1981}
Myerson, R.~B. (1981).
\newblock Optimal auction design.
\newblock {\em Mathematics of Operations Research}, 6(1):58--73.

\bibitem[Naor and Pinkas, 1998]{naor1998secure}
Naor, M. and Pinkas, B. (1998).
\newblock Secure accounting and auditing on the web.
\newblock {\em Computer Networks and ISDN Systems}, 30(1):541--550.

\bibitem[Oentaryo et~al., 2014a]{oentaryo2014detecting}
Oentaryo, R.~J., Lim, E.-P., Finegold, M., Lo, D., Zhu, F., Phua, C., Cheu,
  E.-Y., Yap, G.-E., Sim, K., Nguyen, M.~N., et~al. (2014a).
\newblock Detecting click fraud in online advertising: a data mining approach.
\newblock {\em Journal of Machine Learning Research}, 15(1):99--140.

\bibitem[Oentaryo et~al., 2014b]{oentaryo2014predicting}
Oentaryo, R.~J., Lim, E.-P., Low, J.-W., Lo, D., and Finegold, M. (2014b).
\newblock Predicting response in mobile advertising with hierarchical
  importance-aware factorization machine.
\newblock In {\em Proceedings of the 7th ACM International Conference on Web
  Search and Data Mining}, pages 123--132. ACM.

\bibitem[Osborne and Rubinstein, 1994]{osborne1994course}
Osborne, M.~J. and Rubinstein, A. (1994).
\newblock {\em A course in game theory}.
\newblock MIT press.

\bibitem[Ostrovsky and Schwarz, 2009]{Ostrovsky2009}
Ostrovsky, M. and Schwarz, M. (2009).
\newblock Reserve prices in internet advertising auctions: A field experiment.
\newblock {\em Search}, pages 1--18.

\bibitem[Pan and Yang, 2010]{pan2010survey}
Pan, S.~J. and Yang, Q. (2010).
\newblock A survey on transfer learning.
\newblock {\em IEEE Transactions on Knowledge and Data Engineering},
  22(10):1345--1359.

\bibitem[Perlich et~al., 2012]{perlich2012bid}
Perlich, C., Dalessandro, B., Hook, R., Stitelman, O., Raeder, T., and Provost,
  F. (2012).
\newblock Bid optimizing and inventory scoring in targeted online advertising.
\newblock In {\em Proceedings of the 18th ACM SIGKDD International Conference
  on Knowledge Discovery and Data Mining}, pages 804--812. ACM.

\bibitem[Ponte and Croft, 1998]{ponte1998language}
Ponte, J.~M. and Croft, W.~B. (1998).
\newblock A language modeling approach to information retrieval.
\newblock In {\em Proceedings of the 21st Annual International ACM SIGIR
  Conference on Research and Development in Information Retrieval}, pages
  275--281. ACM.

\bibitem[Qu et~al., 2016]{qu2016product}
Qu, Y., Cai, H., Ren, K., Zhang, W., Yu, Y., Wen, Y., and Wang, J. (2016).
\newblock Product-based neural networks for user response prediction.
\newblock In {\em IEEE 16th International Conference on Data Mining}.

\bibitem[Radlinski et~al., 2008]{Radlinski2008}
Radlinski, F., Broder, A., Ciccolo, P., Gabrilovich, E., Josifovski, V., and
  Riedel, L. (2008).
\newblock Optimizing relevance and revenue in ad search: a query substitution
  approach.
\newblock In {\em Proceedings of the 31st Annual International ACM SIGIR
  Conference on Research and Development in Information Retrieval}, pages
  403--410. ACM.

\bibitem[Raeder et~al., 2012]{Raeder:2012:DPM:2339530.2339740}
Raeder, T., Stitelman, O., Dalessandro, B., Perlich, C., and Provost, F.
  (2012).
\newblock Design principles of massive, robust prediction systems.
\newblock In {\em Proceedings of the 18th ACM SIGKDD International Conference
  on Knowledge Discovery and Data Mining}, pages 1357--1365. ACM.

\bibitem[Ren et~al., 2016]{ren2016user}
Ren, K., Zhang, W., Rong, Y., Zhang, H., Yu, Y., and Wang, J. (2016).
\newblock User response learning for directly optimizing campaign performance
  in display advertising.
\newblock In {\em Proceedings of the 25th ACM International on Conference on
  Information and Knowledge Management}, pages 679--688. ACM.

\bibitem[Rendle, 2010]{rendle2010factorization}
Rendle, S. (2010).
\newblock Factorization machines.
\newblock In {\em IEEE 10th International Conference on Data Mining}, pages
  995--1000. IEEE.

\bibitem[Rendle, 2012]{rendle2012factorization}
Rendle, S. (2012).
\newblock Factorization machines with libfm.
\newblock {\em ACM Transactions on Intelligent Systems and Technology (TIST)},
  3(3):57.

\bibitem[Rendle et~al., 2011]{rendle2011fast}
Rendle, S., Gantner, Z., Freudenthaler, C., and Schmidt-Thieme, L. (2011).
\newblock Fast context-aware recommendations with factorization machines.
\newblock In {\em Proceedings of the 34th International ACM SIGIR Conference on
  Research and Development in Information Retrieval}, pages 635--644. ACM.

\bibitem[Rendle and Schmidt-Thieme, 2010]{rendle2010pairwise}
Rendle, S. and Schmidt-Thieme, L. (2010).
\newblock Pairwise interaction tensor factorization for personalized tag
  recommendation.
\newblock In {\em Proceedings of the Third ACM International Conference on Web
  Search and Data Mining}, pages 81--90. ACM.

\bibitem[Robertson, 1977]{robertson1977probability}
Robertson, S.~E. (1977).
\newblock The probability ranking principle in ir.
\newblock {\em Journal of Documentation}, 33(4):294--304.

\bibitem[Robertson et~al., 1982]{robertson1982unified}
Robertson, S.~E., Maron, M., and Cooper, W.~S. (1982).
\newblock The unified probabilistic model for ir.
\newblock In {\em International Conference on Research and Development in
  Information Retrieval}, pages 108--117. Springer.

\bibitem[Samuelson, 1965]{Samuelson_1965_2}
Samuelson, P. (1965).
\newblock Rational theory of warrant pricing.
\newblock {\em Industrial Management Review}, 6:13--31.

\bibitem[Shao and Li, 2011]{Shao:2011:DMA:2020408.2020453}
Shao, X. and Li, L. (2011).
\newblock Data-driven multi-touch attribution models.
\newblock In {\em Proceedings of the 17th ACM SIGKDD International Conference
  on Knowledge Discovery and Data Mining}, pages 258--264. ACM.

\bibitem[Shapley, 1952]{shapley1952value}
Shapley, L.~S. (1952).
\newblock A value for n-person games.
\newblock Technical report, DTIC Document.

\bibitem[Shen et~al., 2015]{shen20150}
Shen, J., Orten, B., Geyik, S.~C., Liu, D., Shariat, S., Bian, F., and Dasdan,
  A. (2015).
\newblock From 0.5 million to 2.5 million: Efficiently scaling up real-time
  bidding.
\newblock In {\em IEEE International Conference on Data Mining}, pages
  973--978. IEEE.

\bibitem[Shioji and Arai, 2017]{shioji2017neural}
Shioji, E. and Arai, M. (2017).
\newblock Neural feature embedding for user response prediction in real-time
  bidding (rtb).
\newblock {\em arXiv preprint arXiv:1702.00855}.

\bibitem[Sinha et~al., 2014]{sinha2014estimating}
Sinha, R., Saini, S., and Anadhavelu, N. (2014).
\newblock Estimating the incremental effects of interactions for marketing
  attribution.
\newblock In {\em International Conference on Behavior, Economic and Social
  Computing}, pages 1--6. IEEE.

\bibitem[Smeulders et~al., 2000]{smeulders2000content}
Smeulders, A. W.~M., Worring, M., Santini, S., Gupta, A., and Jain, R. (2000).
\newblock Content-based image retrieval at the end of the early years.
\newblock {\em IEEE Transactions on Pattern Analysis and Machine Intelligence},
  22(12):1349--1380.

\bibitem[Soltani et~al., 2010]{soltani2010flash}
Soltani, A., Canty, S., Mayo, Q., Thomas, L., and Hoofnagle, C.~J. (2010).
\newblock Flash cookies and privacy.
\newblock In {\em AAAI Spring Symposium: Intelligent Information Privacy
  Management}, volume 2010, pages 158--163.

\bibitem[Springborn and Barford, 2013]{springborn2013impression}
Springborn, K. and Barford, P. (2013).
\newblock Impression fraud in on-line advertising via pay-per-view networks.
\newblock In {\em USENIX Security}, pages 211--226.

\bibitem[Stitelman et~al., 2013]{stitelman2013using}
Stitelman, O., Perlich, C., Dalessandro, B., Hook, R., Raeder, T., and Provost,
  F. (2013).
\newblock Using co-visitation networks for detecting large scale online display
  advertising exchange fraud.
\newblock In {\em Proceedings of the 19th ACM SIGKDD International Conference
  on Knowledge Discovery and Data Mining}, pages 1240--1248. ACM.

\bibitem[Stone-Gross et~al., 2011]{stone2011understanding}
Stone-Gross, B., Stevens, R., Zarras, A., Kemmerer, R., Kruegel, C., and Vigna,
  G. (2011).
\newblock Understanding fraudulent activities in online ad exchanges.
\newblock In {\em Proceedings of the 2011 ACM SIGCOMM Conference on Internet
  Measurement Conference}, pages 279--294. ACM.

\bibitem[Ta, 2015]{ta2015factorization}
Ta, A.-P. (2015).
\newblock Factorization machines with follow-the-regularized-leader for ctr
  prediction in display advertising.
\newblock In {\em 2015 IEEE International Conference on Big Data}, pages
  2889--2891. IEEE.

\bibitem[Talluri and van Ryzin, 2005]{Talluri_2004}
Talluri, K.~T. and van Ryzin, G.~J. (2005).
\newblock {\em The Theory and Practice of Revenue Management}.
\newblock Springer.

\bibitem[Taylor and Stone, 2009]{taylor2009transfer}
Taylor, M.~E. and Stone, P. (2009).
\newblock Transfer learning for reinforcement learning domains: A survey.
\newblock {\em Journal of Machine Learning Research}, 10:1633--1685.

\bibitem[{The Washington Post}, 2004]{Google2004Dispute}
{The Washington Post} (2004).
\newblock Google ends its dispute with yahoo.
\newblock
  \url{http://www.washingtonpost.com/wp-dyn/articles/A52880-2004Aug9.html}
  (Accessed: 13-Jun-2017).

\bibitem[Thomas et~al., 2015]{thomas2015ad}
Thomas, K., Bursztein, E., Grier, C., Ho, G., Jagpal, N., Kapravelos, A.,
  McCoy, D., Nappa, A., Paxson, V., Pearce, P., Provos, N., and Rajab, M.~A.
  (2015).
\newblock Ad injection at scale: Assessing deceptive advertisement
  modifications.
\newblock In {\em 2015 IEEE Symposium on Security and Privacy}, pages 151--167.
  IEEE.

\bibitem[Tyree et~al., 2011]{Tyree:2011:PBR:1963405.1963461}
Tyree, S., Weinberger, K.~Q., Agrawal, K., and Paykin, J. (2011).
\newblock Parallel boosted regression trees for web search ranking.
\newblock In {\em Proceedings of the 20th International Conference on World
  Wide Web}, pages 387--396. ACM.

\bibitem[Vratonjic et~al., 2010]{vratonjic2010isps}
Vratonjic, N., Manshaei, M.~H., Raya, M., and Hubaux, J.-P. (2010).
\newblock Isps and ad networks against botnet ad fraud.
\newblock In {\em International Conference on Decision and Game Theory for
  Security}, pages 149--167. Springer.

\bibitem[Wang and Chen, 2012]{wang2012selling}
Wang, J. and Chen, B. (2012).
\newblock Selling futures online advertising slots via option contracts.
\newblock In {\em Proceedings of the 21st International Conference on World
  Wide Web}, pages 627--628. ACM.

\bibitem[Wang et~al., 2006]{wang2006unifying}
Wang, J., De~Vries, A.~P., and Reinders, M. J.~T. (2006).
\newblock Unifying user-based and item-based collaborative filtering approaches
  by similarity fusion.
\newblock In {\em Proceedings of the 29th Annual International ACM SIGIR
  Conference on Research and Development in Information Retrieval}, pages
  501--508. ACM.

\bibitem[Wang et~al., 2008]{wang2008unified}
Wang, J., De~Vries, A.~P., and Reinders, M. J.~T. (2008).
\newblock Unified relevance models for rating prediction in collaborative
  filtering.
\newblock {\em ACM Transactions on Information Systems (TOIS)}, 26(3):16.

\bibitem[Wang and Zhu, 2009]{wang2009portfolio}
Wang, J. and Zhu, J. (2009).
\newblock Portfolio theory of information retrieval.
\newblock In {\em Proceedings of the 32nd International ACM SIGIR Conference on
  Research and Development in Information Retrieval}, pages 115--122. ACM.

\bibitem[Weinberger et~al., 2009]{weinberger2009feature}
Weinberger, K., Dasgupta, A., Langford, J., Smola, A., and Attenberg, J.
  (2009).
\newblock Feature hashing for large scale multitask learning.
\newblock In {\em Proceedings of the 26th Annual International Conference on
  Machine Learning}, pages 1113--1120. ACM.

\bibitem[Wooff and Anderson, 2015]{wooff2015time}
Wooff, D.~A. and Anderson, J.~M. (2015).
\newblock Time-weighted multi-touch attribution and channel relevance in the
  customer journey to online purchase.
\newblock {\em Journal of Statistical Theory and Practice}, 9(2):227--249.

\bibitem[Wu et~al., 2015]{wu2015predicting}
Wu, W. C.-H., Yeh, M.-Y., and Chen, M.-S. (2015).
\newblock Predicting winning price in real time bidding with censored data.
\newblock In {\em Proceedings of the 21th ACM SIGKDD International Conference
  on Knowledge Discovery and Data Mining}, pages 1305--1314. ACM.

\bibitem[Xiao et~al., 2009]{Xiao2009}
Xiao, B., Yang, W., and Li, J. (2009).
\newblock Optimal reserve price for the generalized second-price auction in
  sponsored search advertising.
\newblock {\em Journal of Electronic Commerce Research}, 10(3):114--129.

\bibitem[Xu et~al., 2015]{xu2015smart}
Xu, J., Lee, K.-c., Li, W., Qi, H., and Lu, Q. (2015).
\newblock Smart pacing for effective online ad campaign optimization.
\newblock In {\em Proceedings of the 21th ACM SIGKDD International Conference
  on Knowledge Discovery and Data Mining}, pages 2217--2226. ACM.

\bibitem[Xu et~al., 2016]{xu2016lift}
Xu, J., Shao, X., Ma, J., Lee, K.-c., Qi, H., and Lu, Q. (2016).
\newblock Lift-based bidding in ad selection.
\newblock In {\em Proceedings of the 30th AAAI Conference on Artificial
  Intelligence}.

\bibitem[Xu et~al., 2014]{xu2014path}
Xu, L., Duan, J.~A., and Whinston, A. (2014).
\newblock Path to purchase: A mutually exciting point process model for online
  advertising and conversion.
\newblock {\em Management Science}, 60(6):1392--1412.

\bibitem[Yan et~al., 2009]{yan2009much}
Yan, J., Liu, N., Wang, G., Zhang, W., Jiang, Y., and Chen, Z. (2009).
\newblock How much can behavioral targeting help online advertising?
\newblock In {\em Proceedings of the 18th International Conference on World
  Wide Web}, pages 261--270. ACM.

\bibitem[Yuan et~al., 2012]{yuan2012internet}
Yuan, S., Abidin, A.~Z., Sloan, M., and Wang, J. (2012).
\newblock Internet advertising: An interplay among advertisers, online
  publishers, ad exchanges and web users.
\newblock {\em arXiv preprint arXiv:1206.1754}.

\bibitem[Yuan and Wang, 2012]{yuan2012sequential}
Yuan, S. and Wang, J. (2012).
\newblock Sequential selection of correlated ads by pomdps.
\newblock In {\em Proceedings of the 21st ACM international conference on
  Information and knowledge management}, pages 515--524. ACM.

\bibitem[Yuan et~al., 2014]{yuan2014empirical}
Yuan, S., Wang, J., Chen, B., Mason, P., and Seljan, S. (2014).
\newblock An empirical study of reserve price optimisation in real-time
  bidding.
\newblock In {\em Proceedings of the 20th ACM SIGKDD International Conference
  on Knowledge Discovery and Data Mining}, pages 1897--1906. ACM.

\bibitem[Yuan et~al., 2013]{Yuan:2013:RBO:2501040.2501980}
Yuan, S., Wang, J., and Zhao, X. (2013).
\newblock Real-time bidding for online advertising: measurement and analysis.
\newblock In {\em Proceedings of the Seventh International Workshop on Data
  Mining for Online Advertising}, page~3. ACM.

\bibitem[Zhang et~al., 2017]{zhang2017managing}
Zhang, H., Zhang, W., Rong, Y., Ren, K., Li, W., and Wang, J. (2017).
\newblock Managing risk of bidding in display advertising.
\newblock In {\em The Tenth ACM International Conference on Web Search and Data
  Mining}. ACM.

\bibitem[Zhang, 2016]{zhang2016uclthesis}
Zhang, W. (2016).
\newblock {\em Optimal Real-Time Bidding for Display Advertising}.
\newblock PhD thesis, University College London.

\bibitem[Zhang et~al., 2016a]{zhang2016implicit}
Zhang, W., Chen, L., and Wang, J. (2016a).
\newblock Implicit look-alike modelling in display ads: Transfer collaborative
  filtering to ctr estimation.
\newblock In {\em European Conference on Information Retrieval}, pages
  589--601. Springer.

\bibitem[Zhang et~al., 2016b]{zhang2016deep}
Zhang, W., Du, T., and Wang, J. (2016b).
\newblock Deep learning over multi-field categorical data.
\newblock In {\em European Conference on Information Retrieval}, pages 45--57.
  Springer.

\bibitem[Zhang et~al., 2015]{zhang2015empirical}
Zhang, W., Pan, Y., Zhou, T., and Wang, J. (2015).
\newblock An empirical study on display ad impression viewability measurements.
\newblock {\em arXiv preprint arXiv:1505.05788}.

\bibitem[Zhang et~al., 2016c]{zhang2016optimal}
Zhang, W., Ren, K., and Wang, J. (2016c).
\newblock Optimal real-time bidding frameworks discussion.
\newblock {\em arXiv preprint arXiv:1602.01007}.

\bibitem[Zhang et~al., 2016d]{zhang2016feedback}
Zhang, W., Rong, Y., Wang, J., Zhu, T., and Wang, X. (2016d).
\newblock Feedback control of real-time display advertising.
\newblock In {\em Proceedings of the Ninth ACM International Conference on Web
  Search and Data Mining}, pages 407--416. ACM.

\bibitem[Zhang and Wang, 2015]{Zhang:2015}
Zhang, W. and Wang, J. (2015).
\newblock Statistical arbitrage mining for display advertising.
\newblock In {\em Proceedings of the 21th ACM SIGKDD International Conference
  on Knowledge Discovery and Data Mining}, pages 1465--1474. ACM.

\bibitem[Zhang et~al., 2014a]{Zhang:2014:ORB:2623330.2623633}
Zhang, W., Yuan, S., and Wang, J. (2014a).
\newblock Optimal real-time bidding for display advertising.
\newblock In {\em Proceedings of the 20th ACM SIGKDD international conference
  on Knowledge discovery and data mining}, pages 1077--1086. ACM.

\bibitem[Zhang et~al., 2014b]{zhang2014real}
Zhang, W., Yuan, S., Wang, J., and Shen, X. (2014b).
\newblock Real-time bidding benchmarking with ipinyou dataset.
\newblock {\em arXiv preprint arXiv:1407.7073}.

\bibitem[Zhang et~al., 2012]{zhang2012joint}
Zhang, W., Zhang, Y., Gao, B., Yu, Y., Yuan, X., and Liu, T.-Y. (2012).
\newblock Joint optimization of bid and budget allocation in sponsored search.
\newblock In {\em Proceedings of the 18th ACM SIGKDD International Conference
  on Knowledge Discovery and Data Mining}, pages 1177--1185. ACM.

\bibitem[Zhang et~al., 2016e]{zhang2016bidaware}
Zhang, W., Zhou, T., Wang, J., and Xu, J. (2016e).
\newblock Bid-aware gradient descent for unbiased learning with censored data
  in display advertising.
\newblock In {\em Proceedings of the 22nd ACM SIGKDD International Conference
  on Knowledge Discovery and Data Mining}. ACM.

\bibitem[Zhang et~al., 2014c]{zhang2014sequential}
Zhang, Y., Dai, H., Xu, C., Feng, J., Wang, T., Bian, J., Wang, B., and Liu,
  T.-Y. (2014c).
\newblock Sequential click prediction for sponsored search with recurrent
  neural networks.
\newblock In {\em Twenty-Eighth AAAI Conference on Artificial Intelligence}.

\bibitem[Zhang et~al., 2014d]{zhang2014multi}
Zhang, Y., Wei, Y., and Ren, J. (2014d).
\newblock Multi-touch attribution in online advertising with survival theory.
\newblock In {\em IEEE International Conference on Data Mining}, pages
  687--696. IEEE.

\bibitem[Zhang et~al., 2014e]{zhang2014bid}
Zhang, Y., Zhang, W., Gao, B., Yuan, X., and Liu, T.-Y. (2014e).
\newblock Bid keyword suggestion in sponsored search based on competitiveness
  and relevance.
\newblock {\em Information Processing \& Management}, 50(4):508--523.

\bibitem[Zhao et~al., 2013]{zhao2013interactive}
Zhao, X., Zhang, W., and Wang, J. (2013).
\newblock Interactive collaborative filtering.
\newblock In {\em Proceedings of the 22nd ACM International Conference on
  Information \& Knowledge Management}, pages 1411--1420. ACM.

\bibitem[Zhou et~al., 2008]{zhou2008budget}
Zhou, Y., Chakrabarty, D., and Lukose, R. (2008).
\newblock Budget constrained bidding in keyword auctions and online knapsack
  problems.
\newblock In {\em International Workshop on Internet and Network Economics},
  pages 566--576. Springer.

\bibitem[Zhu et~al., 2009a]{zhu2009risky}
Zhu, J., Wang, J., Cox, I.~J., and Taylor, M.~J. (2009a).
\newblock Risky business: modeling and exploiting uncertainty in information
  retrieval.
\newblock In {\em Proceedings of the 32nd international ACM SIGIR conference on
  Research and development in information retrieval}, pages 99--106. ACM.

\bibitem[Zhu, 2004]{zhu2004recall}
Zhu, M. (2004).
\newblock Recall, precision and average precision.
\newblock {\em Department of Statistics and Actuarial Science, University of
  Waterloo, Waterloo}, 2.

\bibitem[Zhu et~al., 2009b]{Zhu2009}
Zhu, Y., Wang, G., Yang, J., Wang, D., Yan, J., Hu, J., and Chen, Z. (2009b).
\newblock Optimizing search engine revenue in sponsored search.
\newblock In {\em Proceedings of the 32nd International ACM SIGIR Conference on
  Research and Development in Information Retrieval}, pages 588--595. ACM.

\end{thebibliography}

\end{document}